\documentclass[letterpaper]{article}

\usepackage{arxiv}

\usepackage[utf8]{inputenc}
\usepackage{float}
\usepackage{enumitem}
\usepackage[english]{babel}
\usepackage{CJKutf8}
\usepackage{amssymb}
\usepackage{amsmath,mathtools}
\usepackage{xcolor}
\usepackage{graphicx}
\usepackage{multirow}
\usepackage{overpic}
\usepackage{psfrag}
\usepackage{bm}
\usepackage{url}
\usepackage{epsfig}
\usepackage{marvosym}
\usepackage{verbatim}
\usepackage{subcaption}
\usepackage{caption}
\usepackage{hyperref}
\hypersetup{colorlinks=true,linkcolor=red,citecolor=blue}
\usepackage{ulem}
\usepackage{lmodern}
\usepackage{booktabs} 
\usepackage{moresize} 
\usepackage{siunitx}
\usepackage{lineno}
\usepackage{tikz}
\usepackage{pgfplots}
\usepgfplotslibrary{colormaps} % LATEX and plain TEX
\usetikzlibrary{pgfplots.colormaps} % LATEX and plain TEX
\pgfplotsset{compat=1.14}
\usepgfplotslibrary{fillbetween}
\usetikzlibrary{shapes, arrows, arrows.meta, calc, positioning, decorations,
decorations.pathmorphing, decorations.text, decorations.markings, fit, plotmarks}
\pgfplotsset{translate gnuplot=true}
\usepackage{etex} 								% extends memory
\usepgfplotslibrary{external} 						% converts each picture as a separate pdf
\tikzsetexternalprefix{ext/}							% put externally compiled figures in ext/	

\definecolor{myblue1}	{RGB}{0,177,234}				% deep cyan
\definecolor{myblue2}	{RGB}{76,200,239}				% lighter cyan
\definecolor{myblue3}	{RGB}{127,215,244}				% lighter cyan
\definecolor{myblue4}	{RGB}{178,231,248}				% lighter cyan
\definecolor{myblue5}	{RGB}{198,251,255}				% lighter cyan
\definecolor{mybluegray1}{RGB}{0,127,167}				% deep cyan/gray mix
\definecolor{mybluegray2}{RGB}{76,165,193}				% lighter cyan/gray mix
\definecolor{mybluegray3}{RGB}{127,191,211}				% lighter cyan/gray mix
\definecolor{mybluegray4}{RGB}{178,216,228}				% lighter cyan/gray mix
\definecolor{mygray1}	{RGB}{76,84,93}				% deep gray
\definecolor{mygray2}	{RGB}{129,135,141}				% lighter gray
\definecolor{mygray3}	{RGB}{165,169,174}				% lighter gray
\definecolor{mygray4}	{RGB}{201,203,206}				% lighter gray
\definecolor{myorange1}	{RGB}{255,126,46}				% deep orange
\definecolor{myorange2}	{RGB}{255,164,108}				% lighter orange
\definecolor{myorange3}	{RGB}{255,190,150}				% lighter orange
\definecolor{myorange4}	{RGB}{255,216,192}				% lighter orange
\definecolor{mypurple1}{RGB}{89,89,171}
\definecolor{mypurple4}{RGB}{189,189,231}

\newcommand\red[1]{\textcolor{black}{#1}}
\newcommand\rred[1]{\textcolor{black}{#1}}

\pgfplotsset{
    colormap={custom_map}{[5pt]
            rgb255(0pt)=(255,126,46);
            rgb255(500pt)=(255,190,150);
            rgb255(1000pt)=(0,177,234);
            rgb255(1500pt)=(127,215,244);
    },
}

\newcommand{\V}[1]{\textbf{#1}}
\newcommand{\GV}[1]{\boldsymbol{#1}}

\newcommand\etal{\textit{et al.}\,\,}
\renewcommand{\emph}[1]{\textit{#1}}

\setenumerate{label=\textcolor{mybluegray1}{$\bm{\diamond}$},itemsep=0.0pt,topsep=5pt}

\pgfplotsset{
	colormap/rdbur/.style={
		colormap={rdbur}{
			rgb255(0cm)=(20,46,97); 
			rgb255(1cm)=(52,100,171); 
			rgb255(2cm)=(83,146,194); 
			rgb255(3cm)=(153,197,222); 
			rgb255(4cm)=(211,229,240);
			rgb255(5cm)=(247,247,247);
			rgb255(6cm)=(250,219,200);
			rgb255(7cm)=(238,165,132);
			rgb255(8cm)=(207,94,80);
			rgb255(9cm)=(171,10,46);
			rgb255(10cm)=(99,0,32);
			}
		}
	}

\newcommand*{\shifttext}[2]{%
  \settowidth{\@tempdima}{#2}%
  \makebox[\@tempdima]{\hspace*{#1}#2}%
}
\title{Graph neural networks for laminar flow prediction around random two-dimensional shapes}

\author{
	J. Chen (\begin{CJK*}{UTF8}{gbsn}陈峻峰\end{CJK*})\\
	MINES ParisTech, CEMEF\\
	PSL - Research University\\
	06904 Sophia Antipolis, France\\
	\texttt{junfeng.chen@mines-paristech.fr}
\And
	E. Hachem\\
	MINES Paristech, CEMEF\\
	PSL - Research University\\
	06904 Sophia Antipolis, France\\
	\texttt{elie.hachem@mines-paristech.fr}
\And
	J. Viquerat\thanks{Corresponding author}\\
	MINES Paristech, CEMEF\\
	PSL - Research University\\
	06904 Sophia Antipolis, France\\
	\texttt{jonathan.viquerat@mines-paristech.fr}
}

\begin{document}
\newgeometry{left=3cm,right=3cm,top=3cm,bottom=3cm}
\maketitle

%%%%%%%%%%%%%%%%%%%%%%%%%%%
\begin{abstract}
In the recent years, the domain of fast flow field prediction has been vastly dominated by pixel-based convolutional neural networks. Yet, the recent advent of graph convolutional neural networks (GCNNs) have attracted a considerable attention in the computational fluid dynamics (CFD) community. In this contribution, we proposed a GCNN structure as a surrogate model for laminar flow prediction around two-dimensional (2D) obstacles. Unlike traditional convolution on image pixels, the graph convolution can be directly applied on body-fitted triangular meshes, hence yielding an easy coupling with CFD solvers. The proposed GCNN model is trained over a data set composed of CFD-computed laminar flows around 2,000 random 2D shapes. Accuracy levels are assessed on reconstructed velocity and pressure fields around out-of-training obstacles, and are compared with that of standard U-net architectures, especially in the boundary layer area. 
\end{abstract}

%%%%%%%%%%%%%%%%%%%%%%%%%%%

%%%%%%%%%%%%%%%%%%%%%%%%%%%
%%%%%%%%%%%%%%%%%%%%%%%%%%%
%%%%%%%%%%%%%%%%%%%%%%%%%%%
\section{Introduction}

During the last few years, the computational fluid dynamics (CFD) community has largely benefited from the fast-paced development of the machine learning (ML) field, and more specifically from that of the neural networks (NN) domain. In many cases, a data-driven surrogate model is constructed by training a neural network on a specific prediction task, in order to counter the high computational cost of CFD solvers. In this context, convolutional neural networks (CNNs) have been widely exploited. \red{Multiple studies used CNNs for steady or unsteady flow prediction, with the flow fields represented by image-like arrays \cite{Guo2016, Jin2018, Lee2019, chen2021twindecoder, Thuerey2020, Fukami2018,patil2021robust}. CNNs are also  used to directly perform lift or drag prediction when the geometric characteristics, Mach number and Reynolds number are at hand \cite{Zhang2018, Viquerat2020}. Under the framework of reduced order modeling, CNNs have been applied to low-dimensional feature extraction of fluid systems \cite{Lee2020, Bukka2020, Gonzalez2018}}. The non-linearity of CNNs brings advantages over traditional linear methods like proper orthogonal decomposition. \red{Recent research also involves the coupling of CNNs with PDE solvers. One interesting axe is to embed CNN corrections into a DNS sovler on coarse grids to obtain more competitive results with remarkable computational speedup \cite{pathak2020using, kochkov2021machine}. CNNs-based surrogate models have also been applied to inverse problems to accelerate the optimization \cite{messner2020convolutional}.}

The use of convolution in neural networks was originally proposed in the context of image processing \cite{LeCun1995-LECCNF}. In practice, CNNs can be used to extract local features from grid-based data (include images) by sliding trainable kernels over an input image and performing convolution at each pixel. Yet, the use of uniform cartesian grids represent a hindrance for CFD applications, due to their poor intrinsic geometrical representation and large associated computational cost. For external and wall-bounded flows, body-fitted or unstructured meshes are usually adopted, in order to respect the geometric characteristics and to accurately capture the physics of the boundary layer. When coupled to CNN-based neural network models, data from the meshes have to be interpolated on \textit{ad-hoc} cartesian grids, before being projected back on the mesh. Gao et al.\cite{GAO2021110079} use elliptic coordinate transformation to transform the irregular physical domain to a regular reference domain. The coordinate transformation allows the authors to use a physics-informed CNN to solve partial differential equations in the reference domain. The physical solution in the irregular domain can then be obtained through an inverse transformation. Compared to fully-connected physics-informed neural networks, the proposed method converges faster and reaches a better accuracy.

A possible alternative consists in designing efficient convolution operations on graph structures, and to apply them directly on unstructured meshes. \red{Extending convolutional neural networks to graphs is of great interest, as many CFD solvers are based on finite element/volume methods, which use unstructured meshes for discretization. In this regard, the design and use of trained graph neural network models can significantly enrich CFD workflows by being used for specific tasks during the resolution process, while still relying on the standard CFD mesh.} Generally speaking, a graph $\textbf{G}(V,E)$ is composed of a set of nodes $V$ and a set of edges $E$, representing the connectivity of nodes. A node-level/edge-level feature vector represents the state of each node/edge. The goal of a graph convolution operation is to update the feature vectors through aggregating neighboring information. Two types of graph convolution are to be distinguished, namely spectral convolutions, and spatial convolutions. By means of graph Fourier transform, spectral convolution turns convolution into product between the entire graph's feature matrix and the convolutional kernel \cite{kipf2017semisupervised}. In \cite{pmlr-v119-de-avila-belbute-peres20a}, the authors combine differentiable PDE solvers and graph convolutional networks (GCN) to perform flow prediction around 2D airfoils. An initial flow field is first obtained by CFD simulation on a coarse mesh, and is then passed to a GCN for a prediction on a refined mesh. This hybrid method provides better results than coarse CFD simulation alone, and is significantly less computationally-intensive than running a full CFD simulation on a refined mesh. However, spectral convolution techniques are based on a global Fourier transform of the entire graph, and therefore require a fixed graph structure. Hence, spectral convolution has thus limited application in real-world CFD sceneries, and are \textit{de facto} excluded from contexts including mesh adaptation.
	
Spatial convolution is more similar to traditional convolution on cartesian data: a convolution kernel operates locally on the node and its neighbours, the kernel parameters being re-used across the entire mesh. A popular implementation of this concept is the message-passing neural network (MPNN), proposed by Gilmer \etal for quantum chemistry \cite{pmlr-v70-gilmer17a}. This method divides the convolution operation into two steps: first, neighbouring nodes and edge features are aggregated into a hidden node state, after what the hidden node state is used to update the node features. Both the aggregation and update steps exploit nonlinear differentiable functions that are represented by small, shallow neural networks, which can be regarded as the convolution kernels. In \cite{ogoke2020graph}, a variant of MPNN called GraphSAGE \cite{NIPS2017_5dd9db5e} is applied to drag force prediction from given laminar velocity field around airfoils. On a dataset with 1550 different airfoils and 21 angles of attack, the graph convolution model has shown better performance than other machine learning methods. Battaglia \etal \cite{battaglia2018relational} propose an alternative method, called graph network (GN) blocks, for general multiple-step graph convolution. A GN block first passes messages from nodes to edges through an edge convolution kernel, before updating the edge features. The new edge features are then aggregated to the nodes, through sum or other permutation invariant operations, as the edge messages. A node convolution kernel then takes the old node features and edges messages to obtain the new node features. In \cite{sanchezgonzalez2020learning}, the authors use graph networks-based simulators to learn physical systems of particles, including fluids, rigid solids and deformable materials. The proposed method shows high performance in reconstructing the results of mesh-free solvers. Pfaff \etal \cite{pfaff2021learning} apply graph networks to mesh-based simulations, including incompressible flow around cylinders and compressible flow around airfoils. The proposed neural network is trained to be an accurate incremental simulator, with capability in adapting the mesh. This method also generalizes well to mesh shapes and mesh sizes which are not present in the training set. A slightly different spatial convolution considers parameterized weight functions to aggregate the neighbouring information \cite{8100059}. The features and relative position of each node pair are used to compute an edge weight through some trainable weight function, and a single weight function is applied to the entire graph inside one convolution. Since the weight function depends on node features, different aggregation patterns would be obtained at different locations of the graph. Xu \etal \cite{doi:10.1063/5.0044093} applied this method to adjoint vector modelling for 2D flow around airfoils. The proposed weight function depends on the edge length and its angle to flow's principle direction. By training a graph convolutional neural network with such weight functions, the authors get similar adjoint vectors to traditional adjoint methods. The predicted adjoint vectors are further used for shape optimization, with resulting shapes being very close to that obtained with direct adjoint methods.

In the present contribution, a graph convolutional neural network(hereafter denoted by GCNN) is implemented for laminar flow prediction around random 2D shapes on triangular meshes. In section \ref{section:dataset} the generation of the dataset used in this paper is presented. In section \ref{section:architecture}, insights about the convolution block and graph neural network architecture are provided. The training details and a global evaluation of the model are provided in section \ref{section:training}. In section \ref{section:cylinder} and \ref{section:naca}, the flow predictions around a cylinder and an airfoil are compared to baseline CFD results. In particular, the velocity profiles and surface pressure distribution are carefully studied. In order to further show the potential of our method, drag forces are computed from the boundary layer information, and compared to the CFD results in section \ref{section:drag force}. A comparison of GCNNs with standard convolutional U-nets in terms of accuracy and computational cost is proposed in section \ref{section:Unet}. Finally, conclusions and future possible developments are proposed in section \ref{section:conclusion}

%%%%%%%%%%%%%%%%%%%%%%%%%%%%%%%%%%%%%%%%%%%%%%%%%%%%%%%%%%
%%%%%%%%%%%%%%%%%%%%%%%%%%%%%%%%%%%%%%%%%%%%%%%%%%%%%%%%%%
%%%%%%%%%%%%%%%%%%%%%%%%%%%%%%%%%%%%%%%%%%%%%%%%%%%%%%%%%%
\section{Dataset construction}
\label{section:dataset}

This section provides insights on the random shape dataset generation used to train networks in the next sections. This dataset was initially used in \cite{Viquerat2020} (section 3.5), thus only the main lines are sketched here. For more details, the reader is referred to \cite{Viquerat2020}. First, we describe the steps to generate arbitrary shapes by means of connected B\'ezier curves. Then, the solving of the Navier-Stokes equations with an immersed method is presented. Finally, the velocity and pressure fields are projected onto triangular meshes of a smaller domain for the training of neural networks.

%%%%%%%%%%%%%%%%%%%%%%%%%%%%%%%%%%%%%%%%%%%%%%%%%%%%%%%%%%
%%%%%%%%%%%%%%%%%%%%%%%%%%%%%%%%%%%%%%%%%%%%%%%%%%%%%%%%%%
\subsection{Random shape generation}

\begin{figure}[p]
\centering
%%%%%%%%%%%%
\def\xaxis{3}
\def\yaxis{3}
\def\sc{1.0}
\def\pOneX{0.7}
\def\pOneY{0.25}
\def\pTwoX{-0.2}
\def\pTwoY{0.5}
\def\pThreeX{-0.7}
\def\pThreeY{-0.3}
\def\pFourX{0.4}
\def\pFourY{-0.6}
\def\rad{0.5}

%%%%%%%%%%%%
%%% Scatter points
%%%%%%%%%%%%
\begin{subfigure}{.45\textwidth}
	\centering
	\includegraphics{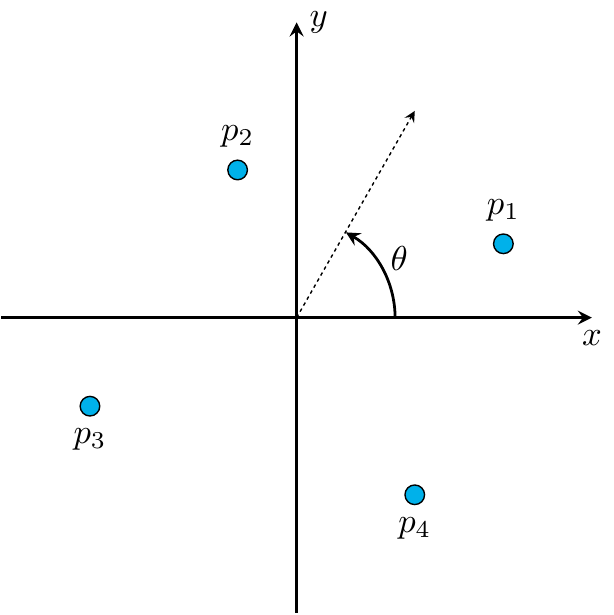}
%	\begin{tikzpicture}[	scale=\sc,
%					axes/.style={thick,->,-stealth},
%					pt/.style={circle,inner sep=0pt,text width=2mm,fill=myblue1,draw=black}]
%
%	%%% Axes 
%	\draw[axes] (-\xaxis,0) -- (\xaxis,0) node[anchor=north] {$x$};
%	\draw[axes] (0,-\yaxis) -- (0,\yaxis) node[anchor=west] {$y$};
%
%	%%% Nodes
%	\node[pt, label=above:$p_1$] 	(p1) at ( \pOneX*\xaxis, 	\pOneY*\yaxis) {};
%	\node[pt, label=above:$p_2$] 	(p2) at (\pTwoX*\xaxis, 	\pTwoY*\yaxis) {};
%	\node[pt, label=below:$p_3$] 	(p3) at (\pThreeX*\xaxis,	\pThreeY*\yaxis) {};
%	\node[pt, label=below:$p_4$] 	(p4) at (\pFourX*\xaxis,	\pFourY*\yaxis) {};
%
%	%%% Increasing angle
%	\draw[->,-stealth,dash pattern=on 1pt] (0,0) -- (0.4*\xaxis,0.7*\yaxis);
%	\draw[->,-stealth,thick] (0,0) ++(0:1.0) arc (0:60:1.0) node[pos=0.45,xshift=0.15cm,yshift=0.15cm] {$\theta$};
%
%	\end{tikzpicture}
	
	\caption{Draw $n_s$ random points, translate them around $(0,0)$ and sort them by ascending trigonometric angle}
	\label{fig:shape_generation_1}
\end{subfigure} \qquad
%%%%%%%%%%%%
%%% Handle angles
%%%%%%%%%%%%
\begin{subfigure}{.45\textwidth}
	\centering
	\includegraphics{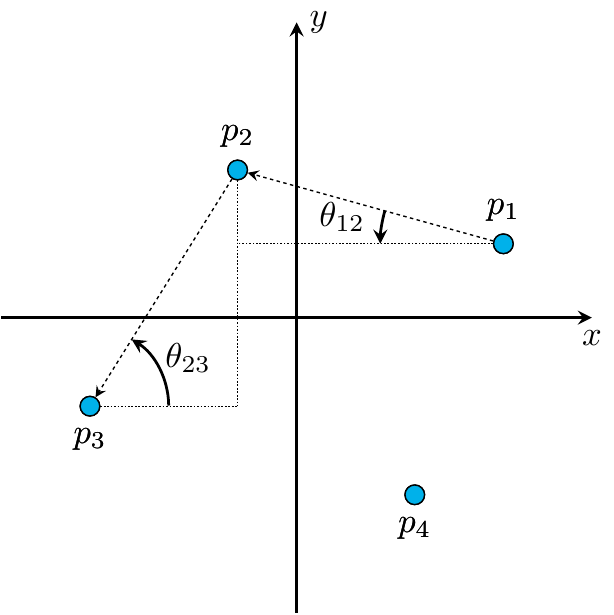}

	\caption{Compute angles between random points, and compute an average angle around each point $\theta^*_i$}
	\label{fig:shape_generation_2}
\end{subfigure}%

\medskip
\medskip
%%%%%%%%%%%%
%%% Draw cubic Bezier curve
%%%%%%%%%%%%
\begin{subfigure}{.45\textwidth}
	\centering
	\includegraphics{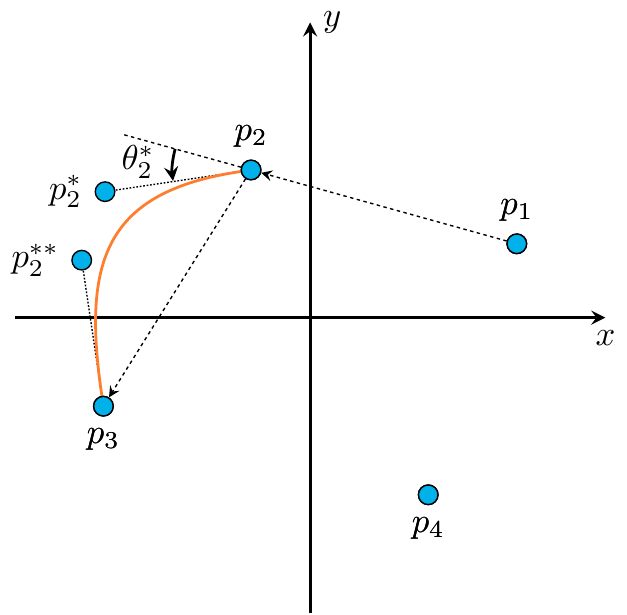}

	\caption{Compute control points coordinates from averaged angles and generate cubic B\'ezier curve}
	\label{fig:shape_generation_3}
\end{subfigure} \qquad
%%%%%%%%%%%%
%%% Draw cubic Bezier curve
%%%%%%%%%%%%
\begin{subfigure}{.45\textwidth}
	\centering
	\includegraphics{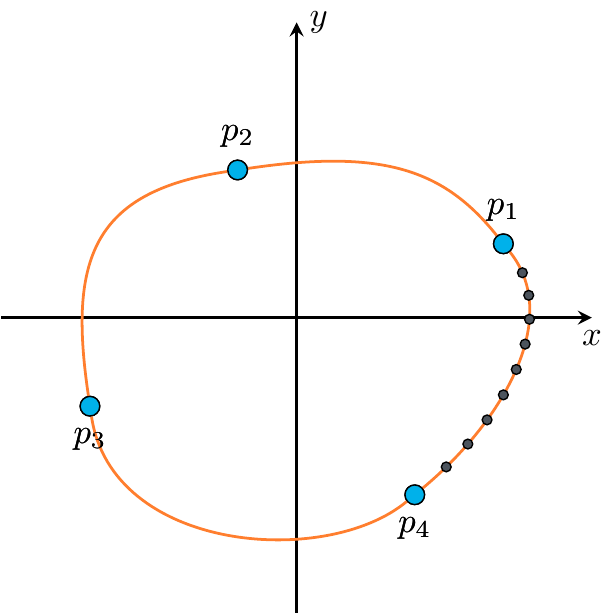}

	\caption{Sample all B\'ezier lines and export for mesh immersion}
	\label{fig:shape_generation_4}
\end{subfigure}% \qquad \qquad
%%%%%%%%%%%%
\caption{\textbf{Random shape generation with cubic B\'ezier curves}. The generation contains four steps, namely (\ref{fig:shape_generation_1}) draw random points, (\ref{fig:shape_generation_2}) compute angles, (\ref{fig:shape_generation_3}) compute control points, (\ref{fig:shape_generation_4}) sample.}
\label{fig:shape_generation}
\end{figure}

\begin{figure}
\centering
\includegraphics[width=.9\textwidth]{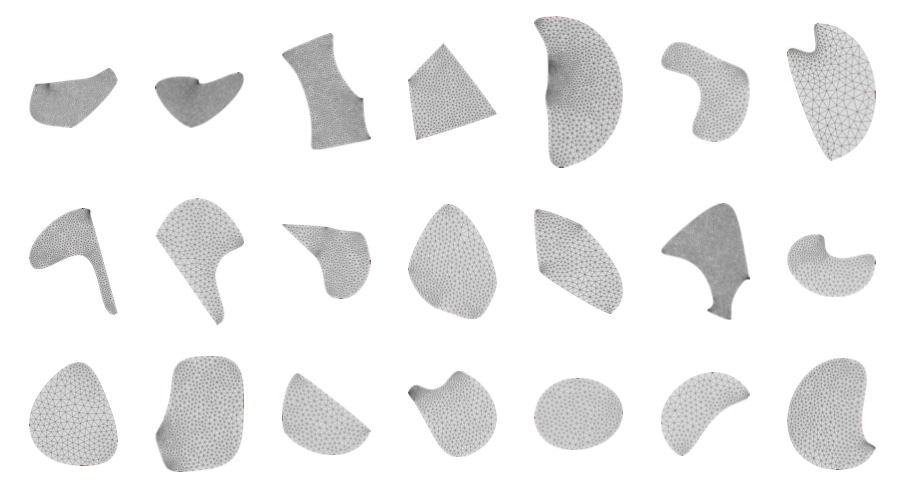} 
%%%%%%%%%%%%
%\def\w{1.5cm}
%\def\arraystretch{4.5}
%\setlength\tabcolsep{9pt}
%\newcolumntype{C}{ >{\centering\arraybackslash} m{1.5cm} }
%
%\makebox[\linewidth]{%
%	\begin{tabular}{CCCCCCC}
%		\includegraphics[width=\w]{shape_1.jpg} 	& 
%		\includegraphics[width=\w]{shape_2.jpg}	& 
%		\includegraphics[width=\w]{shape_3.jpg}	&
%		\includegraphics[width=\w]{shape_4.jpg}	&
%		\includegraphics[width=\w]{shape_5.jpg}	&
%		\includegraphics[width=\w]{shape_6.jpg}	&
%		\includegraphics[width=\w]{shape_8.jpg}\\
%		\includegraphics[width=\w]{shape_9.jpg} 	& 
%		\includegraphics[width=\w]{shape_10.jpg}	&
%		\includegraphics[width=\w]{shape_11.jpg}	&
%		\includegraphics[width=\w]{shape_12.jpg}	&
%		\includegraphics[width=\w]{shape_13.jpg}	&
%		\includegraphics[width=\w]{shape_14.jpg}	&
%		\includegraphics[width=\w]{shape_15.jpg}\\
%		\includegraphics[width=\w]{shape_16.jpg} 	& 
%		\includegraphics[width=\w]{shape_17.jpg}	&
%		\includegraphics[width=\w]{shape_18.jpg}	&
%		\includegraphics[width=\w]{shape_19.jpg}	&
%		\includegraphics[width=\w]{shape_20.jpg}	&
%		\includegraphics[width=\w]{shape_21.jpg}	&
%		\includegraphics[width=\w]{shape_22.jpg}%
%\end{tabular}}

\caption{\textbf{Shape examples drawn from the dataset.} A wide variety of shape is obtained using a restrained number of points ($n_s \in \left[ 4, 6 \right]$), as well as a local curvature $r$ and averaging parameter $\alpha$.}
\label{fig:shape_examples}
\end{figure}

In the first step, $n_s$ random points are drawn in $\left[ 0, 1 \right]^2$, and translated so their center of mass is positioned in $(0,0)$. An ascending trigonometric angle sort is then performed (see figure \ref{fig:shape_generation_1}), and the angles between consecutive random points are then computed. An average angle is then computed around each point (see figure \ref{fig:shape_generation_2}) using:

\begin{equation*}
	\theta^*_i = \alpha \theta_{i-1,i} + (1 - \alpha) \theta_{i,i+1},
\end{equation*}

\noindent with $\alpha \in \left[0,1\right]$. The averaging parameter $\alpha$ allows to alter the sharpness of the curve locally, maximum smoothness being obtained for $\alpha = 0.5$. Then, each pair of points is joined using a cubic B\'ezier curve, defined by four points: the first and last points, $p_i$ and $p_{i+1}$, are part of the curve, while the second and third ones, $p^*_i$ and $p^{**}_i$, are control points that define the tangent of the curve at $p_i$ and $p_{i+1}$. The tangents at $p_i$ and $p_{i+1}$ are respectively controlled by $\theta^*_i$ and $\theta^*_{i+1}$ (see figure \ref{fig:shape_generation_3}). A final sampling of the successive B\'ezier curves leads to a boundary description of the shape (figure \ref{fig:shape_generation_4}). Using this method, a wide variety of shapes can be attained, as shown in figure \ref{fig:shape_examples}.

%%%%%%%%%%%%
%%%%%%%%%%%%

%%%%%%%%%%%%
%%%%%%%%%%%%

%%%%%%%%%%%%%%%%%%%%%%%%%%%%%%%%%%%%%%%%%%%%%%%%%%%%%%%%%%
%%%%%%%%%%%%%%%%%%%%%%%%%%%%%%%%%%%%%%%%%%%%%%%%%%%%%%%%%%
\subsection{Numerical resolution of Navier-Stokes equations}
\label{section:NS}

The flow motion of incompressible newtonian fluids is described by the Navier-Stokes (NS) equations: 

\begin{equation} \label{eq:ns_equation1}
	\left\{
	\begin{aligned}
		\rho\ (\partial_{t} \V{v} + \V{v} \cdot \nabla \V{v}) -\nabla \cdot \left( 2 \eta \GV{\epsilon}(\V{v}) - p \V{I} \right)  & = \V{f}, \\
		\nabla \cdot \V{v} &= 0,
	\end{aligned}
	\right.
\end{equation}

\noindent where $t \in [0,T]$ is the time, $\V{v}(x,t)$ the velocity, $p(x,t)$ the pressure, $\rho$ the fluid density, $\eta$ the dynamic viscosity and $\V{I}$ the identity tensor. In order to efficiently construct the dataset, an immersed boundary method is used for resolution instead of the usual body-fitted method, avoiding a systematic re-meshing of the whole domain for each shape. This method rely on a unified fluid-solid eulerian formulation based on level-set description of the geometry \cite{Bruchon2009}, and leads to the following set of modified equations: 

\begin{equation} \label{eq:ns_equation2}
	\left\{
	\begin{aligned}
		\rho^* (\partial_{t} \V{v} + \V{v} \cdot \nabla \V{v}) -\nabla \cdot \left( 2 \eta \GV{\epsilon}(\V{v}) + \GV{\tau} - p \V{I} \right)  & = \V{f}, \\
		\nabla \cdot \V{v} &= 0,
	\end{aligned}
	\right.
\end{equation}

\noindent where we have introduced the following mixed quantities:

\begin{equation*}
	\begin{aligned}
		\GV{\tau} & = H(\alpha) \GV{\tau}_{\text{s}},\\
		\rho^* & = H(\alpha) \rho_{\text{s}} + (1-H(\alpha)) \rho_{\text{f}},
	\end{aligned}
\end{equation*}

\noindent where the subscripts $f$ and $s$ respectively refer to the fluid and the solid, and $H(\alpha)$ is the Heaviside function:

\begin{equation} \label{eq:heavyside31}
	H(\alpha) = \left\{
	\begin{aligned}
		1 & \text{ if}\ \alpha > 0,\\
		0 & \text{ if}\ \alpha < 0.
	\end{aligned}
	\right.
\end{equation}

\noindent The reader is referred to \cite{Hachem2013} for additional details about formulation (\ref{eq:ns_equation2}). 
\red{The shapes are placed at the origin of a rectangular domain $[-5,10]\times[-5,5]$ meters. The inlet velocity is \num{1}m/s. With density $\rho=1\text{kg}/\text{m}^3$ and kinematic viscosity $\nu=0.1\text{m}^2/\text{s}$, the resulting Reynolds number is \num{10}. At the top and bottom boundaries, $v_y=0$ and $\frac{\partial v_x}{\partial y}=0$ conditions are applied, while at the outlet we apply $\frac{\partial v_x}{\partial x}=\frac{\partial v_y}{\partial x}=0$ and $p=0$. Finally, no-slip condition $v_x=v_y=0$ is imposed on the solid interface. The mesh size around the interface is set at \num{0.01}m.} Eventually, the modified equations (\ref{eq:ns_equation2}) are cast into a stabilized finite element formulation, and solved using a variational multi-scale (VMS) solver \cite{Tezduyar1992, Bazilevs2007, Takizawa2018, Otoguro2019, Otoguro2020}.

%%%%%%%%%%%%%%%%%%%%%%%%%%%%%%%%%%%%%%%%%%%%%%%%%%%%%%%%%%
%%%%%%%%%%%%%%%%%%%%%%%%%%%%%%%%%%%%%%%%%%%%%%%%%%%%%%%%%%
\subsection{Dataset for inference}

\begin{figure}
\centering

\setlength{\fboxsep}{0pt}%
\setlength{\fboxrule}{1pt}%

\begin{subfigure}[t]{.25\textwidth}
	\centering%
	\fbox{\includegraphics[width=.85\linewidth]{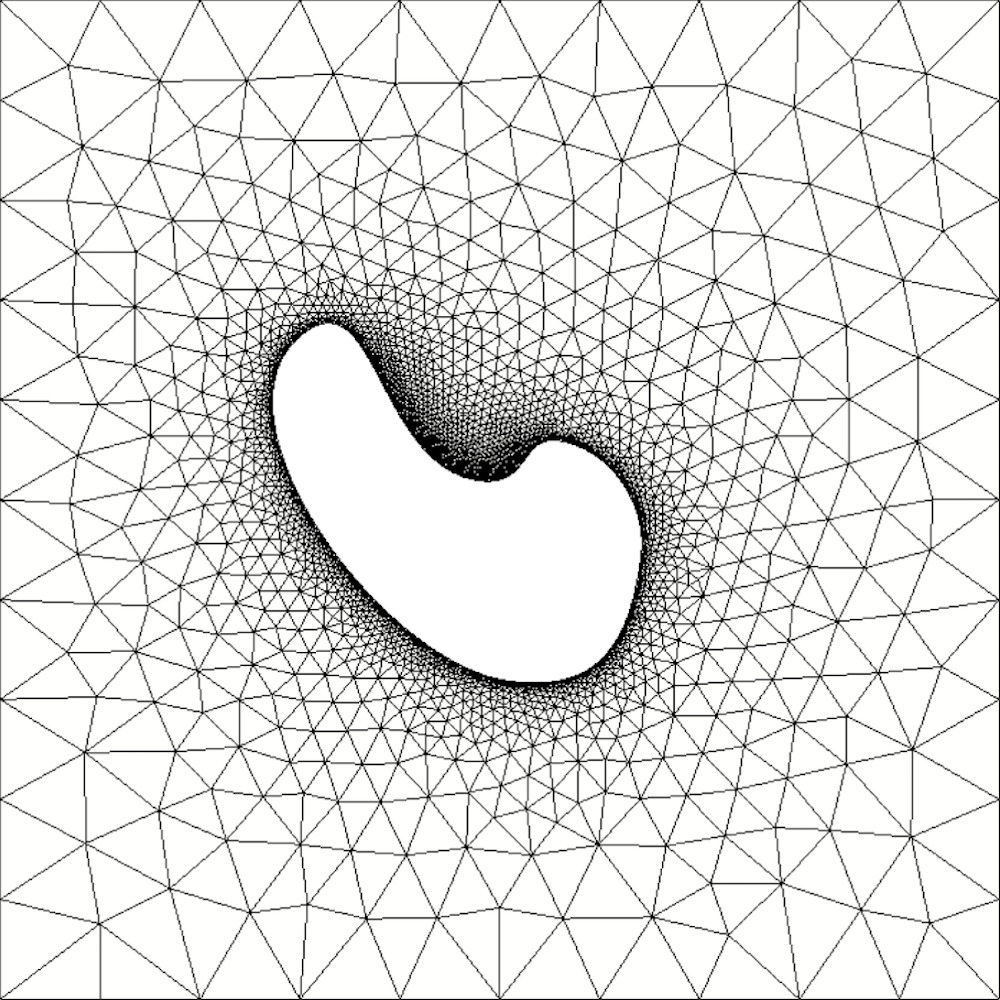}}%
	\caption{Network input}%
	\label{fig:mesh_example}
\end{subfigure} \quad
\begin{subfigure}[t]{.25\textwidth}
	\centering%
	\fbox{\includegraphics[angle=90, width=.85\linewidth]{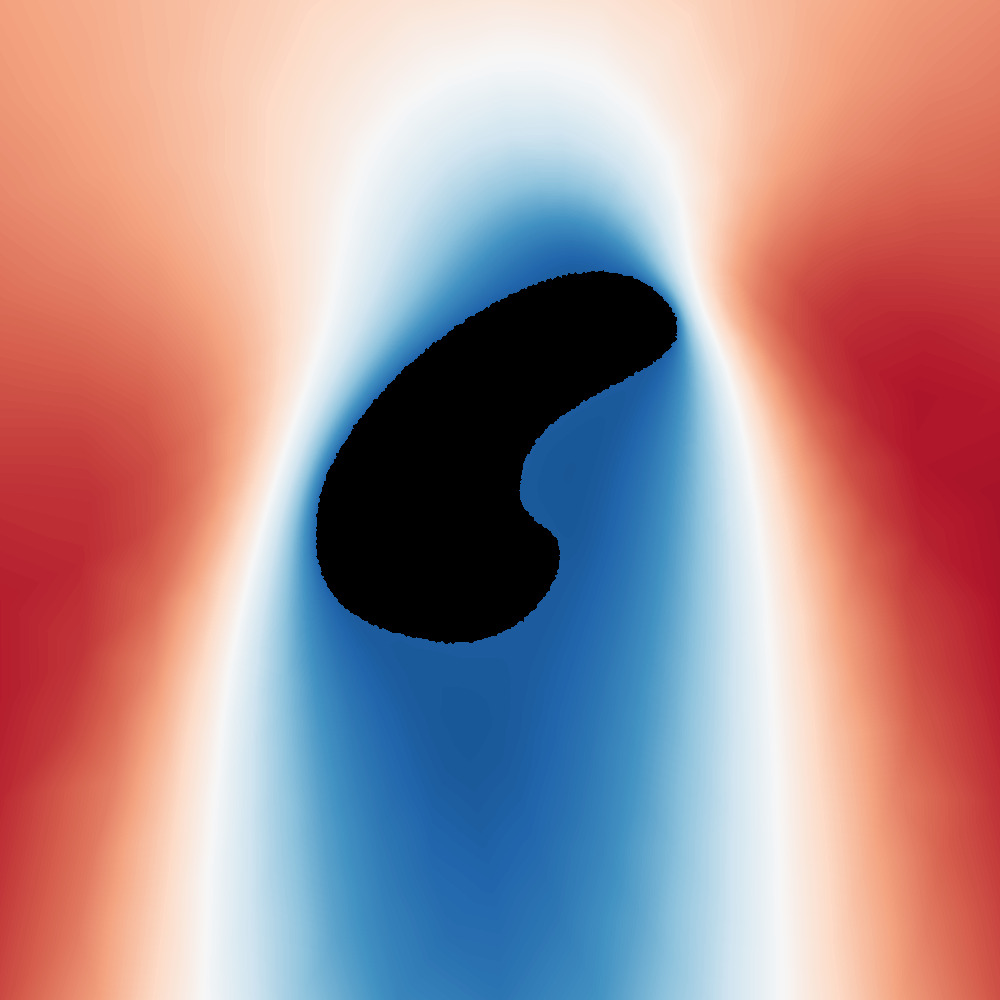}}%
	\caption{Velocity field - $x$ component}%
	\label{fig:u_example}
\end{subfigure}
\begin{subfigure}[t]{.1\textwidth}
	\centering%
	\raisebox{-2.5mm}{\includegraphics{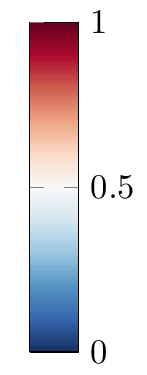}}
	%\raisebox{-2.5mm}{\colorbarc{0}{1}{3.35}}%
\end{subfigure}

\medskip
\medskip

\begin{subfigure}[t]{.25\textwidth}
	\centering
	\fbox{\includegraphics[angle=90, width=.85\linewidth]{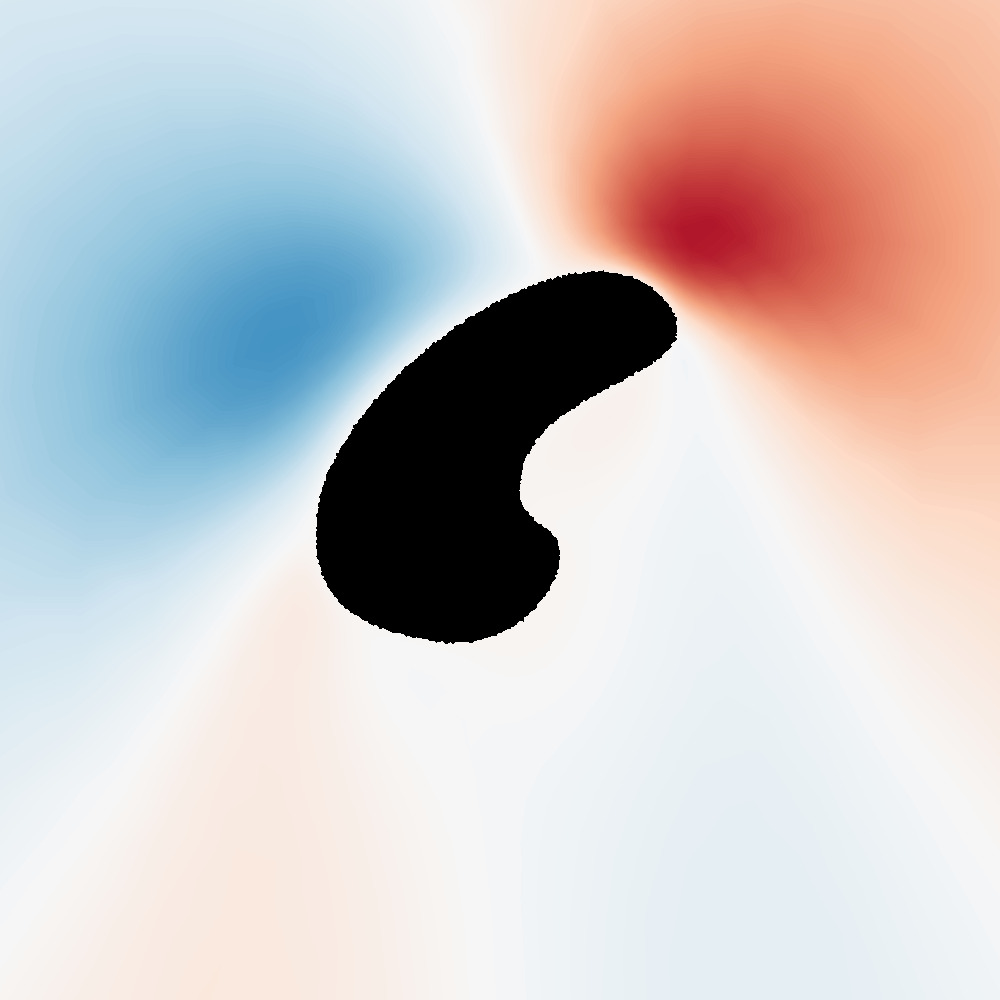}}
	\caption{Velocity field - $y$ component}
	\label{fig:v_example}
\end{subfigure} \quad
\begin{subfigure}[t]{.25\textwidth}
	\centering
	\fbox{\includegraphics[angle=90, width=.85\linewidth]{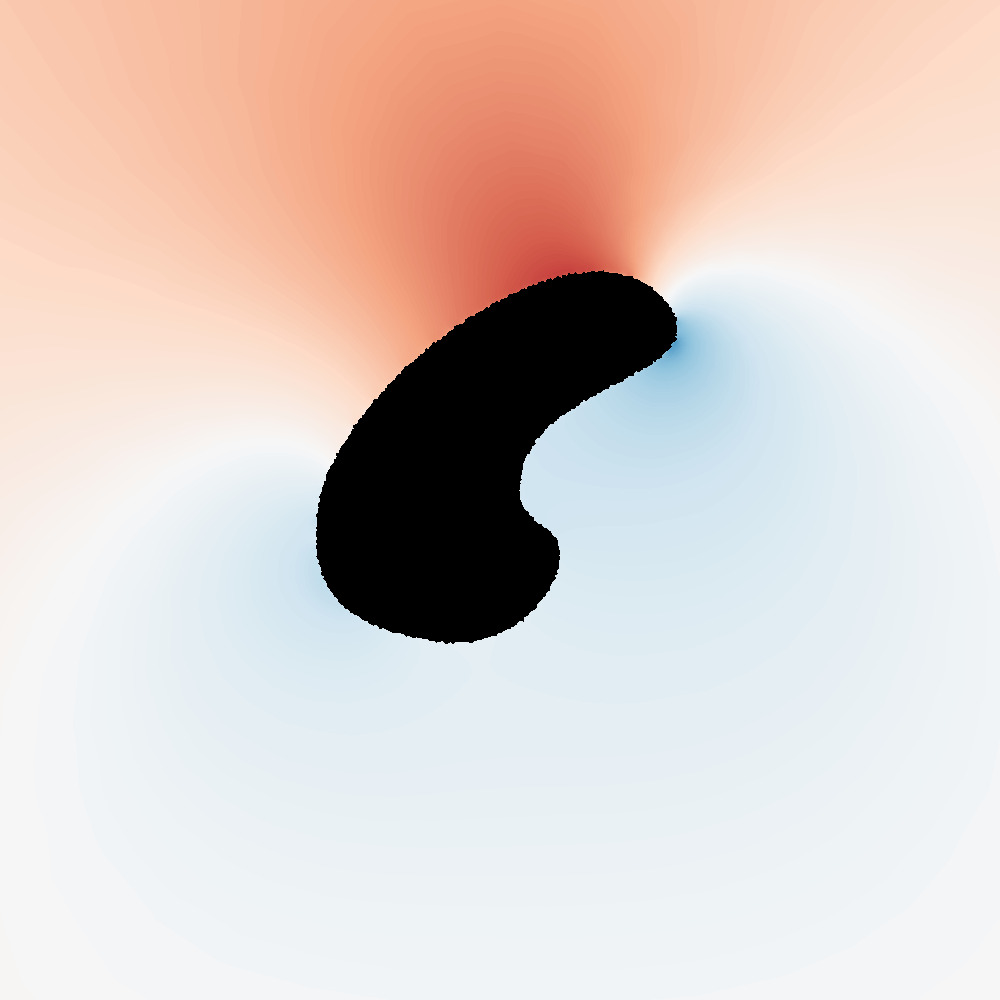}}
	\caption{Pressure field}
	\label{fig:p_example}
\end{subfigure}
\begin{subfigure}[t]{.1\textwidth}
	\centering
	\raisebox{-2.5mm}{\includegraphics{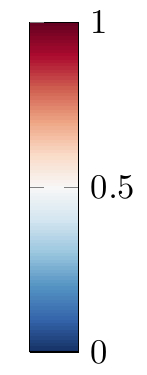}}
	%\raisebox{-2.5mm}{\colorbarc{0}{1}{3.35}}
\end{subfigure}

\caption{\textbf{Triangular mesh, velocity field and pressure field for a dataset element.} Body-fitted triangular meshes in a smaller domain (\ref{fig:mesh_example}) are used for training the neural network, along with the interpolated velocity field (\ref{fig:u_example}, \ref{fig:v_example}) and pressure field (\ref{fig:p_example}).}
\label{fig:dataset_example}
\end{figure}

The dataset\footnote{\rred{{This dataset was already exploited in previous studies, see \cite{Viquerat2020, chen2021twindecoder}.}}} is composed of 2.000 shapes, along with their steady-state velocity and pressure fields at $Re=10$. All the labels were computed following the methods exposed in section \ref{section:NS} using CimLib \cite{Hachem2013}, \rred{an industrial-level CFD solver}. As we are more concerned with the flow patterns near the obstacles, a smaller domain containing the obstacle is used for the neural networks's inference task. For each shape, a body-fitted triangular mesh is generated for the domain $[-2,2]\text{m}\times [-2,2]\text{m}$. The mesh size is $0.01\text{m}$ on the obstacle's surface, which is the same as the immersed meshes used for CFD simulation. The element size grows to $0.4\text{m}$ at the border of the domain. Depending on the complexity of the shapes, the number of nodes varies from $2,000$ to $3,000$ in such a mesh. The velocity and pressure fields from CFD simulations are projected onto these meshes through linear interpolation (see figure \ref{fig:dataset_example}). \rred{For the learning phase,} normalization over the entire data set is applied, mapping separately the two coordinates components, the two velocity components and the pressure into the $[0,1]$ range. \rred{at the end of the prediction phase, the predicted fields are mapped back to their physical ranges, and non-dimensionalized by the inlet velocity and fluid density, both being equal to one.} For additional details about the distribution of the elements in the dataset, the reader is referred to \cite{Viquerat2020} (section $3.5$). The whole data set is split into a training set with $1600$ shapes, a validation set and a test set both with $200$ shapes.

%%%%%%%%%%%%
%%%%%%%%%%%%

%%%%%%%%%%%%%%%%%%%%%%%%%%%%%%%%%%%%%%%%%%%%%%%%%%%%%%%%%%
%%%%%%%%%%%%%%%%%%%%%%%%%%%%%%%%%%%%%%%%%%%%%%%%%%%%%%%%%%
%%%%%%%%%%%%%%%%%%%%%%%%%%%%%%%%%%%%%%%%%%%%%%%%%%%%%%%%%%
\section{\red{Network architecture}}
\label{section:architecture}

%%%%%%%%%%%%%%%%%%%%%%%%%%%%%%%%%%%%%%%%%%%%%%%%%%%%%%%%%%
%%%%%%%%%%%%%%%%%%%%%%%%%%%%%%%%%%%%%%%%%%%%%%%%%%%%%%%%%%
\subsection{Convolution block}
\label{section:graph convolution}

The basic component of the graph convolutional neural network is the convolution block. Here, the proposed convolution block has two components: a two-step graph convolutional layer adapted from the literature \cite{battaglia2018relational,sanchezgonzalez2020learning,pfaff2021learning}, followed by a two-step smoothing layer. In this section, the feature matrices on the nodes and the edges are respectively denoted by $\textbf{X}_V\in \textbf{R}^{N_V\times d_V}$ and $\textbf{X}_E\in \textbf{R}^{N_E\times d_E}$, with $N_V$ and $N_E$ being the number of nodes and edges. $d_V$ and $d_E$ are the dimensions of feature vectors on nodes and edges, which are denoted by $\textbf{x}_v\in \textbf{R}^{d_V}$ and $\textbf{x}_e\in \textbf{R}^{d_E}$.

\begin{figure}
\centering
\includegraphics{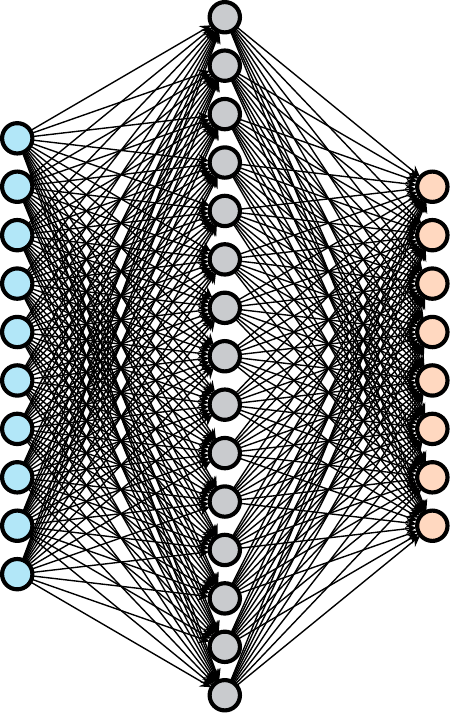}
%%%
%\def\sc{0.4}
%\begin{tikzpicture}[	scale=\sc,
%				arrow/.style=		{thick,color=mybluegray1,rounded corners},
%				netnode/.style=		{scale=\sc,circle, inner sep=0pt, text width=22pt, align=center, very thick},
%				inputnode/.style=	{netnode, fill=myblue4, draw=black},
%				hiddennode/.style=	{netnode, fill=mygray4, draw=black},
%				outputnode/.style=	{netnode, fill=myorange4, 	draw=black},
%				signal/.style=		{arrows={-stealth},draw=black}]
%	\def\nodedist{35pt}
%	\def\layerdist{150pt}
%    
%	\foreach \y in {1,...,10}
%		\node[inputnode] (I\y) at (0,-\y*\nodedist-2*\nodedist) {};  
%	\foreach \y in {1,...,15}
%		\node[hiddennode] (H1\y) at ($(\layerdist,-\y*\nodedist) +(0, 0.5*\nodedist)$) {};
%	\foreach \y in {1,...,8}
%		\node[outputnode] (O\y) at ($0.5*(H11) + 0.5*(H12) + (\layerdist, -\y*\nodedist-2*\nodedist)$) {};
%
%	\foreach \dest in {1,...,15}
%		\foreach \source in {1,...,10}
%			\draw[signal] (I\source) -- (H1\dest);
%	\foreach \dest in {1,...,8}
%		\foreach \source in {1,...,15}
%			\draw[signal] (H1\source) edge (O\dest);
%\end{tikzpicture}
%%%
\caption{\textbf{Fully-connected feedforward neural network used as convolution kernel.} The vectors on the right hand side in equation (\ref{equation:graph_convolution}) are concatenated as the inputs of the convolution kernel. In edge convolution, the input dimension is $2d_V+d_E$, while in node convolution, the input dimension is $d_V+d'_E$. The number of hidden nodes is always 128.}
\label{fig:mlp}
\end{figure}

\paragraph{Convolutional layer} The convolutional layer propagates node-level messages to edges, then aggregates the new edge features and updates the node features according to the following rules:

\begin{equation}
	\begin{aligned}
		\textbf{x}'_e &= f_e \left( \frac{\textbf{x}_{v_1} + \textbf{x}_{v_2}}{2}, \frac{| \textbf{x}_{v_1} - \textbf{x}_{v_2} |}{2}, \textbf{x}_e \right),\\
		\textbf{x}'_v &= f_v \left( \textbf{x}_v, \sum_{e_i\in N(v)}{\textbf{x}'_{e_i}} \right),
	\end{aligned}
\label{equation:graph_convolution}
\end{equation}

where $v_1$ and $v_2$ are the two nodes connected by edge $e$, and $N(v)$ is the set of neighbouring edges around node $v$. For both convolution kernels $f_e$ and $f_v$, a fully-connected feedforward neural network with one hidden layer is used (see figure \ref{fig:mlp}), the number of neurons in the hidden layer being set to $128$. The number of neurons in the output layer remains modifiable, so the output dimension of the kernel can be customized. All the neurons are activated using the swish function\cite{ramachandran2018}, which performs better than ReLU and hyperbolic tangent functions. The weights and bias of the kernels are initialized using the Glorot-normal distribution \cite{pmlr-v9-glorot10a}.

%%%%%%%%%%%%
%%%%%%%%%%%%

During the edge convolution step, symmetric node features $\frac{\textbf{x}_{v_1} + \textbf{x}_{v_2}}{2}$ and $\frac{| \textbf{x}_{v_1} - \textbf{x}_{v_2} |}{2}$ are preferred over the raw features $\textbf{x}_{v_1}$ and $\textbf{x}_{v_2}$ in order to preserve permutation invariance. The summation in the node convolution step is also invariant to permutation of neighbouring edges. As a consequence, the proposed convolution steps (\ref{equation:graph_convolution}) are insensitive to nodes and edges indexing.

\paragraph{Smoothing layer} The smoothing layer perform an average operation on the output graph. The averaging kernel implemented on triangular meshes is divided into two steps:

\begin{equation}
	\begin{aligned}
		\textbf{x}''_e &= \frac{\textbf{x}'_{v1} + \textbf{x}'_{v2} }{2},\\
		\textbf{x}''_v &= \frac{1}{\text{card}(N(v))}\sum_{e_i\in N(v)}{\textbf{x}''_{e_i}}
	\end{aligned}
\label{equation:graph_smoothing}
\end{equation}

The smoothing layer is technically also a convolutional layer. The motivation of adding this layer is not for message propagation, but to reduce the spatial variability of the node features. It weighs down the feature map of a convolutional layer through neighbouring compensation. A sequence of convolution-smoothing blocks can hence regularize the output of the networks to reduce local minima. The smoothing layer is beneficial to further reducing the training loss and producing continuous flow fields. A similar idea is found in the regularized pooling operation used for convolutional layers on images \cite{10.1007/978-3-030-61616-8_20}. The final output of a convolution block is the new edge feature matrix $\textbf{X}'_E \in \textbf{R}^{N_E\times d'_E}$ and the smoothed new node feature matrix $\textbf{X}''_V\in \textbf{R}^{N_V\times d'_V}$

%%%%%%%%%%%%%%%%%%%%%%%%%%%%%%%%%%%%%%%%%%%%%%%%%%%%%%%%%%
%%%%%%%%%%%%%%%%%%%%%%%%%%%%%%%%%%%%%%%%%%%%%%%%%%%%%%%%%%
\subsection{Network architecture}
\label{section:network_architecture}

\begin{figure}
\centering
\includegraphics{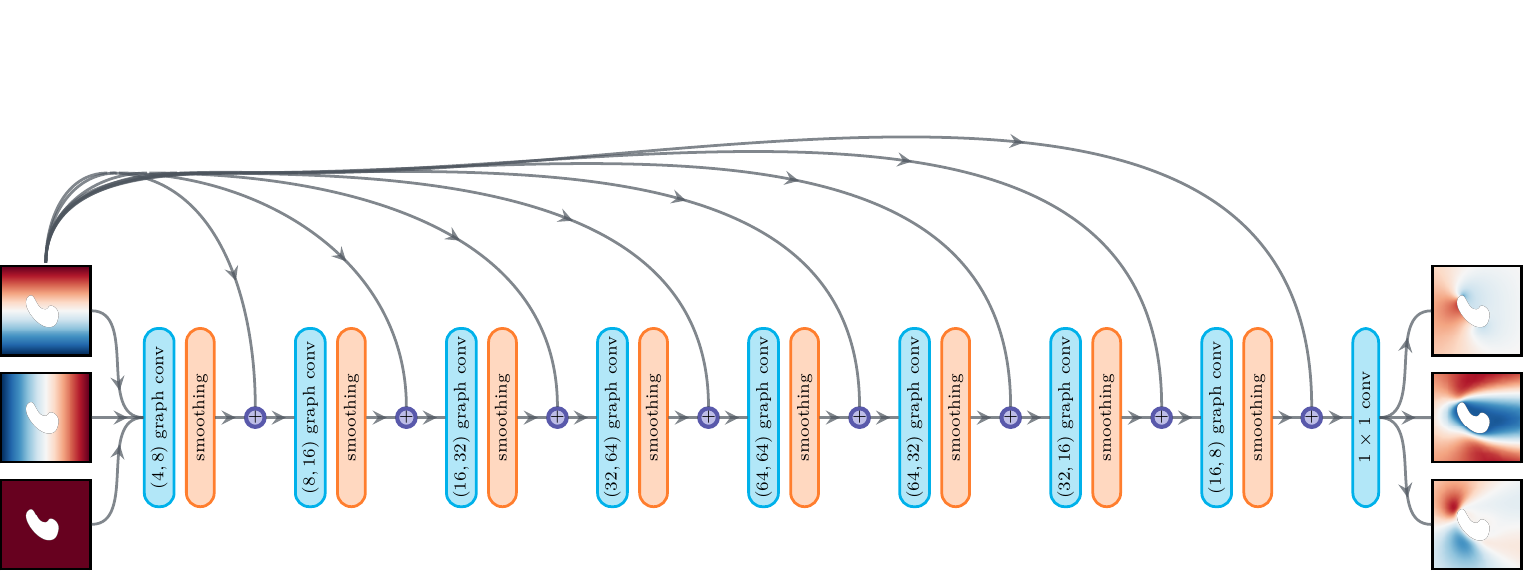}
\caption{\textbf{Proposed network architecture.} The provided input is composed of three images respectively representing (i) the obstacle boundary, (ii) the $x$ coordinates and (iii) the $y$ coordinates. The network has an encoder-decoder structure: in the encoder, the amount of features is doubled after each smoothing step, while in the decoder, it is halved after each smoothing step. For each graph convolution layer, the number of edge and node features are indicated in parenthesis. The network is terminated with a $1\times 1$ convolutional layer to obtain the three output layers, holding respectively the $u$ and $v$ components of the velocity, and the pressure field $p$. The skip connections allow to inject coordinate informations from the input layer at each edge convolution step of the architecture.}
\label{fig:architecture}
\end{figure}

The proposed architecture is presented in figure \ref{fig:architecture}. As can be seen, the input is composed of three images respectively representing (i) the obstacle boundary, (ii) the $x$ coordinates and (iii) the $y$ coordinates. Eight convolution blocks/smoothing layers are stacked to form the graph convolutional neural network, followed by a $1\times 1$ convolution as the output layer. Unless stated otherwise, the intermediate edge and node features dimensions in each graph convolution layer are respectively $(4,8,16,32,64,64,32,16)$ and $(8,16,32,64,64,32,16,8)$, as indicated in parenthesis in the graph convolution layers of figure \ref{fig:architecture}, but a study of the impact of the features dimensions can be found in section \ref{section:Unet}. The final $1\times 1$ convolutional layer transforms node features in $\textbf{R}^8$ into vectors in $\textbf{R}^3$, which represent the predicted velocity and pressure. This symmetric architecture is similar to the encoder-decoder structure used in \cite{chen2021twindecoder}.

An important component of the proposed architecture is the skip connections from the input graph to the convolutional blocks. The coordinates of the nodes are concatenated to the node features after each smoothing layer. These skip connections provide spatial informations to the edge convolution steps in equation (\ref{equation:graph_convolution}). The edge convolution hence takes into account the position of the edge center and relative position of the node pairs, allowing the convolution kernel to display different patterns at different locations in the domain.

%%%%%%%%%%%%
%%%%%%%%%%%%

%%%%%%%%%%%%%%%%%%%%%%%%%%%%%%%%%%%%%%%%%%%%%%%%%%%%%%%%%%
%%%%%%%%%%%%%%%%%%%%%%%%%%%%%%%%%%%%%%%%%%%%%%%%%%%%%%%%%%
%%%%%%%%%%%%%%%%%%%%%%%%%%%%%%%%%%%%%%%%%%%%%%%%%%%%%%%%%%
\section{Results}
\label{section:results}

%%%%%%%%%%%%%%%%%%%%%%%%%%%%%%%%%%%%%%%%%%%%%%%%%%%%%%%%%%
%%%%%%%%%%%%%%%%%%%%%%%%%%%%%%%%%%%%%%%%%%%%%%%%%%%%%%%%%%
\subsection{Training history}
\label{section:training}

With architecture described in section \ref{section:architecture}, the proposed network has $217,853$ trainable parameters. The parameters are calibrated by minimizing a loss function defined on the predicted and the reference flow fields. The mean absolute error (MAE) is adopted as the loss function, and is defined as:

\begin{equation*}
	\mathcal{L} = \frac{1}{N_V}\sum_{i=1}^{N_V} \left( \left| u_i^\text{pred} - u_i^\text{ref} \right| + \left| v_i^\text{pred} - v_i^\text{ref} \right| + \left| p_i^\text{pred} - p_i^\text{ref} \right| \right),
\end{equation*}

with the considered fields being normalized to $[0,1]$, as stated earlier. An in-house TensorFlow implementation is used for the network architecture and training. The network is trained using the Adam optimizer \cite{kingma2017adam} for at most \num{1000} epochs, with mini-batches of size \num{64}\footnote{One mini-batch is a large graph containing \num{64} disjoint subgraphs. For more details please refer to the code page \url{https://github.com/jviquerat/gnn_laminar_flow}.} to limit the required computational resources. The learning rate follows a pre-defined decay schedule described as:

\begin{equation}
\label{equation:decay}
	l_r^e = \frac{l_r^0}{1 + \gamma e},
\end{equation}

where $l_r^e$ is the learning rate used during epoch number $e$, and $\gamma$ is the decay rate. In the experiments, both the initial learning rate and the decay rate were set equal to \num{0.002}. To prevent overfitting, an early stopping criterion is applied: the training is terminated if the average MAE on the validation set is not improved in 60 epochs. 

\begin{figure}
\centering
\includegraphics{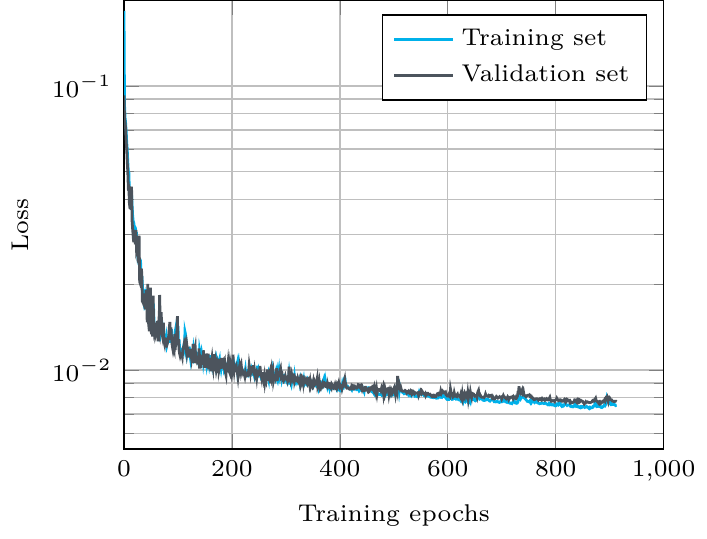}
%\begin{tikzpicture}[trim axis left, trim axis right]
%	\begin{axis}[	scale=0.8,transform shape,
%				label style={font=\scriptsize}, tick label style={font=\scriptsize}, legend style={font=\scriptsize},
%				ymin=0.0, ymax=0.20, xmin=0, xmax=1000,
%				xlabel=Training epochs,ylabel=Loss,ymode=log,
%				legend pos=north east,
%				legend cell align={left},
%				clip=true, grid=both
%				]
%		\legend{Training set, Validation set}
%		\addplot[mark=none,thick,myblue1] table[x index=0,y index=1] {smoothing.csv};
%		\addplot[mark=none,thick,mygray1] table[x index=0,y index=2] {smoothing.csv};
%	\end{axis}	
%\end{tikzpicture}
\caption{\textbf{Training history of the model.} In fewer than \num{1000} training epochs, the training loss and validation converge. No obvious overfitting is observed.}
\label{fig:training_history}
\end{figure}

\begin{figure}
\centering
\includegraphics{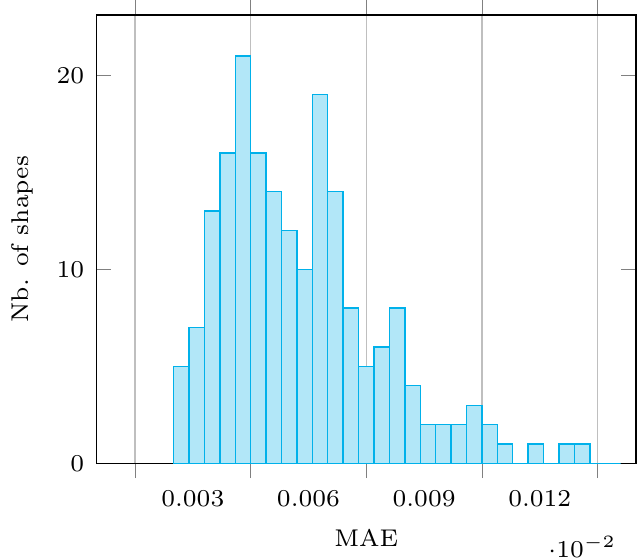}
%\begin{tikzpicture}[trim axis left, trim axis right]
%	\begin{axis}[	scale=0.8,transform shape,
%				label style={font=\scriptsize}, tick label style={font=\scriptsize}, legend style={font=\scriptsize},
%				xmin=0.002, xmax=0.016, ymin=0, ybar interval, 
%				xtick={0,0.003 ,0.006,0.009,0.012,0.015},
%				xticklabels={0,0.003 ,0.006,0.009,0.012,0.015},
%				xlabel=MAE,ylabel= Nb. of shapes]
%	\addplot[draw=myblue1,fill=myblue4,opacity=1.0] table [x index=0, y index=1] {histogram.csv};
%	\end{axis}
%\end{tikzpicture}
\caption{\textbf{Distribution of the MAE over the test set.} The test set contains \num{200} random B\'ezier shapes. Only $13\%$ shapes are with MAE over \num{0.01}.}
\label{fig:mae}
\end{figure}

After \num{914} training epochs\footnote{This error level can be attained by following the weight initialization method in the published codes.}, the training and validation losses converge respectively to \num{7.43e-3} and \num{7.61e-3}, while the MAE over the test set amounts to \num{7.70e-3}, as shown in figure \ref{fig:training_history}. From the MAE distribution over the test set, shown in figure \ref{fig:mae}, it appears that most elements in the test set are accurately predicted, with only a negligible amount of shapes exceeding a MAE of \num{0.01}. This underlines the capabilities of the model to generalize on out-of-training data. Under the training settings described in section \ref{section:training}, the proposed model requires approximately \num{11.96} seconds per epoch on a Tesla V100 GPU card.
%%%%%%%%%%%%
%%%%%%%%%%%%

%%%%%%%%%%%%
%%%%%%%%%%%%

%%%%%%%%%%%%%%%%%%%%%%%%%%%%%%%%%%%%%%%%%%%%%%%%%%%%%%%%%%
%%%%%%%%%%%%%%%%%%%%%%%%%%%%%%%%%%%%%%%%%%%%%%%%%%%%%%%%%%
\subsection{Field prediction on a cylinder}
\label{section:cylinder}

\begin{figure}[p]
\centering
\begin{subfigure}[t]{.25\linewidth}
	\centering
	\fbox{\includegraphics[width=.8\linewidth]{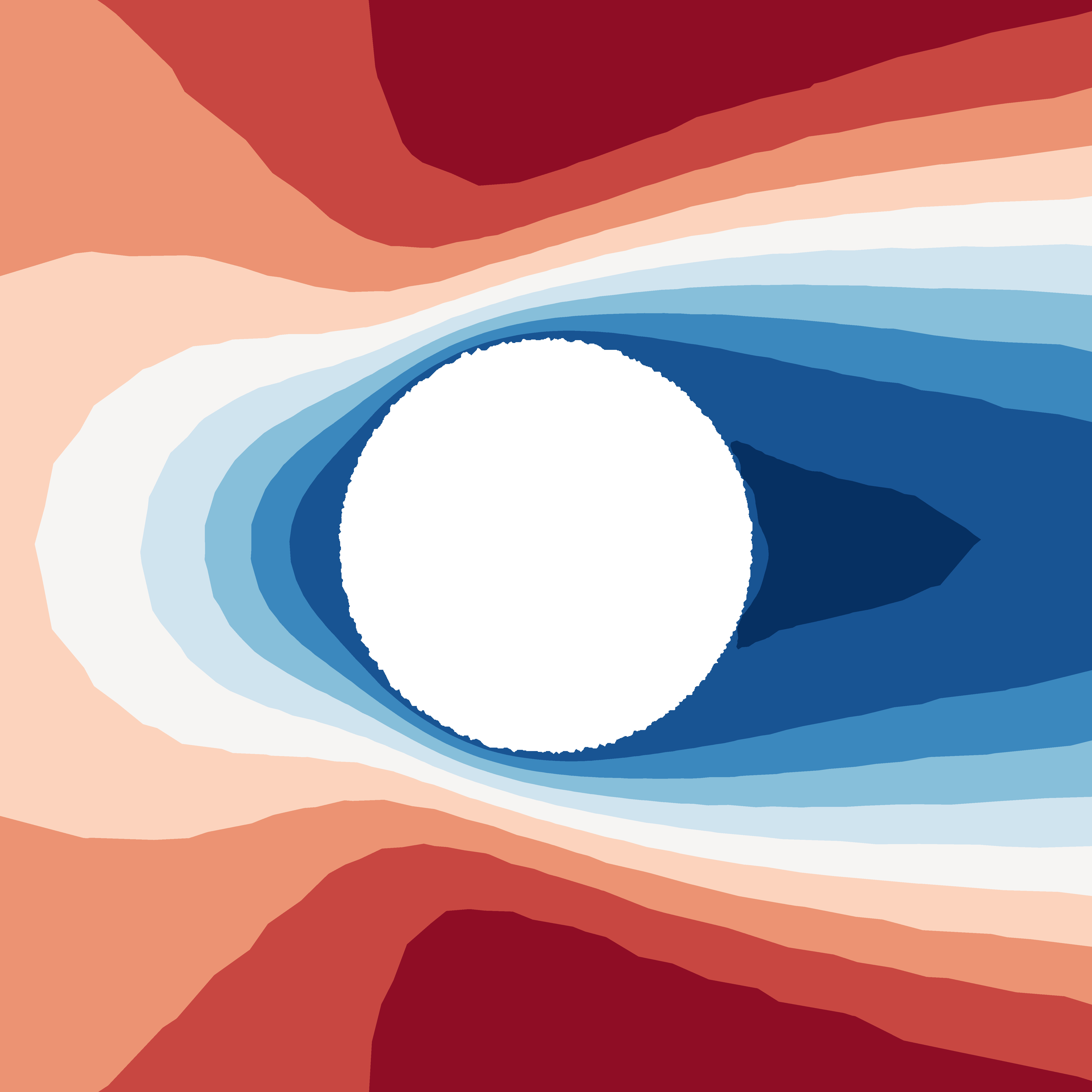}}
	\caption{$u$, reference}
	\label{fig:cylinder_u_ref}
\end{subfigure}%
\begin{subfigure}[t]{.25\linewidth}
	\centering
	\fbox{\includegraphics[width=.8\linewidth]{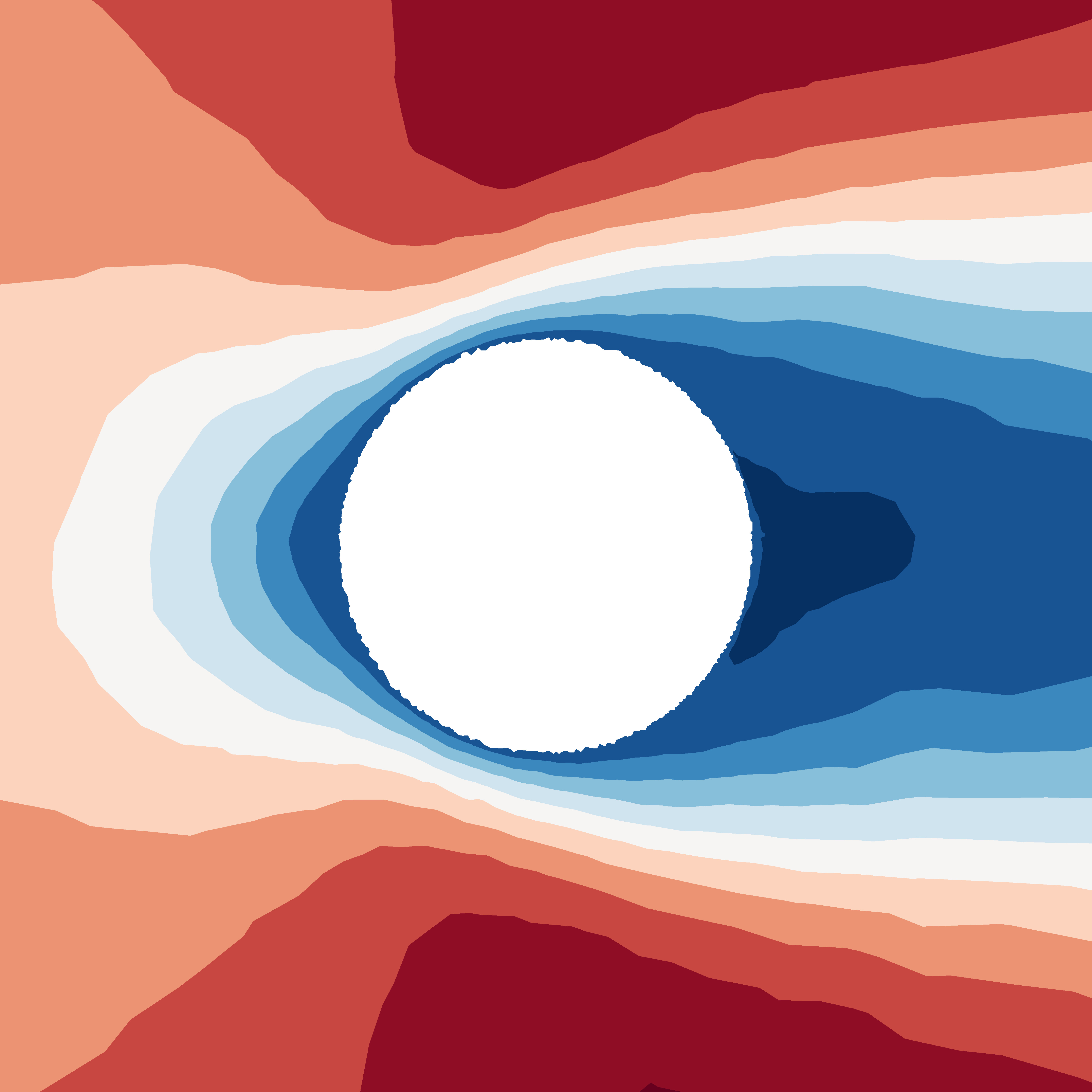}}
	\caption{$u$, predicted}
	\label{fig:cylinder_u_pred}
\end{subfigure}%
\begin{subfigure}[t]{.1\textwidth}
	\centering
	\raisebox{-2.7mm}{\includegraphics{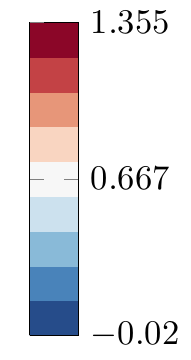}}
	%\raisebox{-2.7mm}{\colorbar{-0.020144}{1.354915}{3.18}}
\end{subfigure}\qquad%
\begin{subfigure}[t]{.25\linewidth}
	\centering
	\fbox{\includegraphics[width=.8\linewidth]{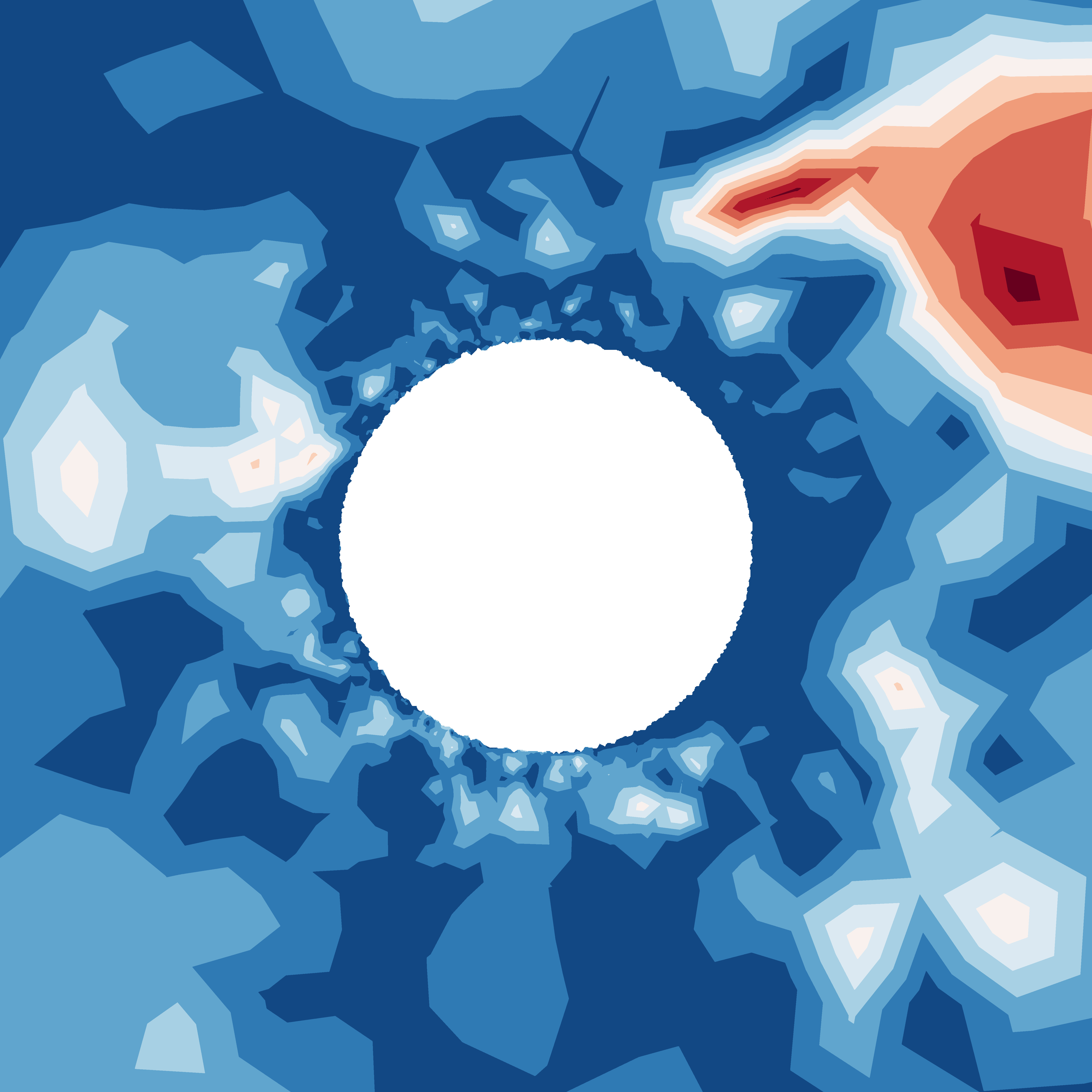}}
	\caption{$u$, absolute error}
	\label{fig:cylinder_u_err}
\end{subfigure}%
\begin{subfigure}[t]{.1\textwidth}
	\centering
	\raisebox{-2.7mm}{\includegraphics{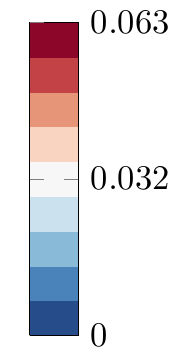}}
	%\raisebox{-2.7mm}{\colorbar{0}{0.063093}{3.18}}
\end{subfigure}%

\medskip

\begin{subfigure}[t]{.25\linewidth}
	\centering
	\fbox{\includegraphics[width=.8\linewidth]{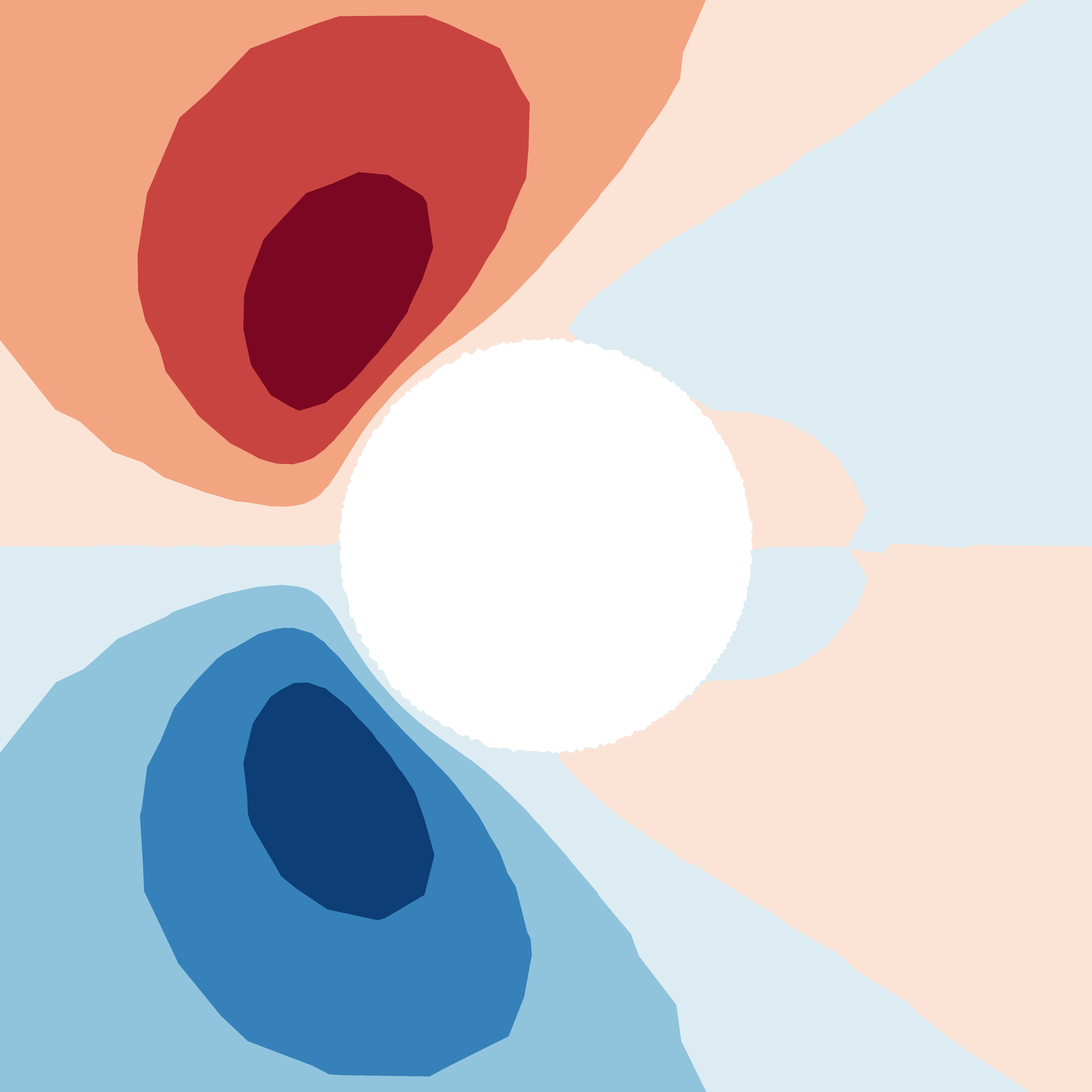}}
	\caption{$v$, reference}
	\label{fig:cylinder_v_ref}
\end{subfigure}%
\begin{subfigure}[t]{.25\linewidth}
	\centering
	\fbox{\includegraphics[width=.8\linewidth]{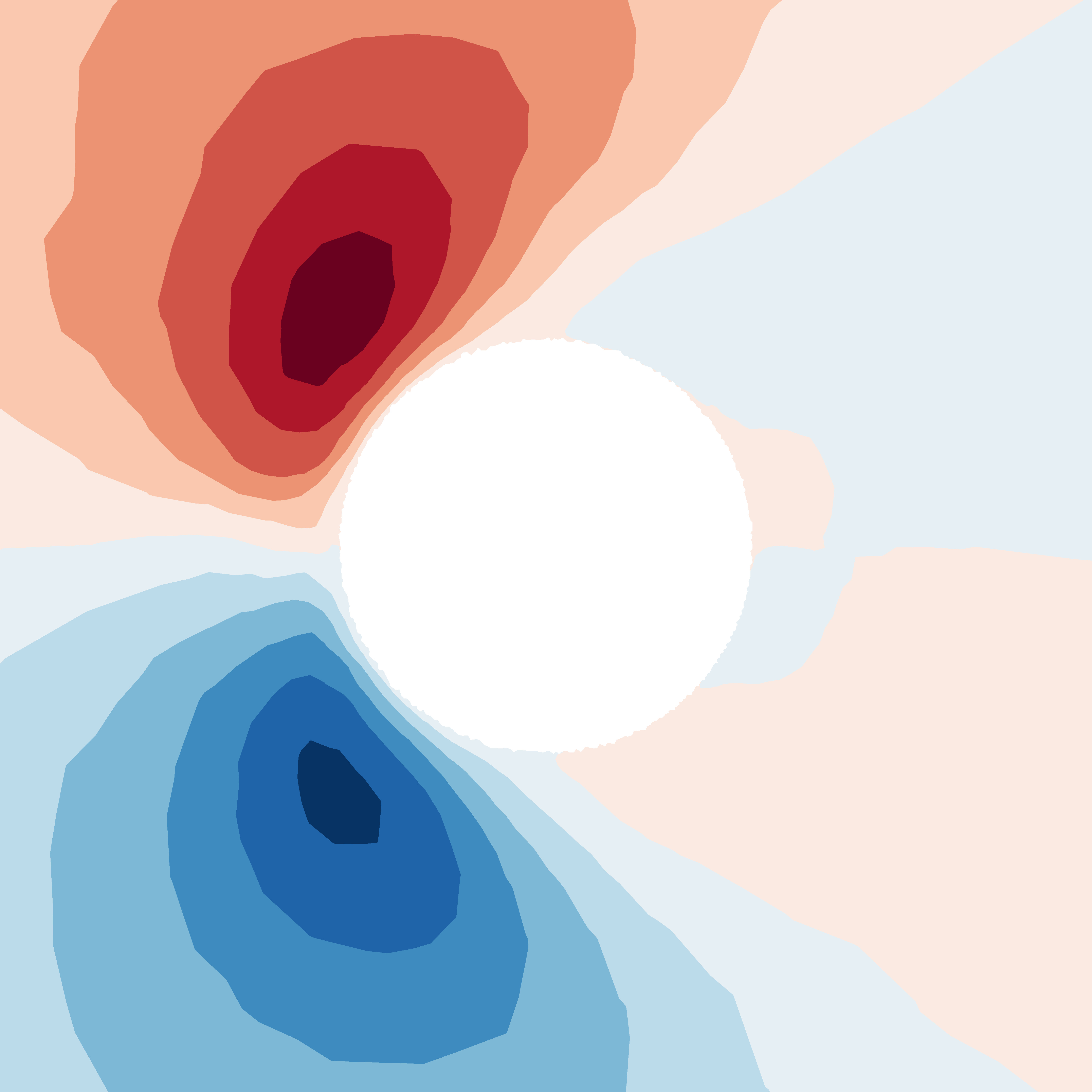}}
	\caption{$v$, predicted}
	\label{fig:cylinder_v_pred}
\end{subfigure}%
\begin{subfigure}[t]{.1\textwidth}
	\centering
	\raisebox{-2.7mm}{\includegraphics{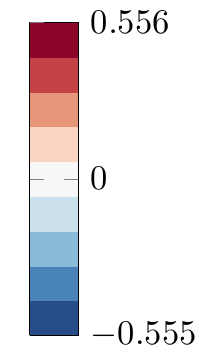}}
	%\raisebox{-2.7mm}{\colorbar{-0.555419}{0.555920}{3.18}}
\end{subfigure}\qquad%
\begin{subfigure}[t]{.25\linewidth}
	\centering
	\fbox{\includegraphics[width=.8\linewidth]{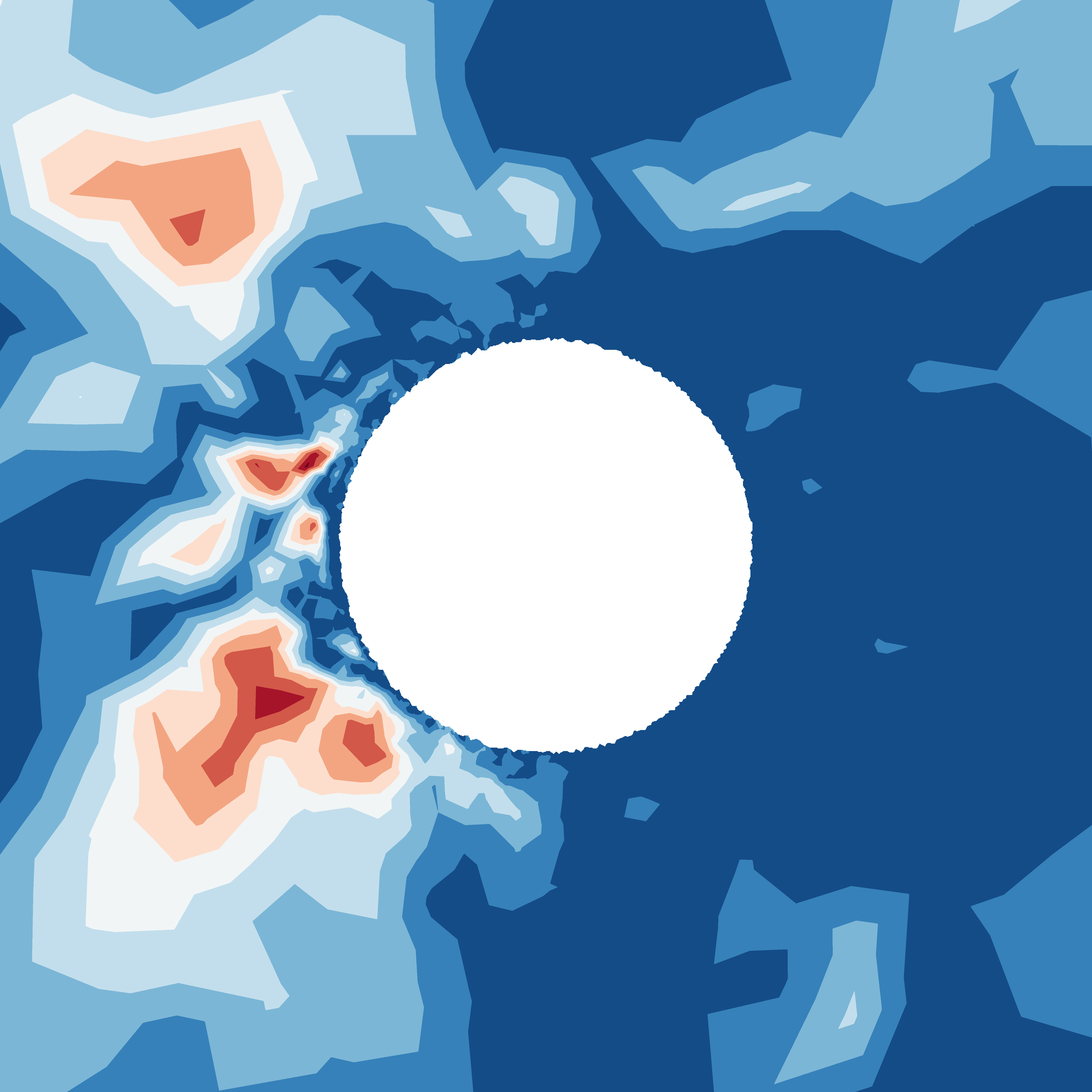}}
	\caption{$v$, absolute error}
	\label{fig:cylinder_v_err}
\end{subfigure}%
\begin{subfigure}[t]{.1\textwidth}
	\centering
	\raisebox{-2.7mm}{\includegraphics{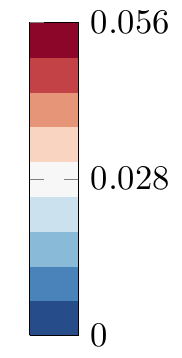}}
	%\raisebox{-2.7mm}{\colorbar{0}{0.055634}{3.18}}
\end{subfigure}%

\medskip

\begin{subfigure}[t]{.25\linewidth}
	\centering
	\fbox{\includegraphics[width=.8\linewidth]{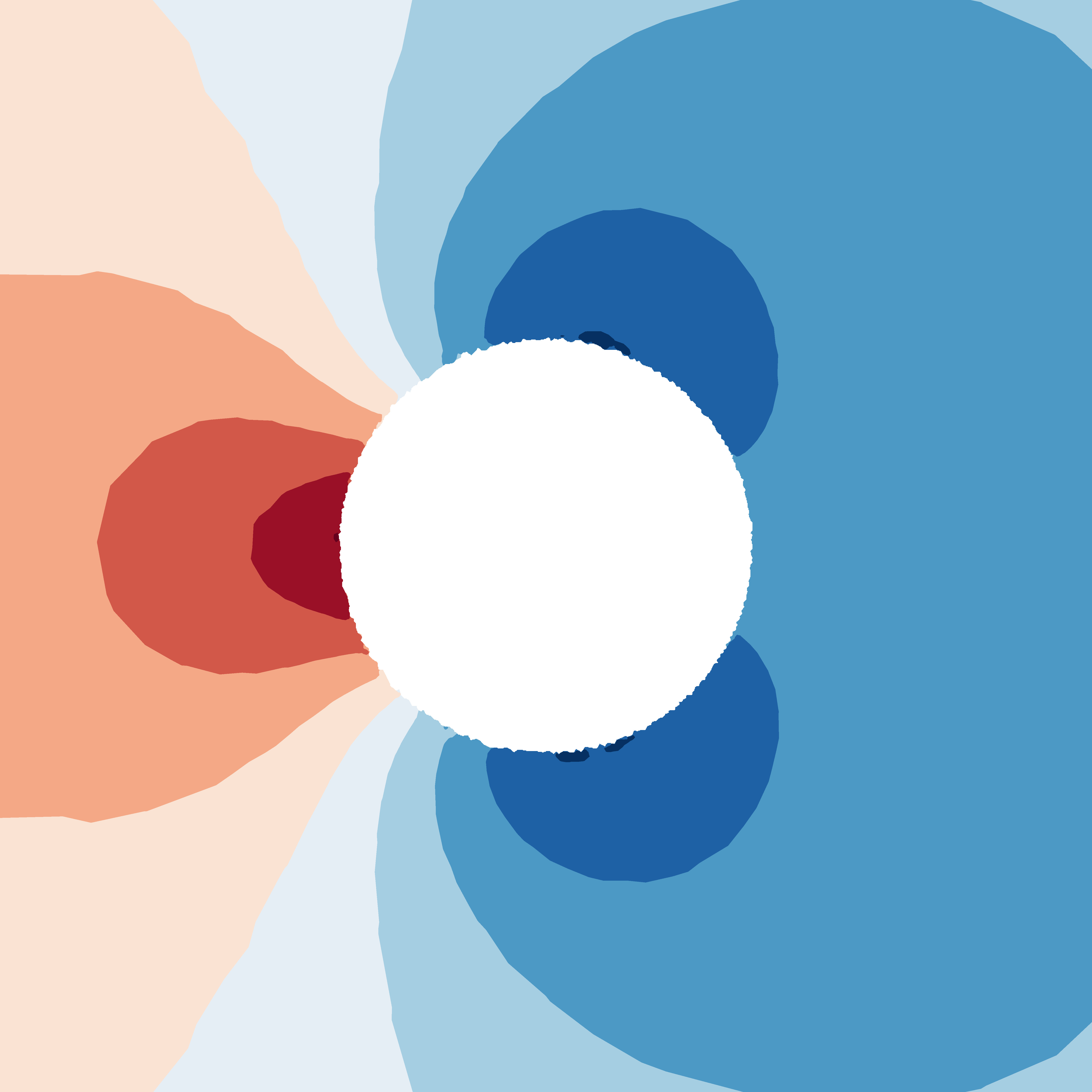}}
	\caption{$p$, reference}
	\label{fig:cylinder_p_ref}
\end{subfigure}%
\begin{subfigure}[t]{.25\linewidth}
	\centering
	\fbox{\includegraphics[width=.8\linewidth]{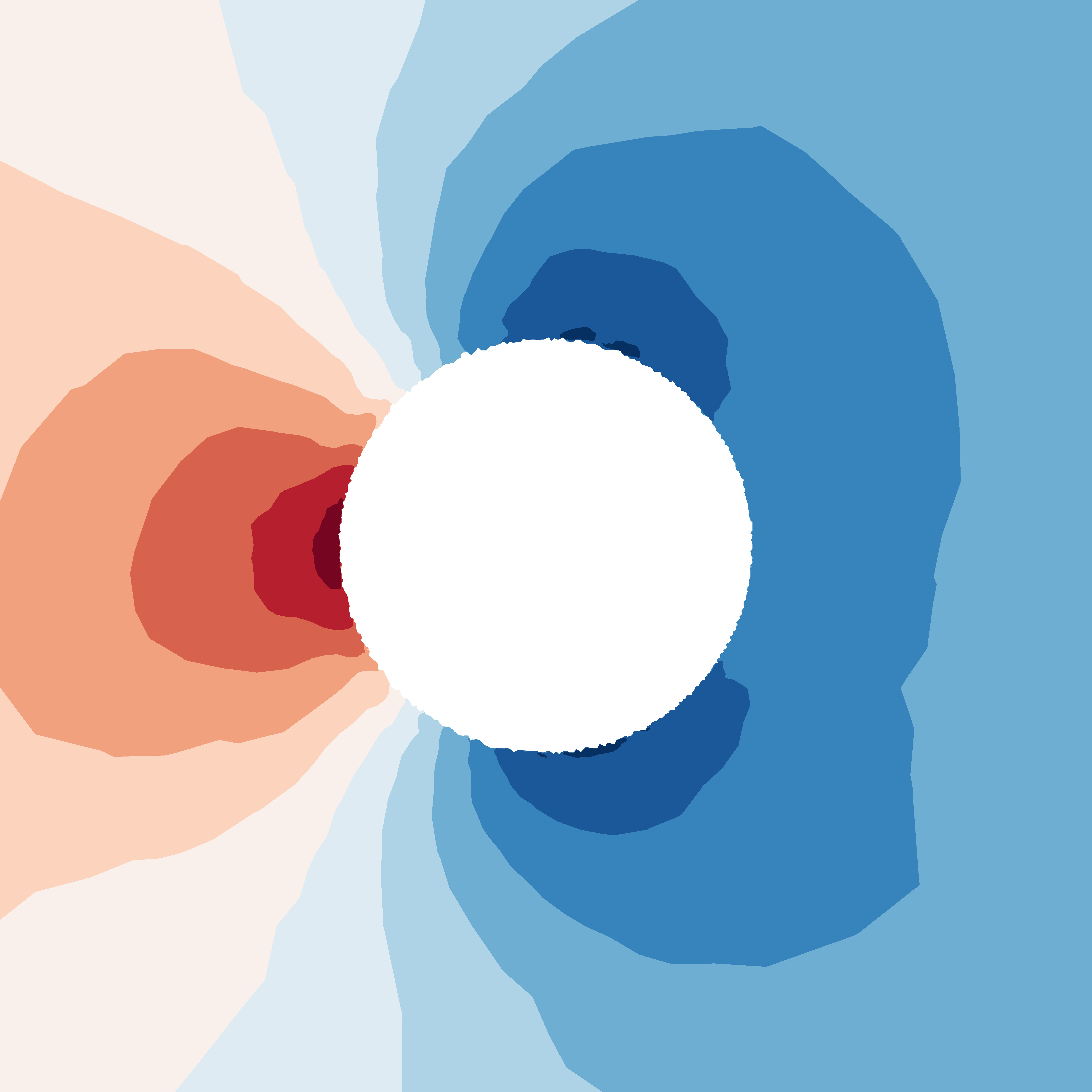}}
	\caption{$p$, predicted}
	\label{fig:cylinder_p_pred}
\end{subfigure}%
\begin{subfigure}[t]{.1\textwidth}
	\centering
	\raisebox{-2.7mm}{\includegraphics{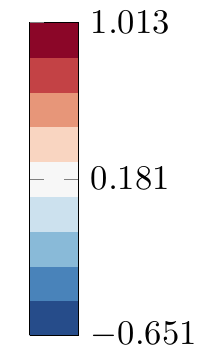}}
	%\raisebox{-2.7mm}{\colorbar{-0.651327}{1.01348}{3.18}}
\end{subfigure}\qquad%
\begin{subfigure}[t]{.25\linewidth}
	\centering
	\fbox{\includegraphics[width=.8\linewidth]{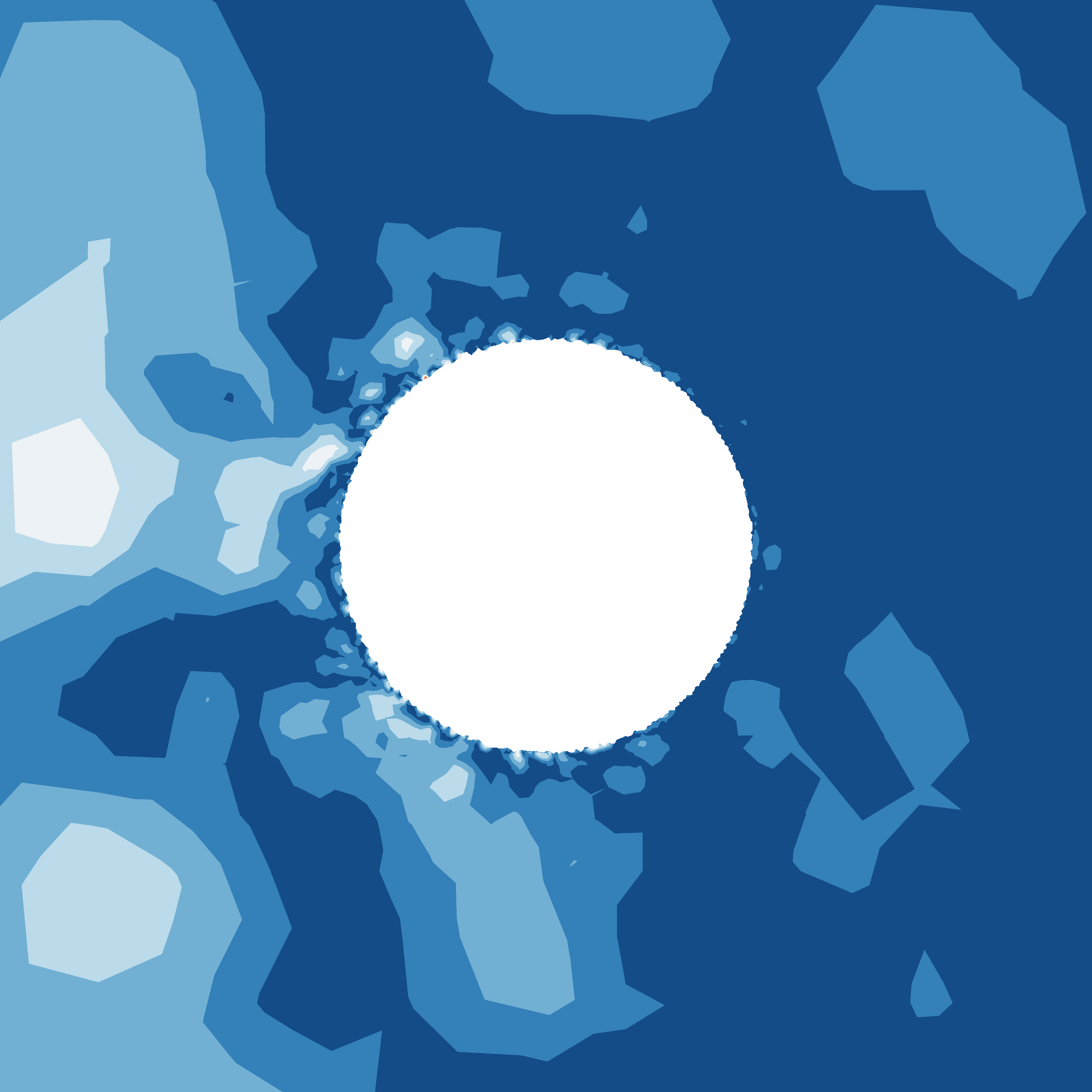}}
	\caption{$p$, absolute error}
	\label{fig:cylinder_p_err}
\end{subfigure}%
\begin{subfigure}[t]{.1\textwidth}
	\centering
	\raisebox{-2.7mm}{\includegraphics{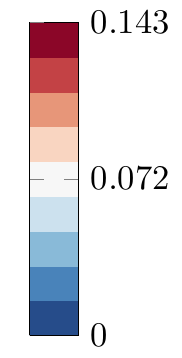}}
	%\raisebox{-2.7mm}{\colorbar{0}{0.143430}{3.18}}
\end{subfigure}%

\caption{\textbf{Flow and shape predictions around a cylinder using a GCNN model.} The normalized velocity and pressure fields outputted by the network are rescaled to their physical ranges \rred{and non-dimensionalized} for better demonstration. The reference (\ref{fig:cylinder_u_ref}, \ref{fig:cylinder_v_ref}, \ref{fig:cylinder_p_ref}) and predicted fields (\ref{fig:cylinder_u_pred}, \ref{fig:cylinder_v_pred}, \ref{fig:cylinder_p_pred}) are very close, with no specific pattern in the MAE maps (\ref{fig:cylinder_u_err}, \ref{fig:cylinder_v_err}, \ref{fig:cylinder_p_err}).}
\label{fig:cylinder_visulization}
\end{figure}

\begin{figure}[p]
\centering
\begin{subfigure}[t]{.4\linewidth}
	\centering
	\shifttext{-18mm}{\includegraphics{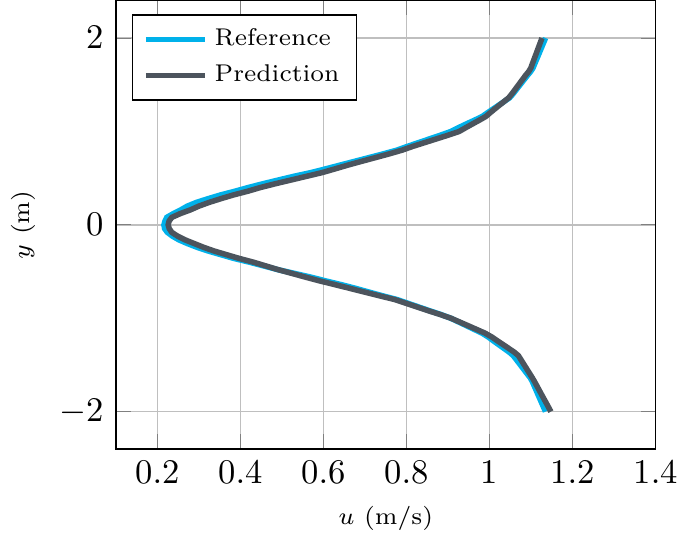}}
%	\begin{tikzpicture}[	trim axis left, trim axis right]
%		\begin{axis}[	scale=0.8, ylabel=$y$ (m),xlabel=$u$ (m/s),
%					label style={font=\scriptsize}, 
%					tick label style={/pgf/number format/.cd, fixed, precision=3, /tikz/.cd}, 
%					legend style={font=\scriptsize},
%					xmin=0.1, xmax=1.4, grid=major,
%					legend pos=north west,
%					legend cell align=left
%					]
%		\legend{Reference, Prediction}
%		\addplot[mark=none,ultra thick,myblue1] table[x index=1,y index=0] {cylindre_x=-1.csv};
%		\addplot[mark=none,ultra thick,mygray1] table[x index=4,y index=0] {cylindre_x=-1.csv};
%		\end{axis}
%	\end{tikzpicture}
\caption{$u$ profile at $x=-1$}
\label{fig:cylinder_u_left}
\end{subfigure}\qquad
\begin{subfigure}[t]{.4\linewidth}
	\centering
	\shifttext{-6mm}{\includegraphics{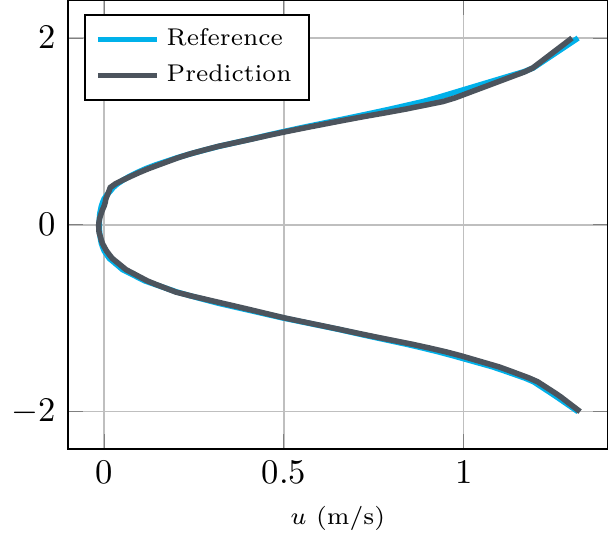}}
%	\begin{tikzpicture}[	trim axis left, trim axis right]
%		\begin{axis}[	scale=0.8, xlabel=$u$ (m/s), %ylabel=$y$ (m),
%					label style={font=\scriptsize}, 
%					tick label style={/pgf/number format/.cd, fixed, precision=3, /tikz/.cd}, 
%					legend style={font=\scriptsize},
%					xmin=-0.1, xmax=1.4,grid=major,
%					legend pos=north west,
%					legend cell align=left
%					]
%		\legend{Reference, Prediction}
%		\addplot[mark=none,ultra thick,myblue1] table[x index=1,y index=0] {cylindre_x=1.csv};
%		\addplot[mark=none,ultra thick,mygray1] table[x index=4,y index=0] {cylindre_x=1.csv};
%		\end{axis}
%	\end{tikzpicture}
\caption{$u$ profile at $x=1$}
\label{fig:cylinder_u_right}
\end{subfigure}

\medskip
\medskip

\begin{subfigure}[t]{.4\linewidth}
	\centering
	\shifttext{-17mm}{\includegraphics{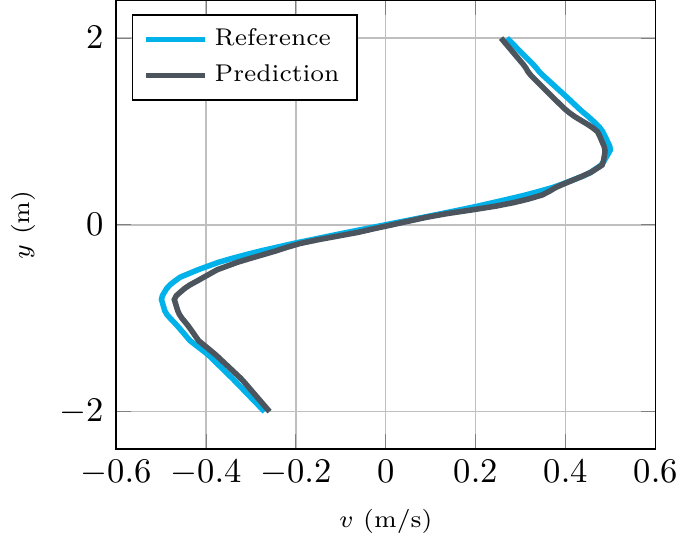}}
%	\begin{tikzpicture}[	trim axis left, trim axis right]
%		\begin{axis}[	scale=0.8, ylabel=$y$ (m),xlabel=$v$ (m/s),
%					label style={font=\scriptsize}, 
%					tick label style={/pgf/number format/.cd, fixed, precision=3, /tikz/.cd}, 
%					legend style={font=\scriptsize},
%					xmin=-0.6, xmax=0.6,grid=major,
%					legend pos=north west,
%					legend cell align=left
%					]
%		\legend{Reference, Prediction}
%		\addplot[mark=none,ultra thick,myblue1] table[x index=2,y index=0] {cylindre_x=-1.csv};
%		\addplot[mark=none,ultra thick,mygray1] table[x index=5,y index=0] {cylindre_x=-1.csv};
%		\end{axis}
%	\end{tikzpicture}
\caption{$v$ profile at $x=-1$}
\label{fig:cylinder_v_left}
\end{subfigure}\qquad
\begin{subfigure}[t]{.4\linewidth}
	\centering
	\shifttext{-5mm}{\includegraphics{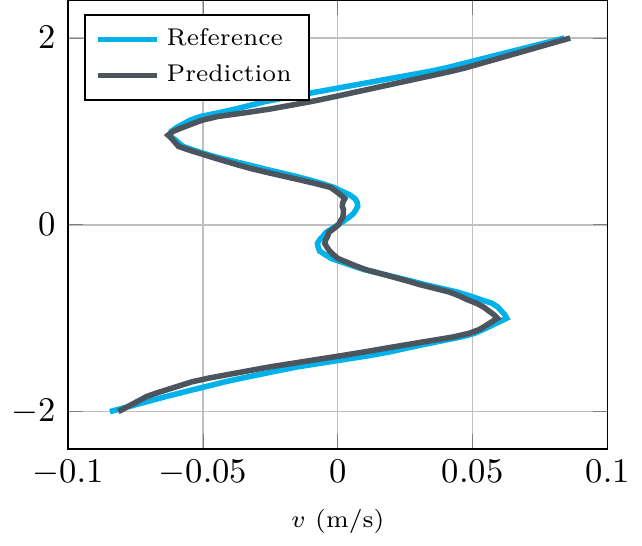}}
%	\begin{tikzpicture}[	trim axis left, trim axis right]
%		\begin{axis}[	scale=0.8, xlabel=$v$ (m/s), %ylabel=$y$ (m),
%					label style={font=\scriptsize}, 
%					tick label style={/pgf/number format/.cd, fixed, precision=3, /tikz/.cd}, 
%					legend style={font=\scriptsize},
%					xmin=-0.1, xmax=0.1,grid=major,
%					legend pos=north west,
%					legend cell align=left
%					]
%		\legend{Reference, Prediction}
%		\addplot[mark=none,ultra thick,myblue1] table[x index=2,y index=0] {cylindre_x=1.csv};
%		\addplot[mark=none,ultra thick,mygray1] table[x index=5,y index=0] {cylindre_x=1.csv};
%		\end{axis}
%	\end{tikzpicture}
\caption{$v$ profile at $x=1$}
\label{fig:cylinder_v_right}
\end{subfigure}

\medskip
\medskip

\begin{subfigure}[t]{.4\linewidth}
	\centering
	\shifttext{-22mm}{\includegraphics{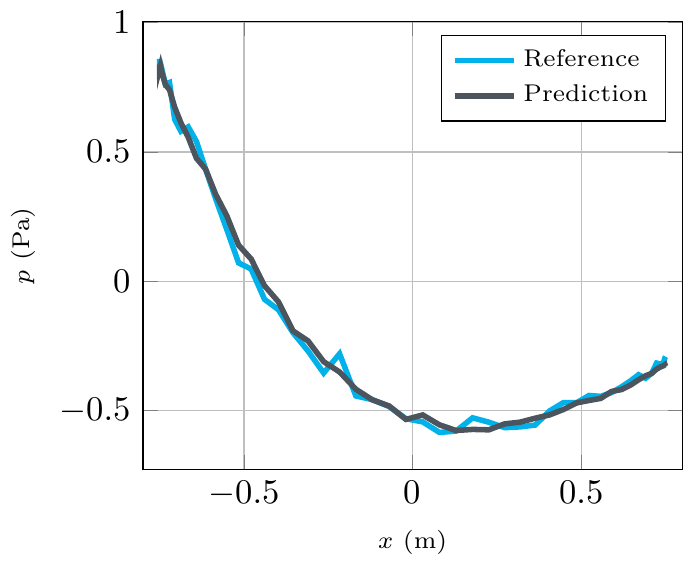}}
%	\begin{tikzpicture}[	trim axis left, trim axis right]
%		\begin{axis}[	scale=0.8, xlabel=$x$ (m), ylabel=$p$ (Pa),
%					label style={font=\scriptsize}, 
%					tick label style={/pgf/number format/.cd, fixed, precision=3, /tikz/.cd}, 
%					legend style={font=\scriptsize},
%					xmin=-0.8, xmax=0.8,grid=major,
%					legend pos=north east,
%					legend cell align=left
%					]
%		\legend{Reference, Prediction}
%		\addplot[mark=none,ultra thick,myblue1] table[x index=0,y index=1] {pressure_low.csv};
%		\addplot[mark=none,ultra thick,mygray1] table[x index=0,y index=2] {pressure_low.csv};
%		\end{axis}
%	\end{tikzpicture}
\caption{Pressure distribution on the lower surface}
\label{fig:cylinder_p_low}
\end{subfigure}\qquad
\begin{subfigure}[t]{.4\linewidth}
	\centering
	\shifttext{-10mm}{\includegraphics{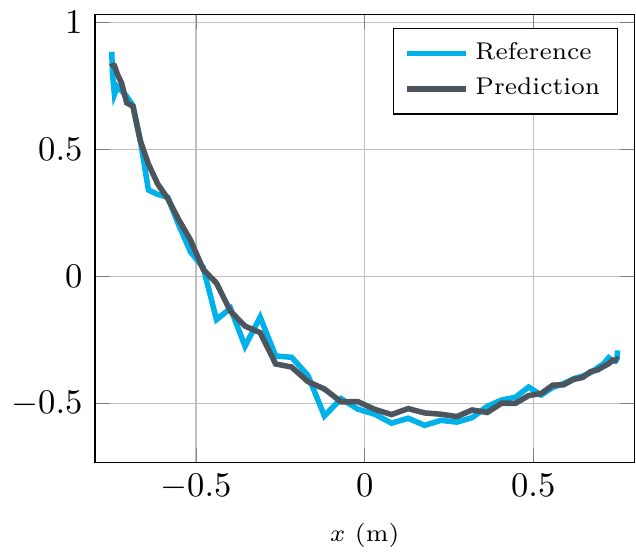}}
%	\begin{tikzpicture}[	trim axis left, trim axis right]
%		\begin{axis}[	scale=0.8, xlabel=$x$ (m),% ylabel=p($Pa$),
%					label style={font=\scriptsize}, 
%					tick label style={/pgf/number format/.cd, fixed, precision=3, /tikz/.cd}, 
%					legend style={font=\scriptsize},
%					xmin=-0.8, xmax=0.8,grid=major,
%					legend pos=north east,
%					legend cell align=left
%					]
%		\legend{Reference, Prediction}
%		\addplot[mark=none,ultra thick,myblue1] table[x index=0,y index=1] {pressure_up.csv};
%		\addplot[mark=none,ultra thick,mygray1] table[x index=0,y index=2] {pressure_up.csv};
%		\end{axis}
%	\end{tikzpicture}
\caption{Pressure distribution on the upper surface}
\label{fig:cylinder_p_up}
\end{subfigure}
\caption{\textbf{Velocity and pressure profiles around the cylinder.} The velocities are recorded at $x=-1$ (\ref{fig:cylinder_u_left}, \ref{fig:cylinder_v_left}) and $x=1$ (\ref{fig:cylinder_u_right}, \ref{fig:cylinder_v_right}), and display the expected symmetry with respect to the $x$ axis. The pressure is recorded on the lower (\ref{fig:cylinder_p_low}) and upper surfaces (\ref{fig:cylinder_p_up}) of the cylinder. \rred{The wiggles in the referential pressure profiles are caused by obstacle boundary resolution by the immersed boundary method, which employs an element size equal to \num{0.01}m. These numerical artefacts could be reduced using smaller boundary elements, although we find they do not impact the neural networks training.}}
\label{fig:cylinder_profiles}
\end{figure}

To test the trained GCNN model on an out-of-training input, we consider the velocity and pressure predictions performed on a 2D cylinder with radius $0.75$m placed at the origin, and embedded in a body-fitted triangular mesh with \num{2447} nodes, with a $0.01$m mesh size on the cylinder. The MAE of the predicted velocity and pressure fields is \num{4.3e-3}, as the geometric characteristic of the cylinder are close to that of the B\'ezier shapes from the training set. In figure \ref{fig:cylinder_visulization}, we observe almost symmetric velocity fields with respect to the $x$-axis for all three fields. Although the shapes in the training set are not necessarily symmetric, the trained GCNN is aware of the symmetry of flow around a cylinder. The velocity profiles and surface pressure distribution displayed in figure \ref{fig:cylinder_profiles} further confirm the symmetric predictions around a cylinder.

%%%%%%%%%%%%
%%%%%%%%%%%% 

%%%%%%%%%%%%
%%%%%%%%%%%%

%%%%%%%%%%%%%%%%%%%%%%%%%%%%%%%%%%%%%%%%%%%%%%%%%%%%%%%%%%
%%%%%%%%%%%%%%%%%%%%%%%%%%%%%%%%%%%%%%%%%%%%%%%%%%%%%%%%%%
\subsection{Prediction on a NACA0012 airfoil}
\label{section:naca}

\begin{figure}[p]
\centering
\begin{subfigure}[t]{.25\linewidth}
	\centering
	\fbox{\includegraphics[width=.8\linewidth]{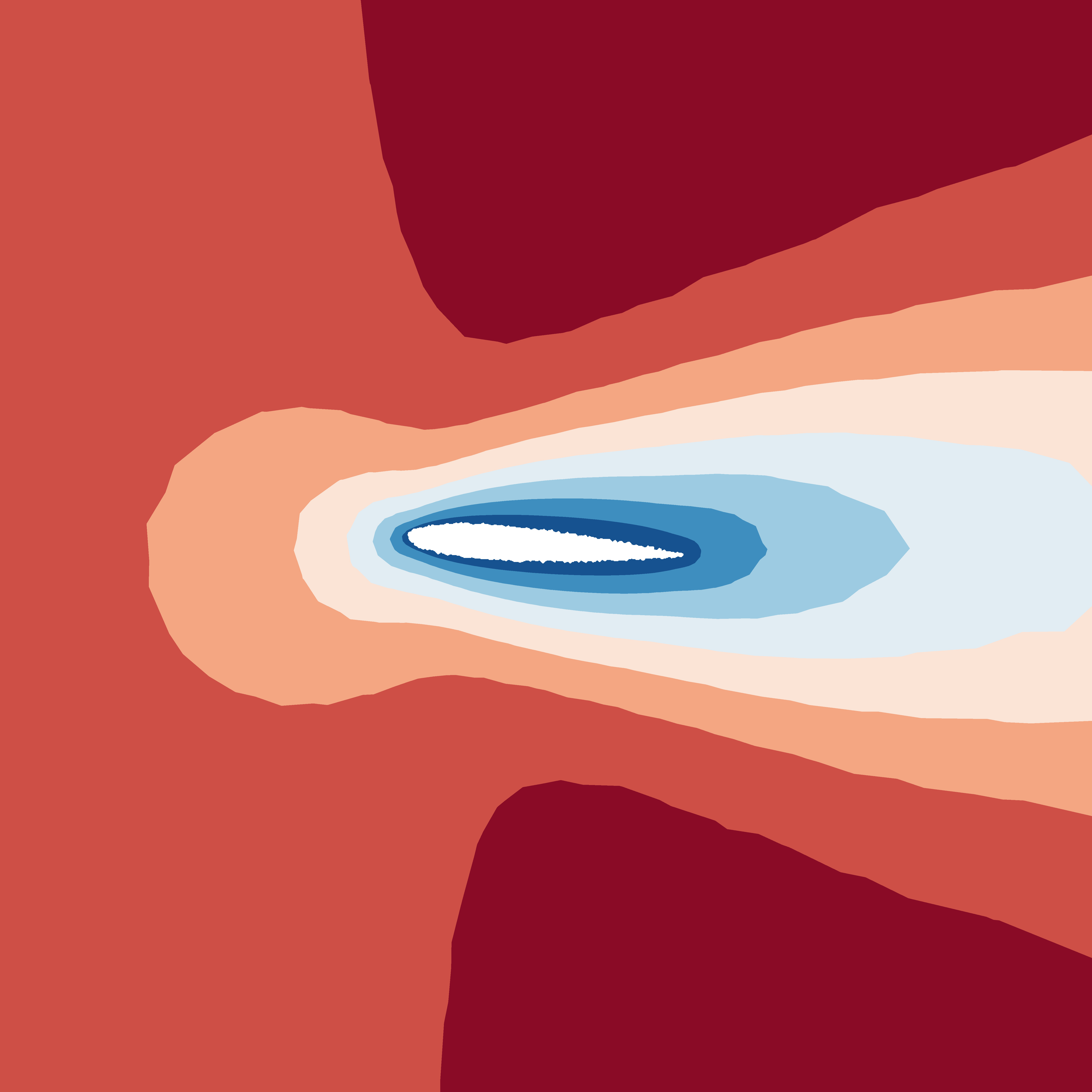}}
	\caption{$u$, reference}
	\label{fig:naca_u_ref}
\end{subfigure}%
\begin{subfigure}[t]{.25\linewidth}
	\centering
	\fbox{\includegraphics[width=.8\linewidth]{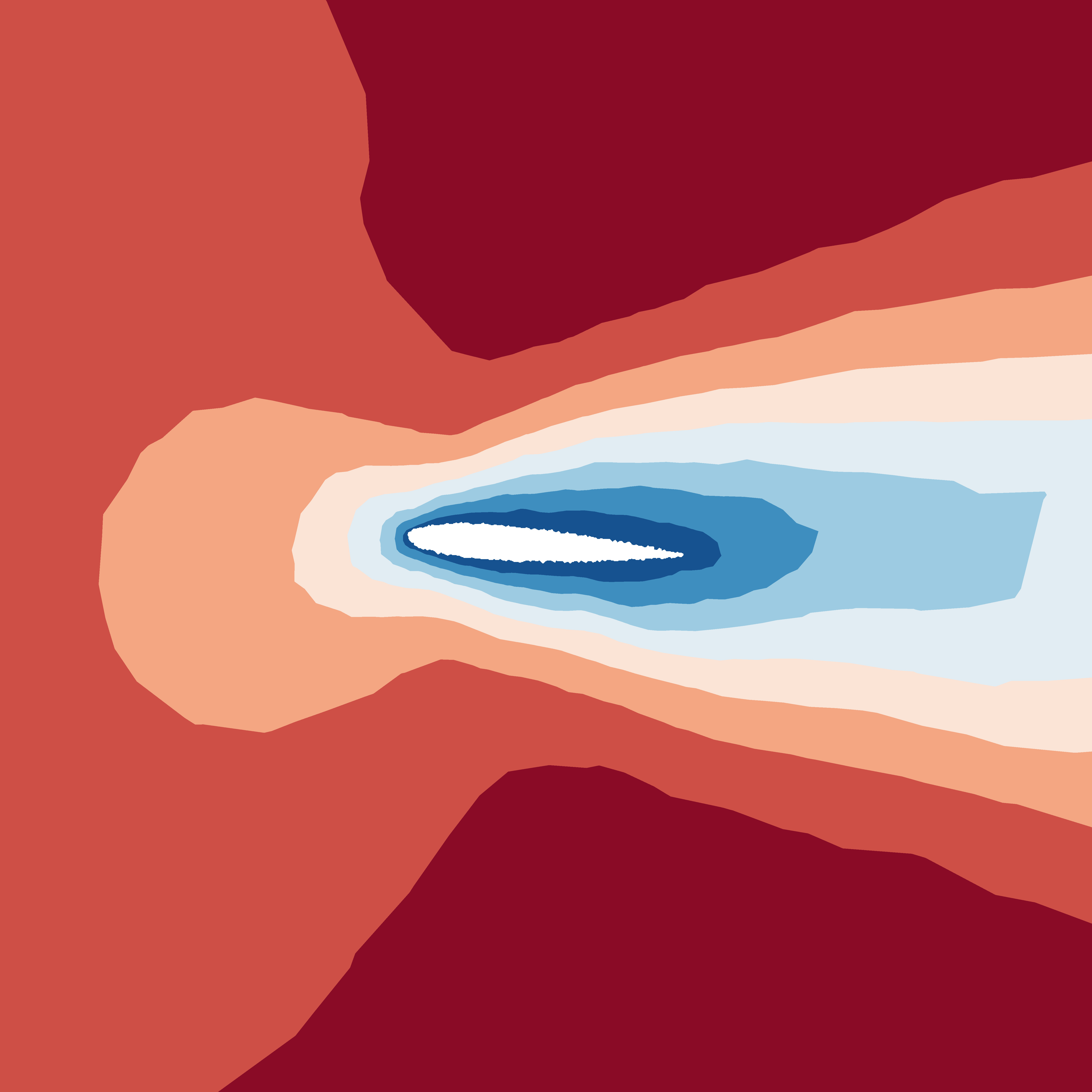}}
	\caption{$u$, predicted}
	\label{fig:naca_u_pred}
\end{subfigure}%
\begin{subfigure}[t]{.1\textwidth}
	\centering
	\raisebox{-2.7mm}{\includegraphics{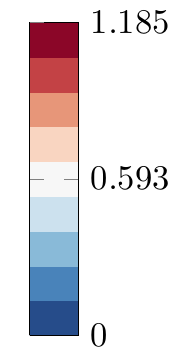}}
	%\raisebox{-2.7mm}{\colorbar{0}{1.18507}{3.18}}
\end{subfigure}\qquad%
\begin{subfigure}[t]{.25\linewidth}
	\centering
	\fbox{\includegraphics[width=.8\linewidth]{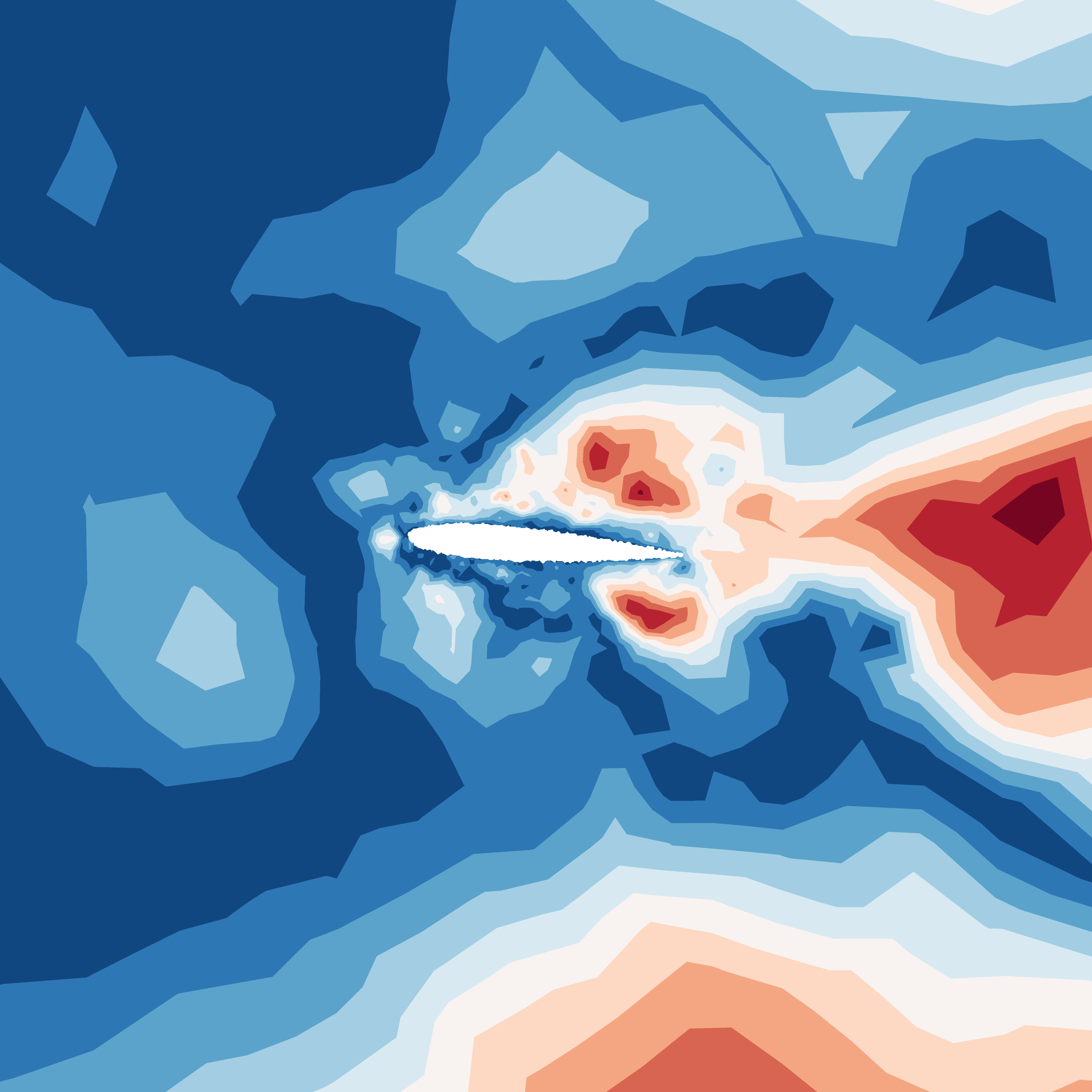}}
	\caption{$u$, absolute error}
	\label{fig:naca_u_err}
\end{subfigure}%
\begin{subfigure}[t]{.1\textwidth}
	\centering
	\raisebox{-2.7mm}{\includegraphics{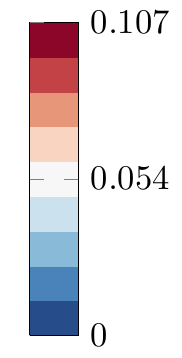}}
	%\raisebox{-2.7mm}{\colorbar{0}{0.107377}{3.18}}
\end{subfigure}%

\medskip

\begin{subfigure}[t]{.25\linewidth}
	\centering
	\fbox{\includegraphics[width=.8\linewidth]{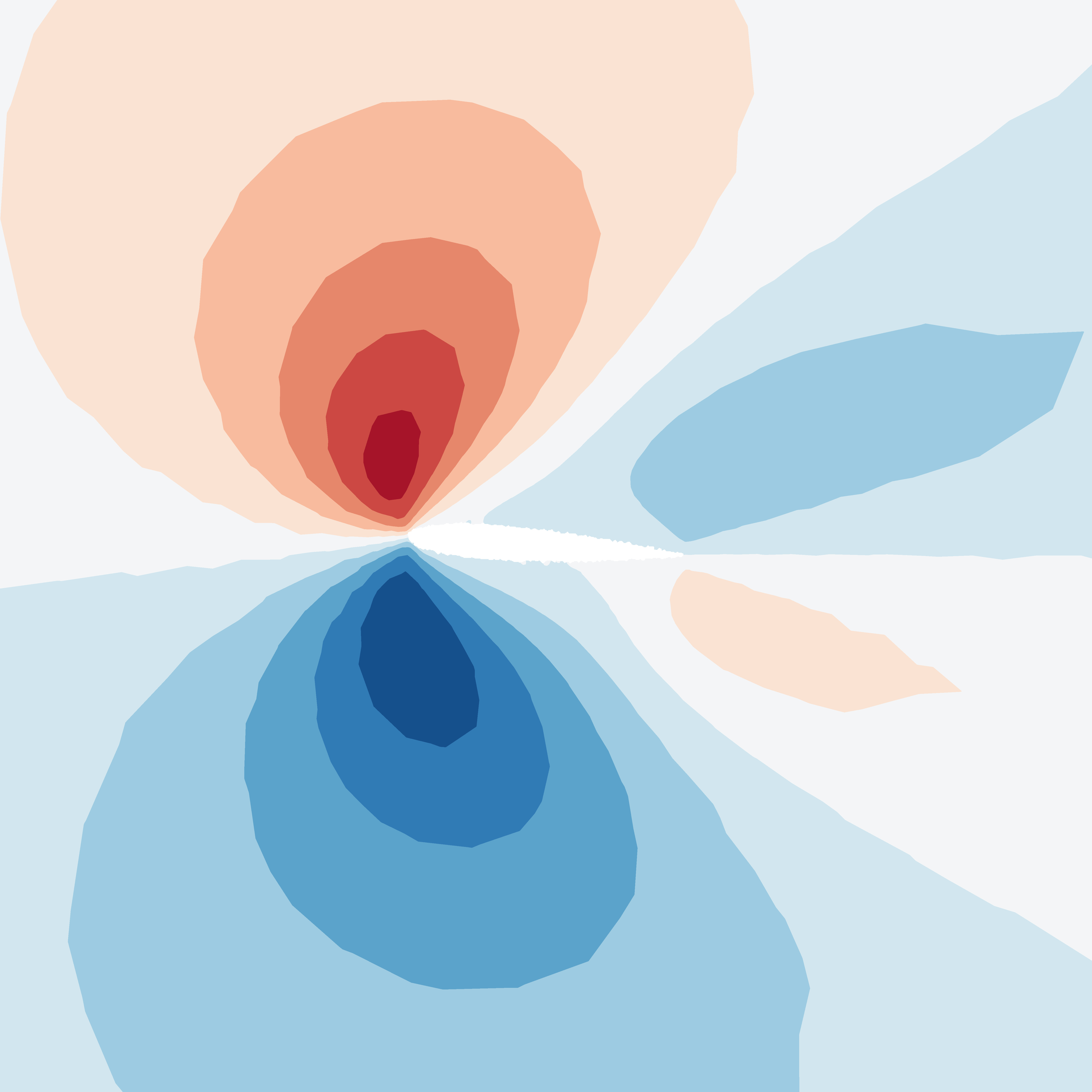}}
	\caption{$v$, reference}
	\label{fig:naca_v_ref}
\end{subfigure}%
\begin{subfigure}[t]{.25\linewidth}
	\centering
	\fbox{\includegraphics[width=.8\linewidth]{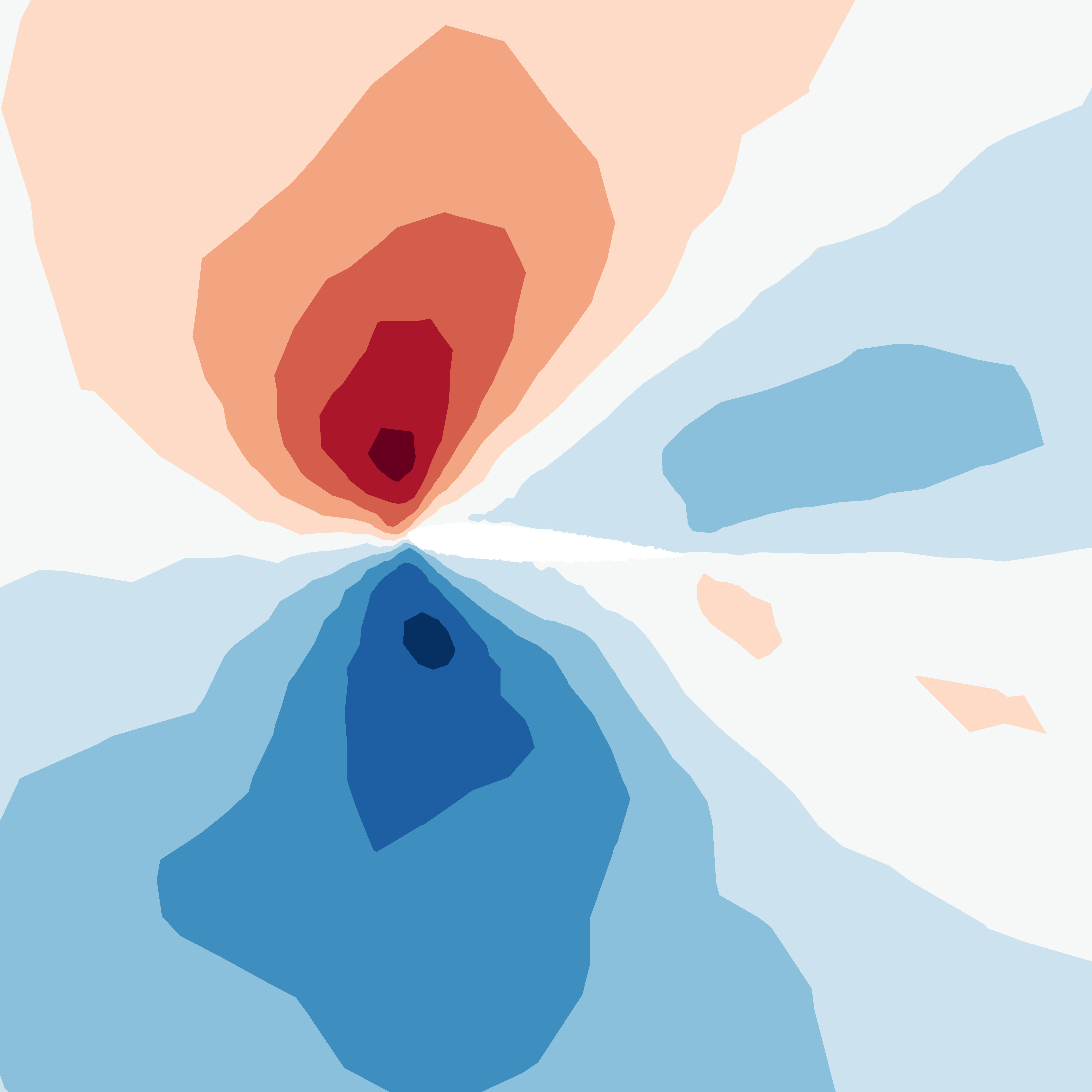}}
	\caption{$v$, predicted}
	\label{fig:naca_v_pred}
\end{subfigure}%
\begin{subfigure}[t]{.1\textwidth}
	\centering
	\raisebox{-2.7mm}{\includegraphics{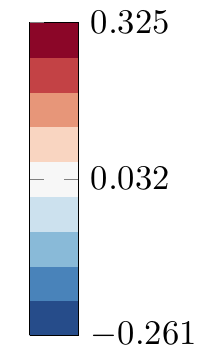}}
	%\raisebox{-2.7mm}{\colorbar{-0.261202}{0.324881}{3.18}}
\end{subfigure}\qquad%
\begin{subfigure}[t]{.25\linewidth}
	\centering
	\fbox{\includegraphics[width=.8\linewidth]{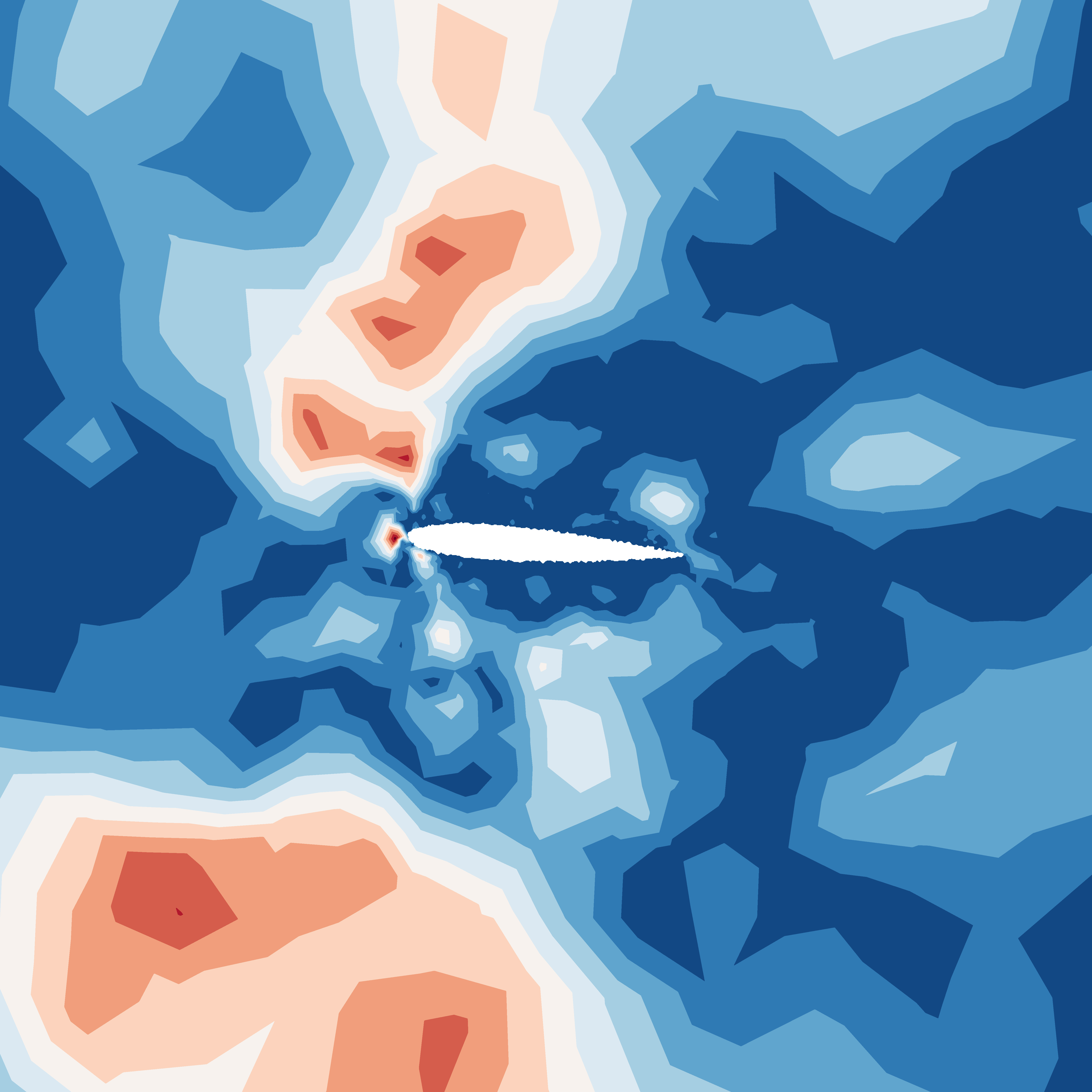}}
	\caption{$v$, absolute error}
	\label{fig:naca_v_err}
\end{subfigure}%
\begin{subfigure}[t]{.1\textwidth}
	\centering
	\raisebox{-2.7mm}{\includegraphics{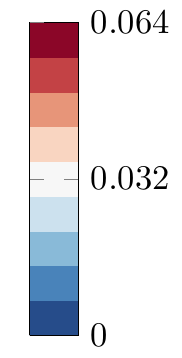}}
	%\raisebox{-2.7mm}{\colorbar{0}{0.063587}{3.18}}
\end{subfigure}%

\medskip

\begin{subfigure}[t]{.25\linewidth}
	\centering
	\fbox{\includegraphics[width=.8\linewidth]{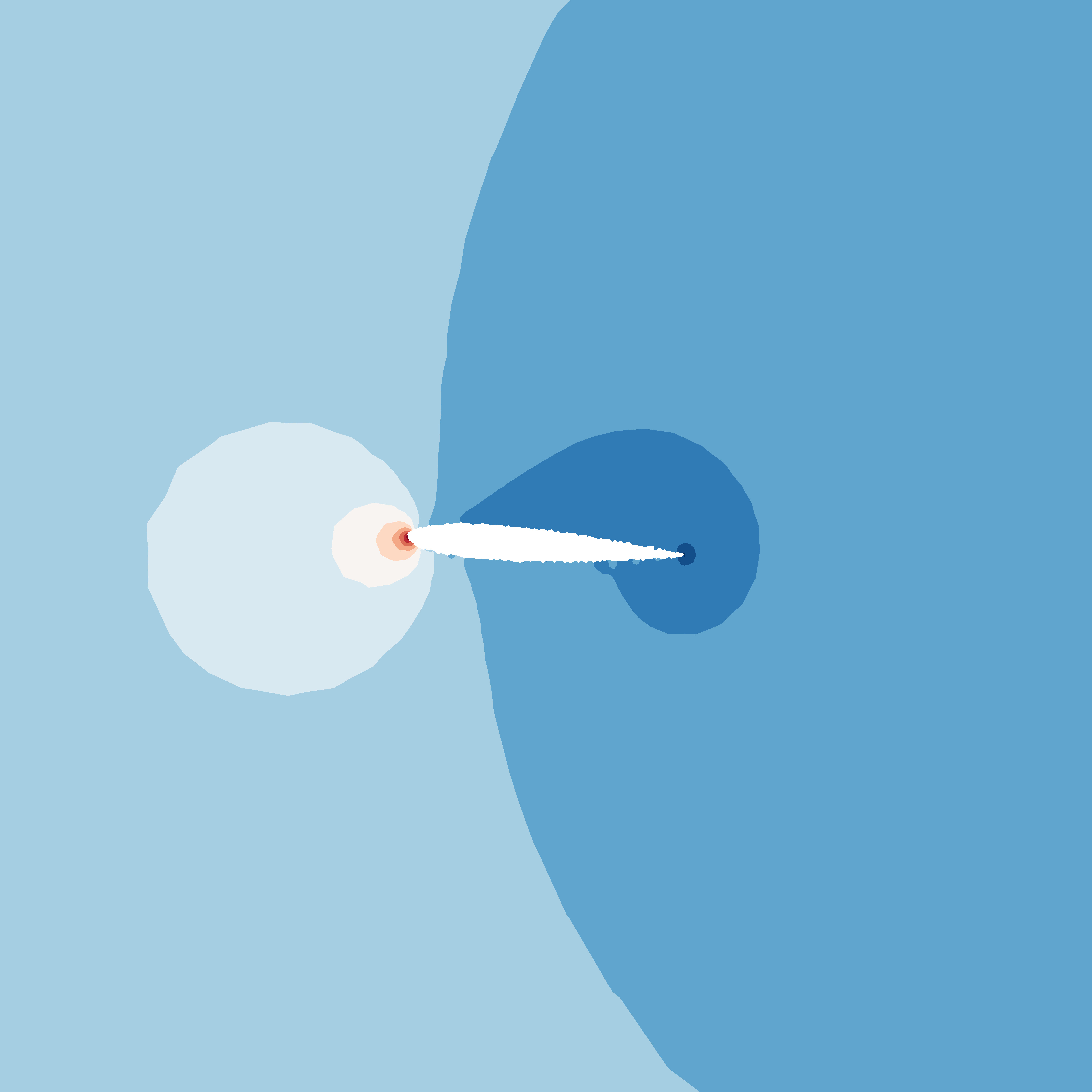}}
	\caption{$p$, reference}
	\label{fig:naca_p_ref}
\end{subfigure}%
\begin{subfigure}[t]{.25\linewidth}
	\centering
	\fbox{\includegraphics[width=.8\linewidth]{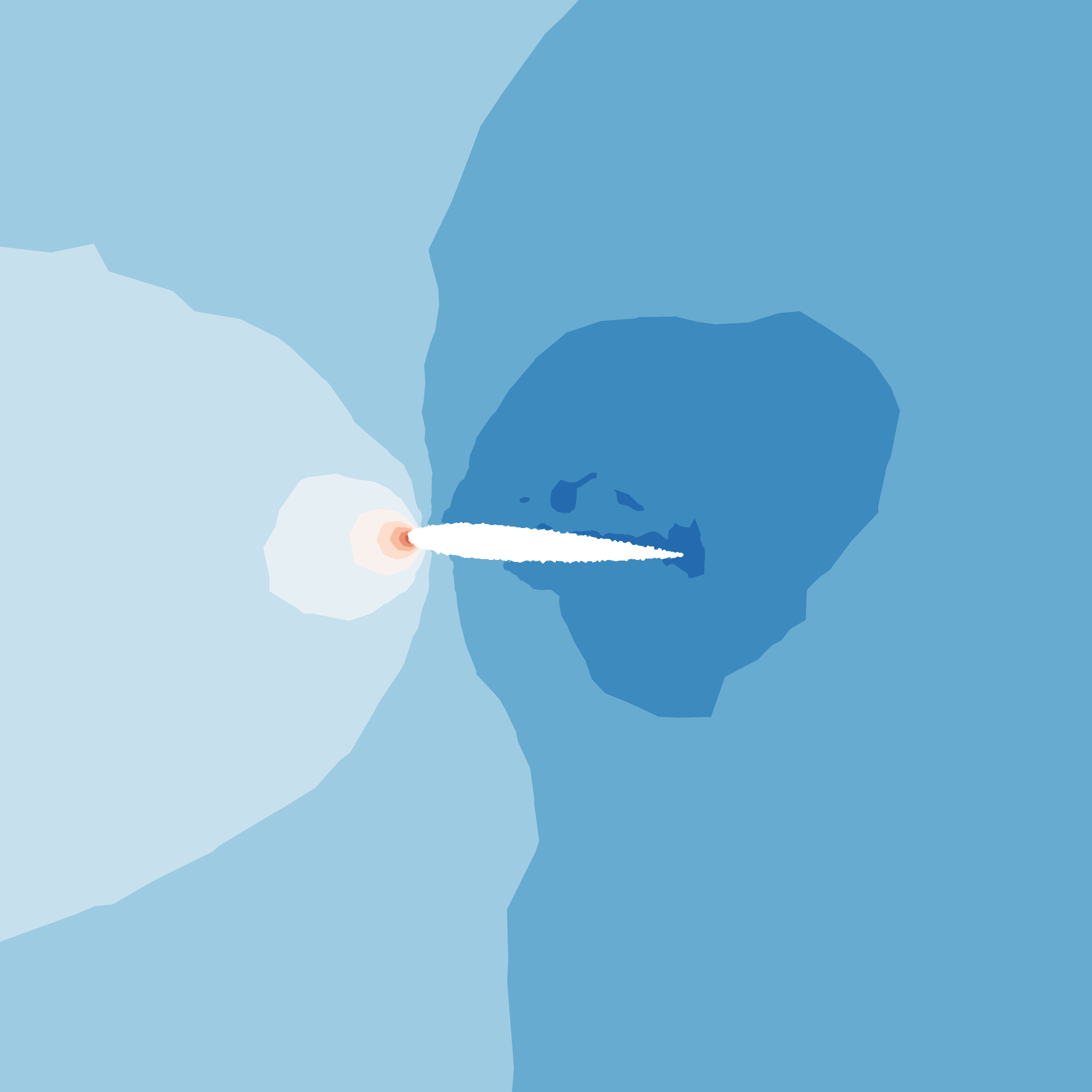}}
	\caption{$p$, predicted}
	\label{fig:naca_p_pred}
\end{subfigure}%
\begin{subfigure}[t]{.1\textwidth}
	\centering
	\raisebox{-2.7mm}{\includegraphics{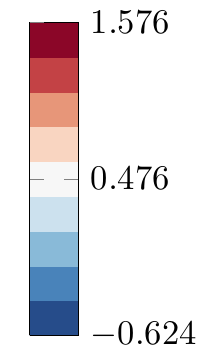}}
	%\raisebox{-2.7mm}{\colorbar{-0.624110}{1.575745}{3.18}}
\end{subfigure}\qquad%
\begin{subfigure}[t]{.25\linewidth}
	\centering
	\fbox{\includegraphics[width=.8\linewidth]{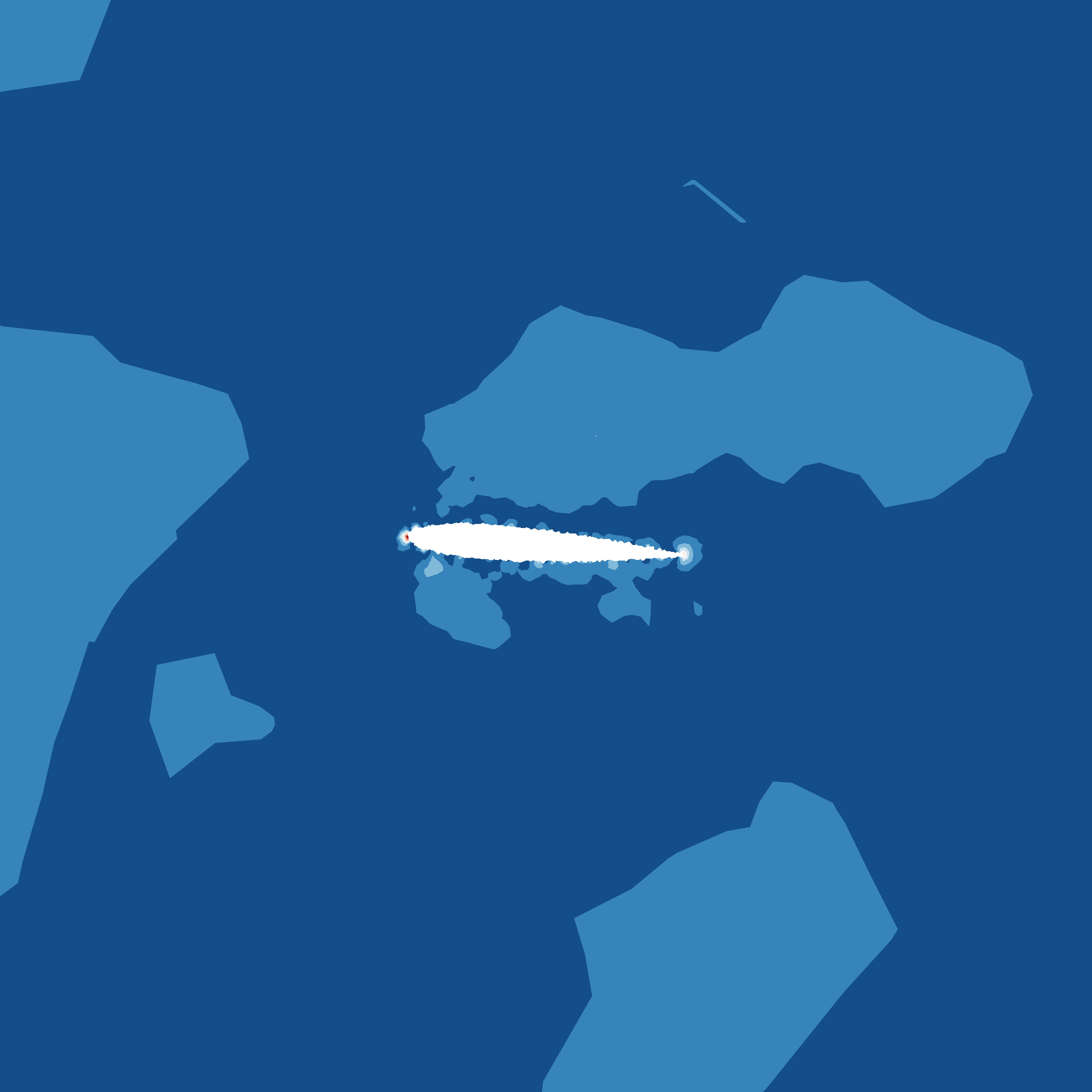}}
	\caption{$p$, absolute error}
	\label{fig:naca_p_err}
\end{subfigure}%
\begin{subfigure}[t]{.1\textwidth}
	\centering
	\raisebox{-2.7mm}{\includegraphics{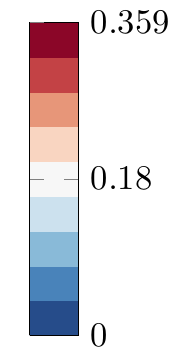}}
	%\raisebox{-2.7mm}{\colorbar{0}{0.35917892}{3.18}}
\end{subfigure}%

\caption{\textbf{Flow predictions on a NACA0012 airfoil using a GCNN model.} Although the field amplitudes are not correctly evaluated, the velocity (\ref{fig:naca_u_ref}, \ref{fig:naca_u_pred}, \ref{fig:naca_u_err}, \ref{fig:naca_v_ref}, \ref{fig:naca_v_pred}, \ref{fig:naca_v_err}) and pressure maps (\ref{fig:naca_p_ref}, \ref{fig:naca_p_pred}, \ref{fig:naca_p_err}) display reasonably physical features, such as sign changes in the vertical velocity or localized high pressure area.}
\label{fig:naca_visualization}
\end{figure}

\begin{figure}[p]
\centering
\begin{subfigure}[t]{.4\linewidth}
	\centering
	\shifttext{-18mm}{\includegraphics{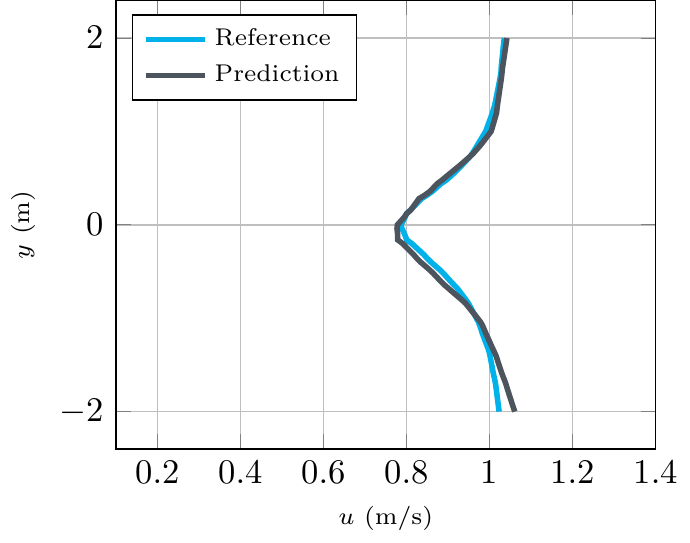}}
%	\begin{tikzpicture}[	trim axis left, trim axis right]
%		\begin{axis}[	scale=0.8, ylabel=$y$ (m),xlabel=$u$ (m/s),
%					label style={font=\scriptsize}, 
%					tick label style={/pgf/number format/.cd, fixed, precision=3, /tikz/.cd}, 
%					legend style={font=\scriptsize},
%					xmin=0.1, xmax=1.4, grid=major,
%					legend pos=north west,
%					legend cell align=left
%					]
%		\legend{Reference, Prediction}
%		\addplot[mark=none,ultra thick,myblue1] table[x index=1,y index=0] {naca_x=-1.csv};
%		\addplot[mark=none,ultra thick,mygray1] table[x index=4,y index=0] {naca_x=-1.csv};
%		\end{axis}
%	\end{tikzpicture}
\caption{$u$ profile at $x=-1$}
\label{fig:naca_u_left}
\end{subfigure}\qquad
\begin{subfigure}[t]{.4\linewidth}
	\centering
	\shifttext{-6mm}{\includegraphics{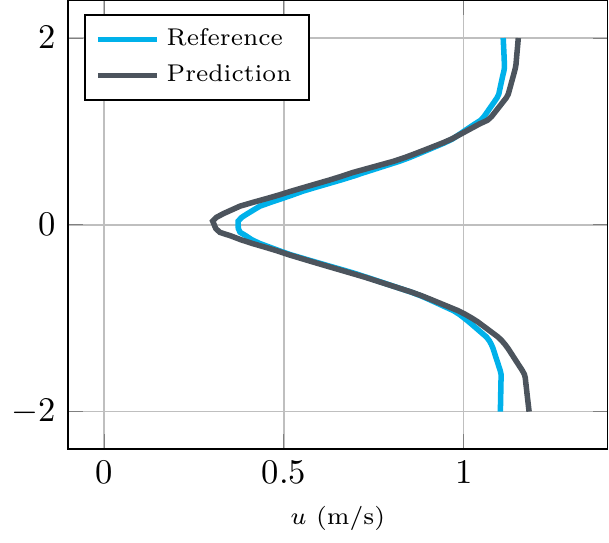}}
%	\begin{tikzpicture}[	trim axis left, trim axis right]
%		\begin{axis}[	scale=0.8, xlabel=$u$ (m/s), %ylabel=$y$ (m),
%					label style={font=\scriptsize}, 
%					tick label style={/pgf/number format/.cd, fixed, precision=3, /tikz/.cd}, 
%					legend style={font=\scriptsize},
%					xmin=-0.1, xmax=1.4,grid=major,
%					legend pos=north west,
%					legend cell align=left
%					]
%		\legend{Reference, Prediction}
%		\addplot[mark=none,ultra thick,myblue1] table[x index=1,y index=0] {naca_x=1.csv};
%		\addplot[mark=none,ultra thick,mygray1] table[x index=4,y index=0] {naca_x=1.csv};
%		\end{axis}
%	\end{tikzpicture}
\caption{$u$ profile at $x=1$}
\label{fig:naca_u_right}
\end{subfigure}

\medskip
\medskip

\begin{subfigure}[t]{.4\linewidth}
	\centering
	\shifttext{-17mm}{\includegraphics{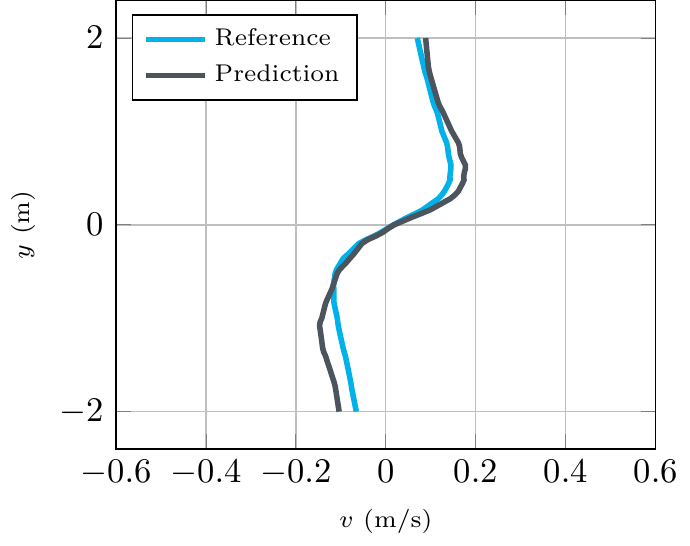}}
%	\begin{tikzpicture}[	trim axis left, trim axis right]
%		\begin{axis}[	scale=0.8, ylabel=$y$ (m),xlabel=$v$ (m/s),
%					label style={font=\scriptsize}, 
%					tick label style={/pgf/number format/.cd, fixed, precision=3, /tikz/.cd}, 
%					legend style={font=\scriptsize},
%					xmin=-0.6, xmax=0.6,grid=major,
%					legend pos=north west,
%					legend cell align=left
%					]
%		\legend{Reference, Prediction}
%		\addplot[mark=none,ultra thick,myblue1] table[x index=2,y index=0] {naca_x=-1.csv};
%		\addplot[mark=none,ultra thick,mygray1] table[x index=5,y index=0] {naca_x=-1.csv};
%		\end{axis}
%	\end{tikzpicture}
\caption{$v$ profile at $x=-1$}
\label{fig:naca_v_left}
\end{subfigure}\qquad
\begin{subfigure}[t]{.4\linewidth}
	\centering
	\shifttext{-5mm}{\includegraphics{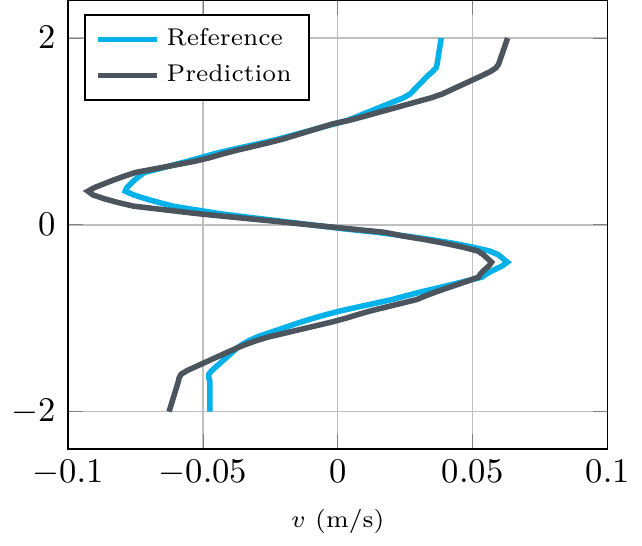}}
%	\begin{tikzpicture}[	trim axis left, trim axis right]
%		\begin{axis}[	scale=0.8, xlabel=$v$ (m/s), %ylabel=$y$ (m),
%					label style={font=\scriptsize}, 
%					tick label style={/pgf/number format/.cd, fixed, precision=3, /tikz/.cd}, 
%					legend style={font=\scriptsize},
%					xmin=-0.1, xmax=0.1,grid=major,
%					legend pos=north west,
%					legend cell align=left
%					]
%		\legend{Reference, Prediction}
%		\addplot[mark=none,ultra thick,myblue1] table[x index=2,y index=0] {naca_x=1.csv};
%		\addplot[mark=none,ultra thick,mygray1] table[x index=5,y index=0] {naca_x=1.csv};
%		\end{axis}
%	\end{tikzpicture}
\caption{$v$ profile at $x=1$}
\label{fig:naca_v_right}
\end{subfigure}

\medskip
\medskip

\begin{subfigure}[t]{.4\linewidth}
	\centering
	\shifttext{-11mm}{\includegraphics{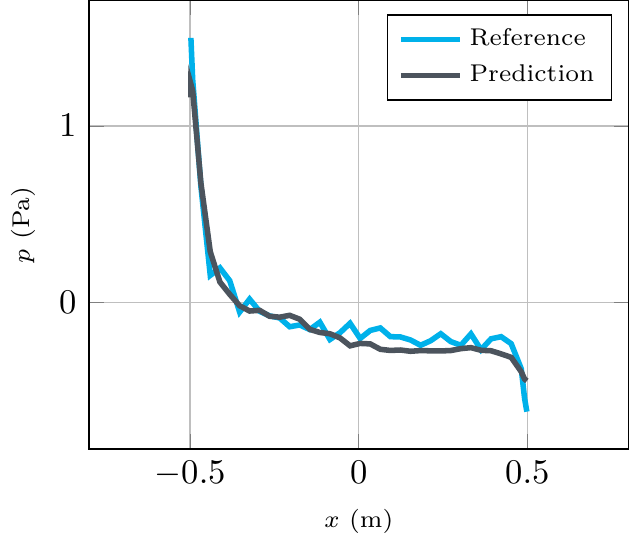}}
%	\begin{tikzpicture}[	trim axis left, trim axis right]
%		\begin{axis}[	scale=0.8, xlabel=$x$ (m), ylabel=$p$ (Pa),
%					label style={font=\scriptsize}, 
%					tick label style={/pgf/number format/.cd, fixed, precision=3, /tikz/.cd}, 
%					legend style={font=\scriptsize},
%					xmin=-0.8, xmax=0.8,grid=major,
%					legend pos=north east,
%					legend cell align=left
%					]
%		\legend{Reference, Prediction}
%		\addplot[mark=none,ultra thick,myblue1] table[x index=0,y index=1] {pressure_low.csv};
%		\addplot[mark=none,ultra thick,mygray1] table[x index=0,y index=2] {pressure_low.csv};
%		\end{axis}
%	\end{tikzpicture}
\caption{Pressure distribution on the lower surface}
\label{fig:naca_p_low}
\end{subfigure}\qquad
\begin{subfigure}[t]{.4\linewidth}
	\centering
	\shifttext{-10mm}{\includegraphics{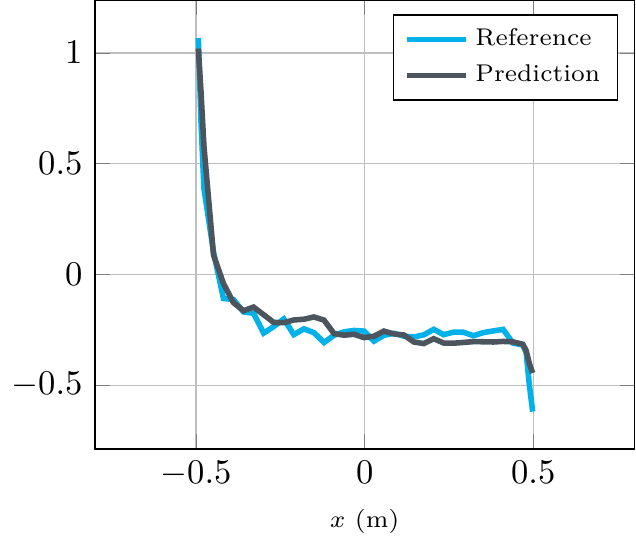}}
%	\begin{tikzpicture}[	trim axis left, trim axis right]
%		\begin{axis}[	scale=0.8, xlabel=$x$ (m),% ylabel=p($Pa$),
%					label style={font=\scriptsize}, 
%					tick label style={/pgf/number format/.cd, fixed, precision=3, /tikz/.cd}, 
%					legend style={font=\scriptsize},
%					xmin=-0.8, xmax=0.8,grid=major,
%					legend pos=north east,
%					legend cell align=left
%					]
%		\legend{Reference, Prediction}
%		\addplot[mark=none,ultra thick,myblue1] table[x index=0,y index=1] {pressure_up.csv};
%		\addplot[mark=none,ultra thick,mygray1] table[x index=0,y index=2] {pressure_up.csv};
%		\end{axis}
%	\end{tikzpicture}
\caption{Pressure distribution on the upper surface}
\label{fig:naca_p_up}
\end{subfigure}
\caption{\textbf{Velocity and pressure profiles around the NACA0012 airfoil.} The velocities are recorded at $x=-1$ (\ref{fig:naca_u_left}, \ref{fig:naca_v_left}) and $x=1$ (\ref{fig:naca_u_right}, \ref{fig:naca_v_right}), and display significant errors around the trailing edge. With significant discrepancy of velocity magnitude, the physical pattern of the velocity profiles is still well recovered. The pressure is recorded on the lower (\ref{fig:naca_p_low}) and upper surfaces (\ref{fig:naca_p_up}) of the airfoil, with \red{remarkable error at the leading and trailing edges}.}
\label{fig:naca_profiles}
\end{figure}

The NACA airfoils from the UIUC database are among the most studied geometries in the CFD community. In the context of this contribution, the prediction on airfoils is regarded as good test of the model generalization capabilities on out-of-training shapes, as their geometry is very different from that of B\'ezier random shapes. The demonstrated airfoil is NACA0012, with an angle of attack equal to $4^{\circ}$ and a chord length equals to $1m$. The airfoil is embedded into a triangular mesh made of 1489 nodes. \red{The referential flow fields are from a CFD resolution with the same configurations as the B\'ezier shapes and the cylinder.} The MAE of the predicted fields obtained from the GCNN model is \num{1.04e-2}, which corresponds to the worst prediction levels observed in the test set. Yet, the predicted fields presented in figure \ref{fig:naca_visualization} display reasonable physical features. Although the model overestimates the velocity magnitude, \red{it} successfully detects the locations where the vertical velocity changes sign or attains local minimum and maximum, as can be further observed in the plots of figure \ref{fig:naca_profiles}. More, the surface pressure distribution is recovered by the model, although with an underestimated pressure magnitude on the leading edge and the trailing edge. The incorrect velocity and pressure magnitude both result from the dissimilarity between flow around the airfoil and flow around the B\'ezier shapes. \red{Since the GCNN model in this contribution is trained under the supervised learning framework, it should be conservatively applied to geometries that are not too different from that of the training dataset.}

%%%%%%%%%%%%
%%%%%%%%%%%% 

%%%%%%%%%%%%
%%%%%%%%%%%%

%%%%%%%%%%%%%%%%%%%%%%%%%%%%%%%%%%%%%%%%%%%%%%%%%%%%%%%%%%
%%%%%%%%%%%%%%%%%%%%%%%%%%%%%%%%%%%%%%%%%%%%%%%%%%%%%%%%%%
\subsection{Drag force prediction}
\label{section:drag force}

\begin{figure}
\centering
\includegraphics{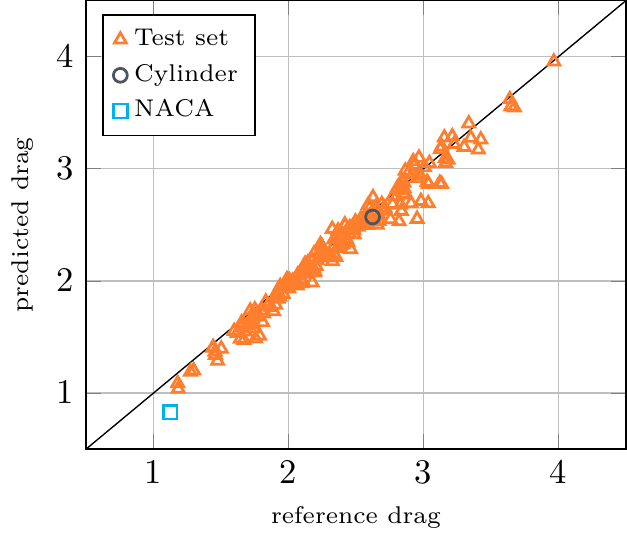}
%\begin{tikzpicture}[trim axis left, trim axis right]
%	\begin{axis}[	scale=0.8, xlabel=reference drag, ylabel=predicted drag,
%				label style={font=\scriptsize}, 
%				tick label style={/pgf/number format/.cd, fixed, precision=3, /tikz/.cd}, 
%				legend style={font=\scriptsize},
%				xmin=0.5, xmax=4.5, ymin=0.5, ymax=4.5, grid=major,
%				legend pos=north west,
%				legend cell align=left
%				]
%	\legend{Test set, Cylinder, NACA}
%	\addplot[only marks,thick, myorange1, mark=triangle] table[x index=0,y index=1] {drag.csv};
%	\addplot[only marks, draw=mygray1, thick, mark=o] coordinates {(2.623,2.570)};
%	\addplot[only marks, draw=myblue1, thick, mark=square] coordinates {(1.122,0.829)}; 
%	\addplot[draw=black, thin] coordinates {(0,0)(4.5,4.5)}; 
%	\end{axis}
%\end{tikzpicture}
\caption{\textbf{Drag force computed from the boundary layer reconstruction obtained with the graph neural network.} The included data contains the \num{200} B\'ezier shapes from the test set, along with the cylinder and the NACA0012 airfoil considered in the previous sections. Excellent agreement is observed on most shapes from the test set, with an average relative error of $3.43\%$. While the cylinder drag is accurately computed, the airfoil drag is off by $26.1\%$ from its reference value.}
\label{fig:drag_force}
\end{figure}

In this section, the accuracy of the drag force prediction is used as a measure of the quality of the predicted boundary layer. The drag force \red{at Re$=10$} is obtained by integrating the edge-wise contribution of pressure and viscous force on the obstacle, as shown in equation (\ref{equation:drag force}).
%%%%%%
\begin{equation}
F_D = \textbf{e}_x \cdot \oint_S \! (p \cdot \textbf{n} + \mu \frac{\partial v_s}{\partial \textbf{s}}) \, \mathrm{d} s,
\label{equation:drag force}
\end{equation}
%%%%%%%
with $\textbf{e}_x$ is the unit vector in the x-direction, $n$ and $s$ repectively the inward- and outward-pointing unit normal vector. $\frac{\partial v_s}{\partial \textbf{s}}$ is the normal gradient of the tangent velocity. The drag forces computed on the \num{200} shapes of the test set are compared with their reference drag values in figure \ref{fig:drag_force}. Overall, a good agreement is observed, with an average relative error of $3.43\%$, and a coefficient of determination equal to \num{0.987}, which is in line with the results of the GraphSage method \cite{ogoke2020graph}. It must be noted that, in the literature, neural networks are usually directly trained on adequate databases to predict drag or lift coefficients. \red{With the drag force computed by reconstructed flow fields on triangular meshes, it is possible to apply the GCNN model to shape optimization in future works.}

%%%%%%%%%%%%
%%%%%%%%%%%%

%%%%%%%%%%%%%%%%%%%%%%%%%%%%%%%%%%%%%%%%%%%%%%%%%%%%%%%%%%
%%%%%%%%%%%%%%%%%%%%%%%%%%%%%%%%%%%%%%%%%%%%%%%%%%%%%%%%%%
\subsection{Comparison with U-nets}
\label{section:Unet}

\begin{figure}
\centering

%% Layer style
%\tikzset{layer/.style={	rectangle, 	thick, rounded corners,
%				fill=myorange4, draw=myorange1,
%				minimum width=\w}}
%\tikzset{down/.style={regular polygon, regular polygon sides=3, 	rotate=-90,
%				very thick, rounded corners,
%				fill=myblue4, draw=myblue1,
%				minimum width=\t}}
%\tikzset{up/.style={	down, rotate=180}}
%\tikzset{inout/.style={	inner sep=0pt, minimum size=0pt}}
%	
%\def\s{0.75}
%\def\t{\s*4cm}
%\def\w{\s*0.4cm}
%\def\lar{\s*2cm}
%\def\med{\s*1.4cm}
%\def\sma{\s*0.8cm}
%\def\io{0.25}

\begin{subfigure}[b]{.4\textwidth}
	\centering
	\includegraphics{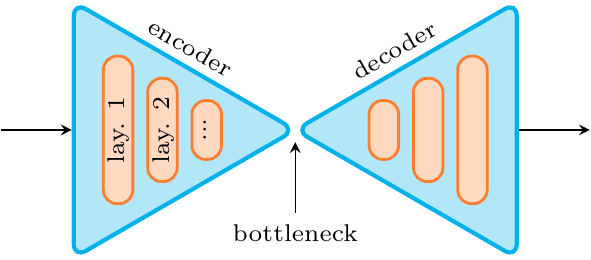}
	
%	\begin{tikzpicture}
%		% Frame
%		\node[down] 	(tri1) at (0.25*\t,0) {};
%		\node[up] 		(tri2) at (\t+0.25*\t,0) {};
%
%		% Input/output
%		\node[inout] (input) at (-\io*\t,0) {};
%		\node[inout] (output) at (1.5*\t+\io*\t,0) {};
%		\draw[-stealth] (input.east) -- (tri1.south);
%		\draw[-stealth] (tri2.south) -- (output.west);
%
%		% Layers
%		\node[layer,minimum height=\lar] 	(down_1) at (0.15*\t,0) {};
%		\node[layer,minimum height=\med] 	(down_2) at (0.3*\t,0) {};
%		\node[layer,minimum height=\sma] 	(down_3) at (0.45*\t,0) {};
%		\node[layer,minimum height=\lar] 	(up_1) at (\t+0.5*\t-0.15*\t,0) {};
%		\node[layer,minimum height=\med] 	(up_2) at (\t+0.5*\t-0.3*\t,0) {};
%		\node[layer,minimum height=\sma] 	(up_3) at (\t+0.5*\t-0.45*\t,0) {};
%		
%		% Texts
%		\node[rotate=90,anchor=north]		at (0.075*\t,0) {\scriptsize lay. 1};
%		\node[rotate=90,anchor=north]		at (0.225*\t,0) {\scriptsize lay. 2};
%		\node[rotate=90,anchor=north]		at (0.4*\t,0) {\scriptsize ...};
%		\node[rotate=-30,anchor=north]		at (0.43*\t,0.34*\t) {\scriptsize encoder};
%		\node[rotate=30,anchor=north]		at (0.5*\t+0.55*\t,0.33*\t) {\scriptsize decoder};
%		
%		% Bottleneck
%		\node[] (bottleneck) at (0.75*\t,0) {};
%		\node[] (bottleneck_txt) at (0.75*\t,-0.35*\t) {\scriptsize bottleneck};
%		\draw[stealth-] (bottleneck.south) to [out=-90,in=90] (bottleneck_txt.north);
%		
%	\end{tikzpicture}
	\caption{Auto-encoder architecture}
	\label{fig:AE}
\end{subfigure} \quad
\begin{subfigure}[b]{.4\textwidth}
	\centering
	\includegraphics{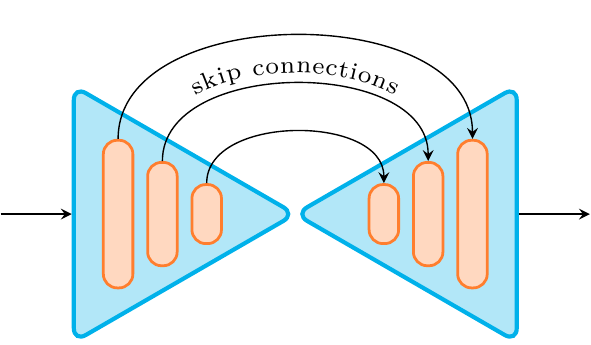}	
%	\begin{tikzpicture}
%		% Frame
%		\node[down] 	(tri1) at (0.25*\t,0) {};
%		\node[up] 		(tri2) at (\t+0.25*\t,0) {};
%
%		% Input/output
%		\node[inout] (input) at (-\io*\t,0) {};
%		\node[inout] (output) at (1.5*\t+\io*\t,0) {};
%		\draw[-stealth] (input.east) -- (tri1.south);
%		\draw[-stealth] (tri2.south) -- (output.west);
%
%		% Layers
%		\node[layer,minimum height=\lar] 	(down_1) at (0.15*\t,0) {};
%		\node[layer,minimum height=\med] 	(down_2) at (0.3*\t,0) {};
%		\node[layer,minimum height=\sma] 	(down_3) at (0.45*\t,0) {};
%		\node[layer,minimum height=\lar] 	(up_1) at (\t+0.5*\t-0.15*\t,0) {};
%		\node[layer,minimum height=\med] 	(up_2) at (\t+0.5*\t-0.3*\t,0) {};
%		\node[layer,minimum height=\sma] 	(up_3) at (\t+0.5*\t-0.45*\t,0) {};
%		
%		% Arrows
%		\def\myshift#1{\raisebox{1ex}}
%		\draw[-stealth] (down_1.north) to [out=90,in=90] (up_1.north);
%		\draw[-stealth, postaction={	decorate,
%								decoration={	text along path,
%											text align=center,
%											text={|\scriptsize\myshift|skip connections}}}] (down_2.north) to [out=90,in=90] (up_2.north);
%		\draw[-stealth] (down_3.north) to [out=90,in=90] (up_3.north);
%	\end{tikzpicture}
	\caption{U-net architecture}
	\label{fig:Unet}
\end{subfigure}

\caption{\textbf{Sketch of auto-encoder and U-net architectures.} Standard auto-encoders (\ref{fig:AE}) are composed of an encoder and a decoder paths, and can be exploited either for end-to-end regression tasks (in a supervised way, with labels), or for the inference of latent space representations (in an unsupervised way, without labels). U-net auto-encoders (\ref{fig:Unet}) are a specific class of AE, in which skip connections are added from the encoder branch to the decoder one in order to mix high-level features from the latent space with low-level one from the contractive path. They usually present a superior level of performance on regression tasks.}
\label{fig:unet}
\end{figure}

U-nets are a popular neural network architecture for end-to-end image prediction. Initially introduced by Ronneberger \etal\cite{UnetRonneberger} in the field of biomedical image segmentaion, they were recently exploited for flow prediction tasks \cite{Thuerey2020,Fukami2018}. U-nets have an auto-encoder-like structure, but introduce skip connections between the corresponding convolutional and de-convolutional layers from the encoder and decoder paths in order to mix high-level features from the latent space with low-level ones from the encoder. A sketch is proposed in figure \ref{fig:unet}. By design, U-nets require grid-based data, such as images. As a consequence, flow fields as well as obstacle boundaries need to be projected onto a regular cartesian grid, implying geometric inaccuracies and a higher amount of data. Here, we consider a cartesian grid with a resolution equal to \num{0.01}, in order to match the local resolution of the boundary layer of the graph network. The domain size is  $[-2,2] \times [-2,2]$, hence each data instance is a $401 \times 401$ 2D array. The U-net architecture is composed of 5 convolution-convolution-maxpooling blocks with $3 \times 3$ convolutional filters and $2 \times 2$ pooling size for the encoder part, and 5 deconvolution-convolution-convolution blocks with $2 \times 2$ deconvolutional filters and $3 \times 3$ convolutional filters for the decoder part. In the encoder branch, the amount of filters is doubled after each block, while in the coder, the amount of filters is halved after each block. The U-net’s architecture is hence symmetric, and scalable through a modification of the initial number of filters in the first block. The relation between the number of initial filters and the number of trainable parameters is shown in table \ref{tab:trainable_parameters}. Similarly, in a GCNN, one can modify the node feature dimension of the first convolutional block while proportionally changing the other intermediate feature dimensions.

%%%%%%%%%%%%
%%%%%%%%%%%%

%%%%%%%%%%%%
\begin{table}[h]
\footnotesize
\centering

\subfloat[U-net]{
\begin{tabular}{ccc}
\toprule
&\textbf{$1^\text{st}$ layer}	&\textbf{Parameters}   \\
\midrule
&\num{7}   &\num{1489162}      \\ 
\midrule
&\num{8}   &\num{1944763}    \\
\midrule
&\num{9}   &\num{2461080}    \\
\midrule
&\num{10}   &\num{3038113}      \\ 
\bottomrule
\label{tab:unet_parameters}
\end{tabular}
}
\quad \quad \quad
\subfloat[GCNN]{
\begin{tabular}{ccc}
\toprule
&\textbf{$1^\text{st}$ layer}	 &\textbf{Parameters} \\
\midrule
&\num{2}  &\num{61190}      \\
\midrule
&\num{4}  &\num{113411}      \\
\midrule
&\num{6}  &\num{165632}      \\
\midrule
&\num{8}   &\num{217853}   \\
\bottomrule
\label{tab:gcnn_parameters}
\end{tabular}
}
\caption{\textbf{Trainable parameters of U-nets and GCNNs}. For simplification, both architectures use symmetric and scalable configurations. Hence, their complexity can be represented by the number of convolutional filters or node feature's dimension in the $1^\text{st}$ convolutional layer.}
\label{tab:trainable_parameters}
\end{table}
%%%%%%%%%%%%

\begin{figure}
\centering
\begin{subfigure}[t]{.4\textwidth}
	\centering
	\shifttext{-6mm}{\includegraphics{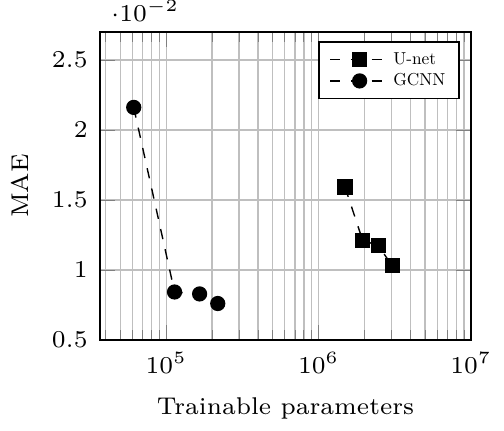}}
%	\begin{tikzpicture}[	trim axis left, trim axis right]
%		\begin{axis}[	scale=0.55,transform shape,
%					label style={font=\scriptsize}, tick label style={font=\scriptsize},
%					ymin=0.005, ymax=0.027, xmin=0, xmax=10000000,
%					xlabel=Trainable parameters,xmode=log,ylabel=MAE,
%					legend style={nodes={scale=0.5, transform shape}},
%					legend pos=north east,
%					legend cell align={left},
%					clip=true, grid=both
%					]
%			\legend{U-net, GCNN}
%			\addplot[dashed, mark=square*, mark options={scale=1,solid}, color=black] table[x index=0,y index=1] {compareUnet.csv};
%			\addplot[dashed, mark=*, mark options={scale=1,solid}, color=black] table[x index=2,y index=3] {compareUnet.csv};
%		\end{axis}	
%	\end{tikzpicture}
	\caption{MAE on the test set.}
	\label{fig:compareUnet}
\end{subfigure}\qquad
\begin{subfigure}[t]{.4\textwidth}
	\centering
	\shifttext{-6mm}{\includegraphics{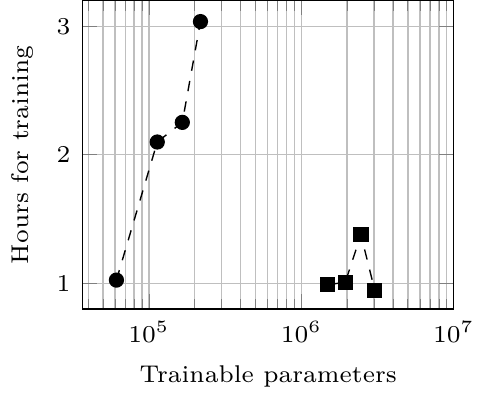}}
%	\begin{tikzpicture}[	trim axis left, trim axis right]
%		\begin{axis}[	scale=0.55,transform shape,
%					label style={font=\scriptsize}, tick label style={font=\scriptsize},
%					ymin=0.8, ymax=3.2, xmin=0, xmax=10000000,
%					xlabel=Trainable parameters,xmode=log,ylabel=Hours for training,
%					clip=true, grid=both
%					]
%			\addplot[dashed, mark=square*, mark options={scale=1,solid}, color=black] table[x index=0,y index=1] {compareAllTime.csv};
%			\addplot[dashed, mark=*, mark options={scale=1,solid}, color=black] table[x index=2,y index=3] {compareAllTime.csv};
%		\end{axis}	
%	\end{tikzpicture}
\caption{Total training time.}
\label{fig:compareAllTime}
\end{subfigure}

\medskip
\medskip

\begin{subfigure}[t]{.4\textwidth}
	\centering
	\shifttext{-6mm}{\includegraphics{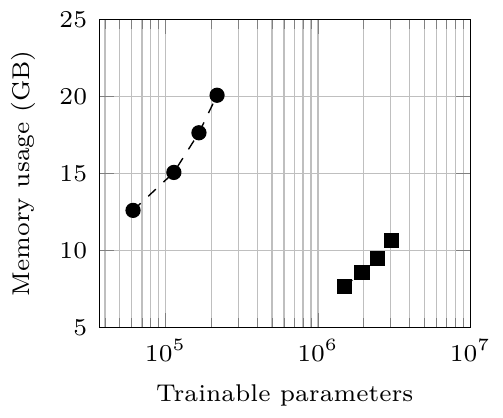}}
%	\begin{tikzpicture}[	trim axis left, trim axis right]
%		\begin{axis}[	scale=0.55,transform shape,
%					label style={font=\scriptsize}, tick label style={font=\scriptsize},
%					ymin=5, ymax=25, xmin=0, xmax=10000000,
%					xlabel=Trainable parameters,xmode=log,ylabel=Memory usage (GB),
%					clip=true, grid=both
%					]
%			\addplot[dashed, mark=square*, mark options={scale=1,solid}, color=black] table[x index=0,y index=1] {compareMemory.csv};
%			\addplot[dashed, mark=*, mark options={scale=1,solid}, color=black] table[x index=2,y index=3] {compareMemory.csv};
%		\end{axis}	
%	\end{tikzpicture}
\caption{Required memory for gradient backpropagation.}
\label{fig:compareMemory}
\end{subfigure}\qquad
\begin{subfigure}[t]{.4\textwidth}
	\centering
	\shifttext{-6mm}{\includegraphics{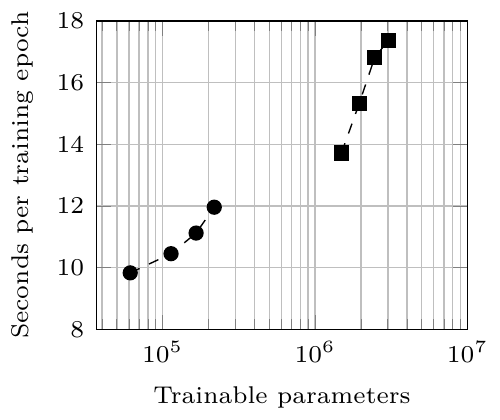}}
%	\begin{tikzpicture}[	trim axis left, trim axis right]
%		\begin{axis}[	scale=0.55,transform shape,
%					label style={font=\scriptsize}, tick label style={font=\scriptsize},
%					ymin=8, ymax=18, xmin=0, xmax=10000000,
%					xlabel=Trainable parameters,xmode=log,ylabel=Seconds per training epoch,
%					clip=true, grid=both
%					]
%			\addplot[dashed, mark=square*, mark options={scale=1,solid}, color=black] table[x index=0,y index=1] {compareTime.csv};
%			\addplot[dashed, mark=*, mark options={scale=1,solid}, color=black] table[x index=2,y index=3] {compareTime.csv};
%		\end{axis}	
%	\end{tikzpicture}
\caption{Training time per epoch.}
\label{fig:compareTime}
\end{subfigure}

\caption{\textbf{Performance comparison between U-net and GCNN} in terms of test set MAE (\ref{fig:compareUnet}), total training time (\ref{fig:compareAllTime}), memory usage (\ref{fig:compareMemory}) and per-epoch training time (\ref{fig:compareTime}). As can be seen, GCNNs require much fewer parameters to reach the same accuracy of U-nets. However, training GCNNs \red{requires more memory to perform gradient backpropagation} and is more time-consuming considering their smaller model size.}
\end{figure}

To perform a performance comparison between U-net and GCNN architectures, the U-nets are implemented and trained in the Keras API with the following configurations: batch size equals to \num{64}, optimized by the Adam optimizer with an initial learning rate equals to \num{0.001}, a learning rate decay equals to 0.0002, early stopping patience is \num{20} epochs. A sample prediction obtained with a U-net holding \num{1944763} free parameters is shown in figure \ref{fig:predUnet}. The prediction on this sample is globally of high quality with an MAE equals to \num{7.09e-3}, but contrarily to GCNN predictions, most of the discrepancies are concentrated on the border of the obstacle. Indeed, the U-net not only predicts the velocity and pressure fields, but also has to reconstruct the obstacle boundary, which cannot be strictly restored due to the staircasing effect of the cartesian grid. To perform a fair quantitative comparison with GCNN prediction, the results obtained with the U-net model were interpolated on the triangular meshes. The average MAE on the test set is used for comparison, and is displayed in figure \ref{fig:compareUnet} as a function of the number of trainable parameters, both for GCNN and U-net. As can be observed, the GCNN require fewer trainable parameters than U-net (approximately 1 order of magnitude) to reach a similar accuracy, highlighting the cost-efficient of GCNN in field reconstruction compared to grid-based techniques. Indeed, the GCNN is able to exploit local grid refinement or coarsening to save degrees of freedom, which is a central strength of finite-element and finite-volume CFD methods. In terms of memory consumption, the GCNN requires 1.05 GB to import the data set (including the nodes coordinates and connectivity, and the velocity and pressure fields), which must be compared to the 13.5 GB required by the U-net architecture for the same task. Hence, the data used for GCNN is more likely to fit into standard GPU cards.

Yet, the main observed drawback of GCNNs is the larger training time required to reached a converged model. Compared to a traditional convolution, the 4-step graph convolutional block presented in equations (\ref{equation:graph_convolution}) and (\ref{equation:graph_smoothing}) requires heavy matrix operations, which significantly increases the time required for a forward calculation. In order to implement automatic differentiation, the generated intermediate variables are not deleted from the memory until one whole back-propagation of the gradients in all the blocks is finished. \red{The required memory to store the intermediate variables are compared in \ref{fig:compareMemory}} when batch size is \num{64}. In spite of the small input data size, GCNNs still require large memory for the intermediate variables in the training process. The limitation of memory thus does not support using large batch size, which also slows the training. In figure \ref{fig:compareAllTime} and \ref{fig:compareTime}, the training time of both methods are compared when batch size is 64. Considering the input size as well as the network's complexity, training GCNNs is more expensive than training U-nets.

%%%%%%%%%%%%
\begin{figure}
\centering
\begin{subfigure}[t]{.25\linewidth}
	\centering
	\fbox{\includegraphics[width=.8\linewidth]{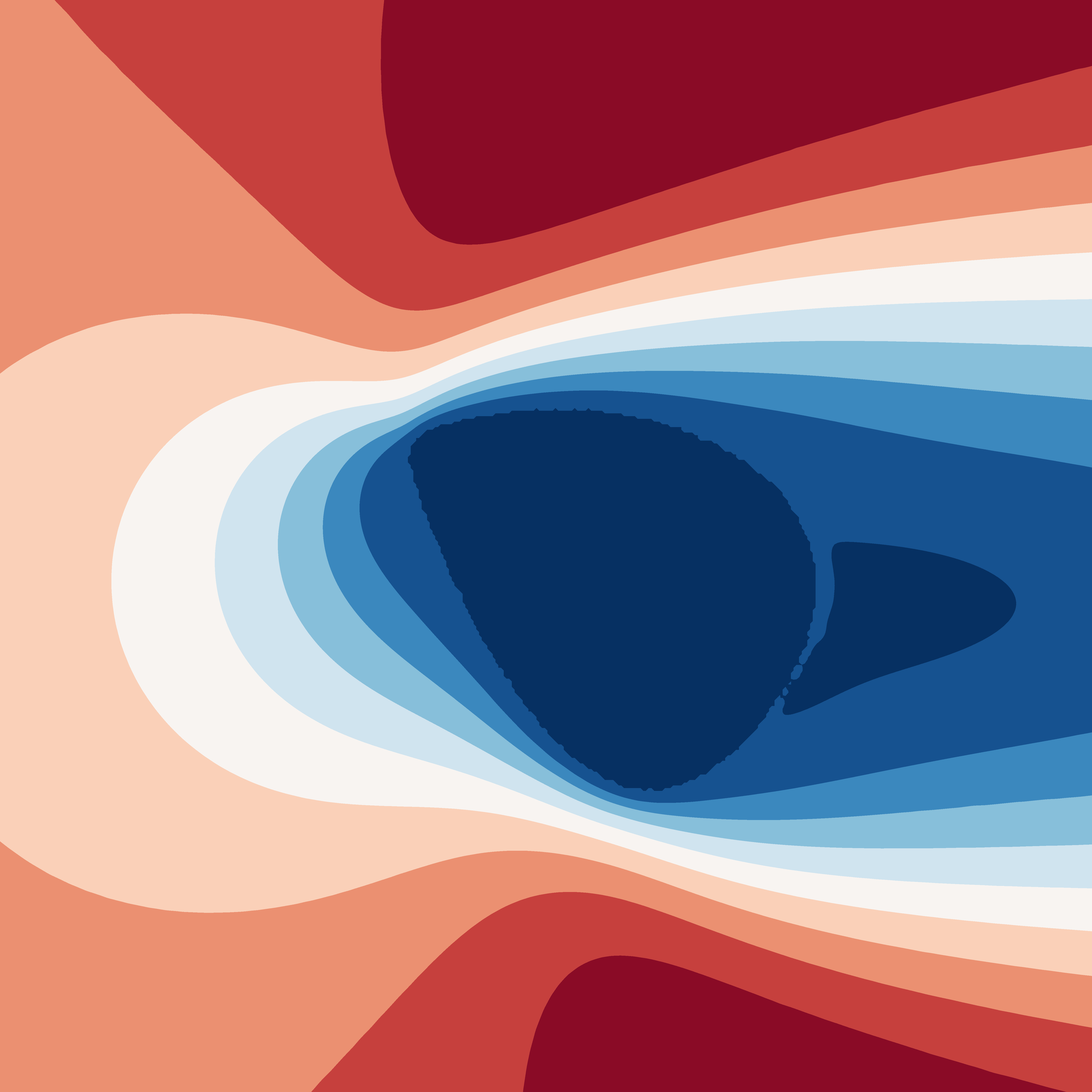}}
	\caption{$u$, reference}
	\label{fig:unet_u_ref}
\end{subfigure}%
\begin{subfigure}[t]{.25\linewidth}
	\centering
	\fbox{\includegraphics[width=.8\linewidth]{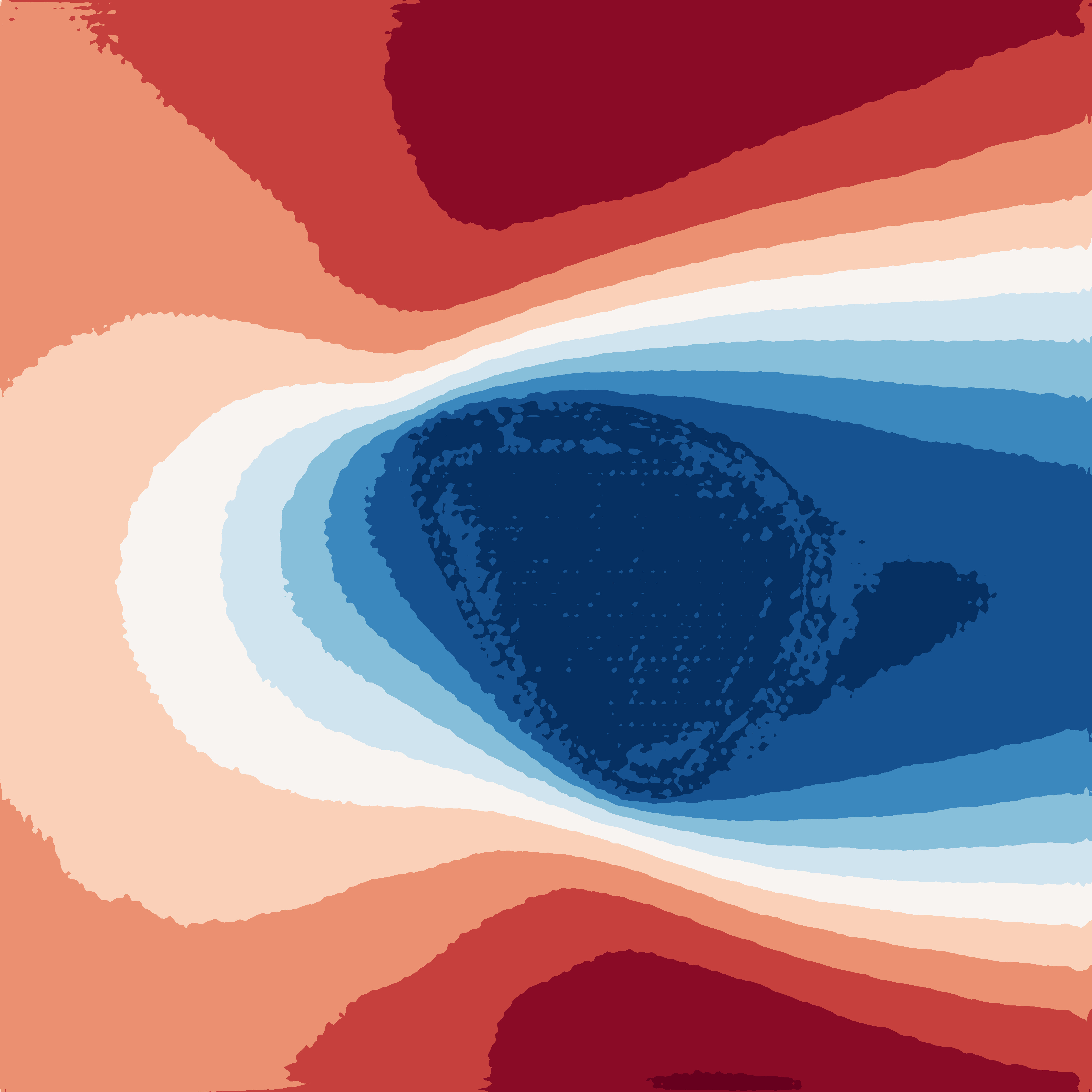}}
	\caption{$u$, predicted}
	\label{fig:unet_u_pred}
\end{subfigure}%
\begin{subfigure}[t]{.1\textwidth}
	\centering
	\raisebox{-2.7mm}{\includegraphics{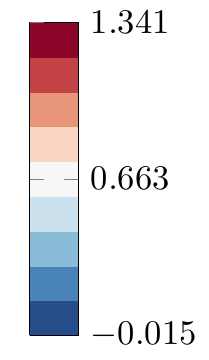}}
	%\raisebox{-2.7mm}{\colorbar{-0.014665}{1.34147}{3.18}}
\end{subfigure}\qquad%
\begin{subfigure}[t]{.25\linewidth}
	\centering
	\fbox{\includegraphics[width=.8\linewidth]{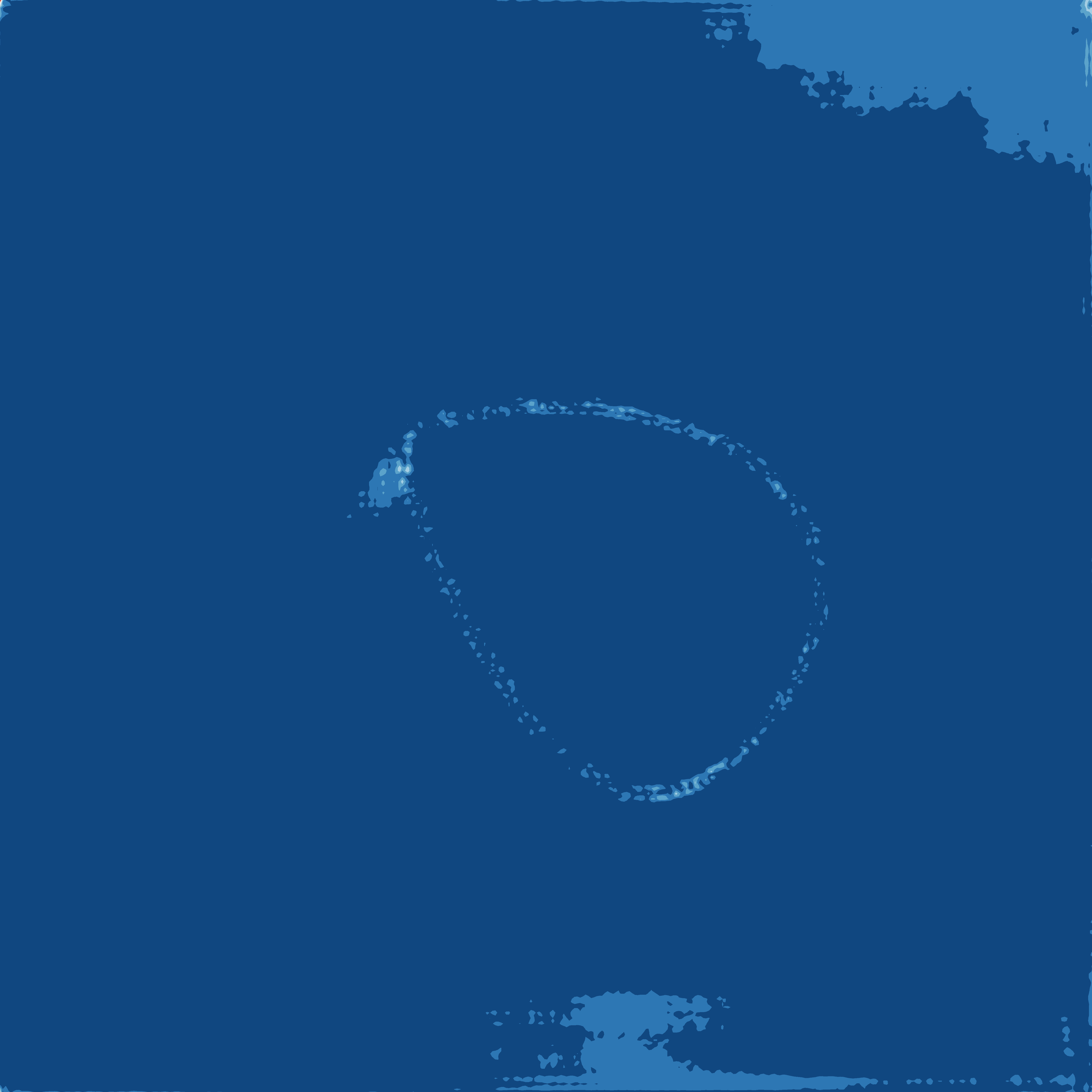}}
	\caption{$u$, absolute error}
	\label{fig:unet_u_err}
\end{subfigure}%
\begin{subfigure}[t]{.1\textwidth}
	\centering
	\raisebox{-2.7mm}{\includegraphics{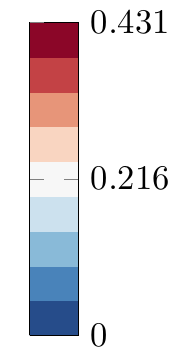}}
	%\raisebox{-2.7mm}{\colorbar{0}{0.431306}{3.18}}
\end{subfigure}%

\medskip

\begin{subfigure}[t]{.25\linewidth}
	\centering
	\fbox{\includegraphics[width=.8\linewidth]{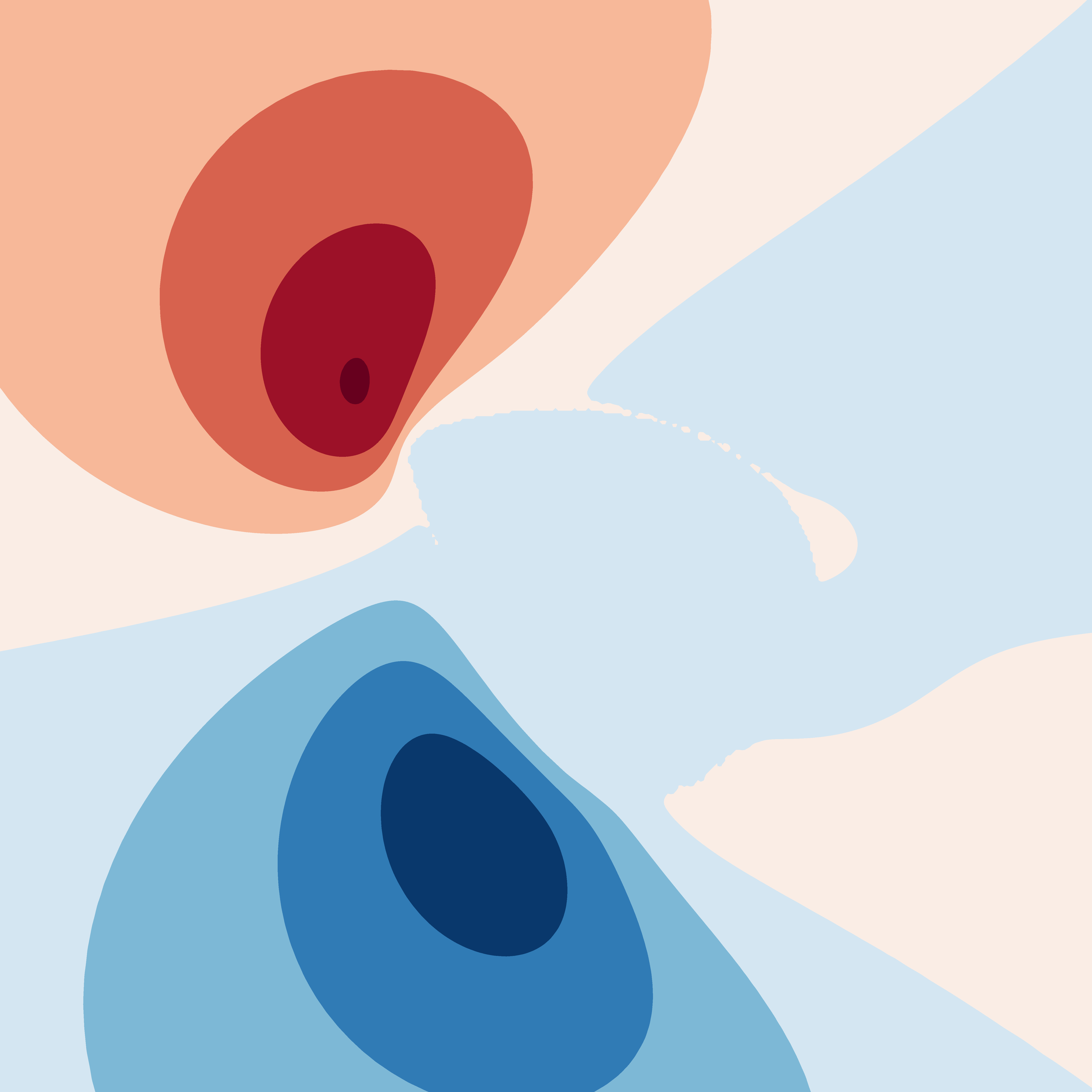}}
	\caption{$v$, reference}
	\label{fig:unet_v_ref}
\end{subfigure}%
\begin{subfigure}[t]{.25\linewidth}
	\centering
	\fbox{\includegraphics[width=.8\linewidth]{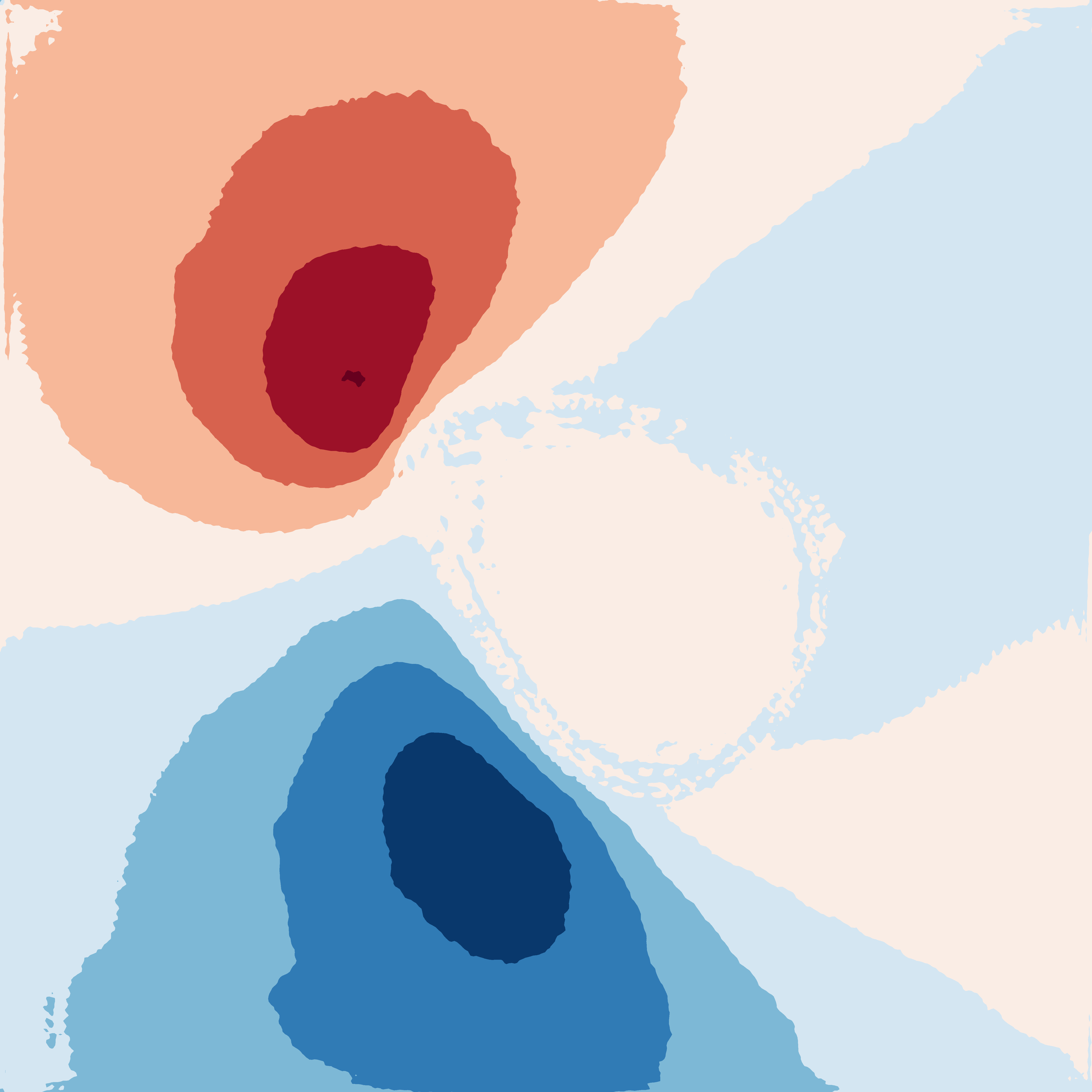}}
	\caption{$v$, predicted}
	\label{fig:unet_v_pred}
\end{subfigure}%
\begin{subfigure}[t]{.1\textwidth}
	\centering
	\raisebox{-2.7mm}{\includegraphics{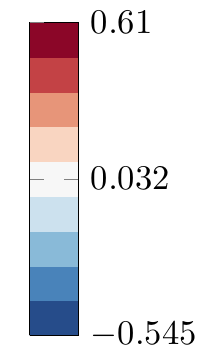}}
	%\raisebox{-2.7mm}{\colorbar{-0.54483}{0.609577}{3.18}}
\end{subfigure}\qquad%
\begin{subfigure}[t]{.25\linewidth}
	\centering
	\fbox{\includegraphics[width=.8\linewidth]{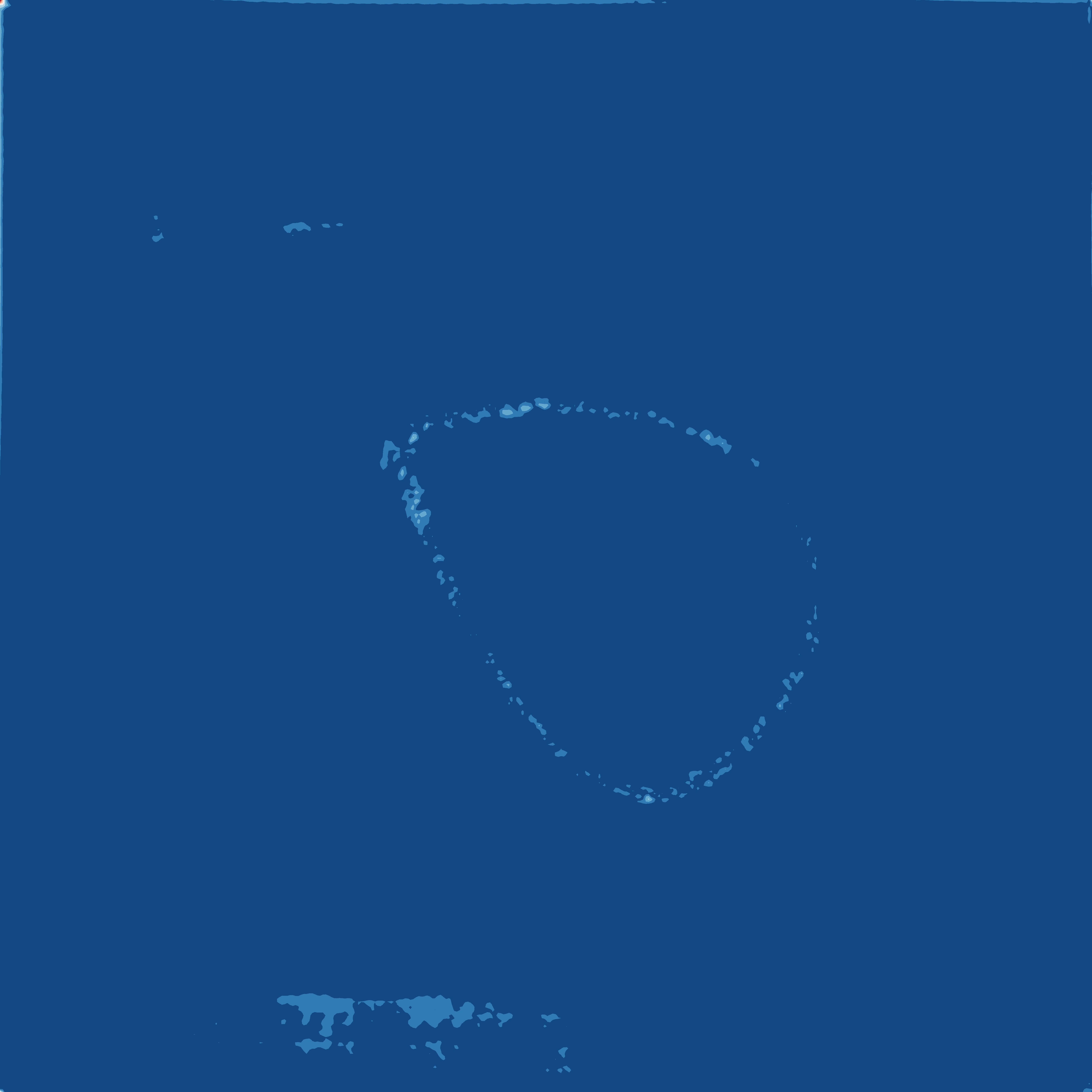}}
	\caption{$v$, absolute error}
	\label{fig:unet_v_err}
\end{subfigure}%
\begin{subfigure}[t]{.1\textwidth}
	\centering
	\raisebox{-2.7mm}{\includegraphics{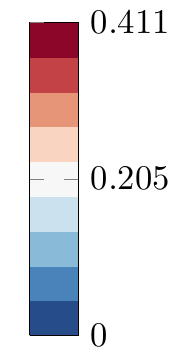}}
	%\raisebox{-2.7mm}{\colorbar{0}{0.410686}{3.18}}
\end{subfigure}%

\medskip

\begin{subfigure}[t]{.25\linewidth}
	\centering
	\fbox{\includegraphics[width=.8\linewidth]{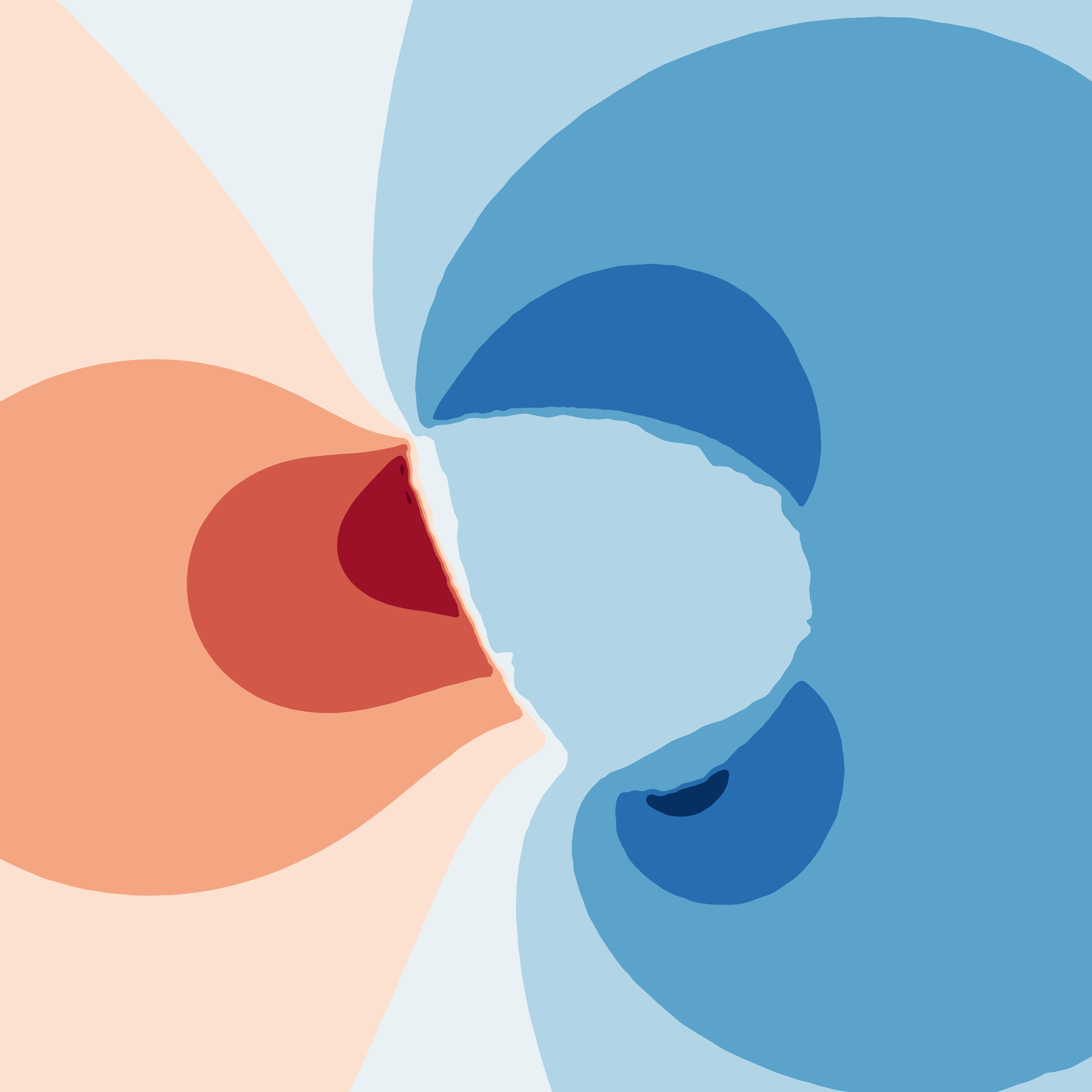}}
	\caption{$p$, reference}
	\label{fig:unet_p_ref}
\end{subfigure}%
\begin{subfigure}[t]{.25\linewidth}
	\centering
	\fbox{\includegraphics[width=.8\linewidth]{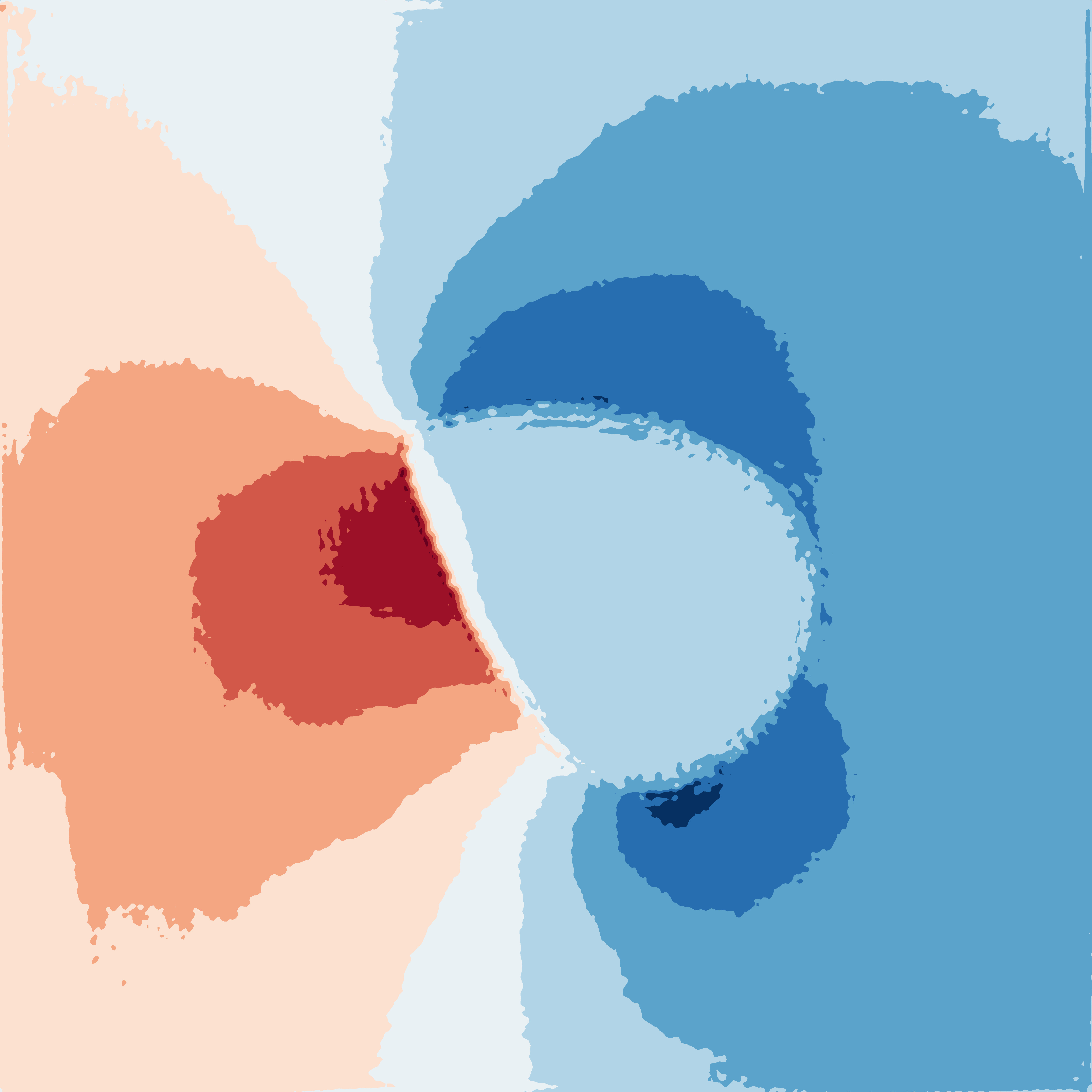}}
	\caption{$p$, predicted}
	\label{fig:unet_p_pred}
\end{subfigure}%
\begin{subfigure}[t]{.1\textwidth}
	\centering
	\raisebox{-2.7mm}{\includegraphics{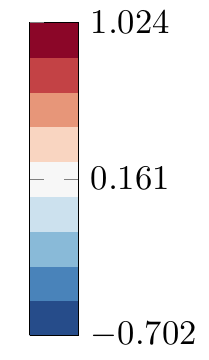}}
	%\raisebox{-2.7mm}{\colorbar{-0.70241}{1.023732}{3.18}}
\end{subfigure}\qquad%
\begin{subfigure}[t]{.25\linewidth}
	\centering
	\fbox{\includegraphics[width=.8\linewidth]{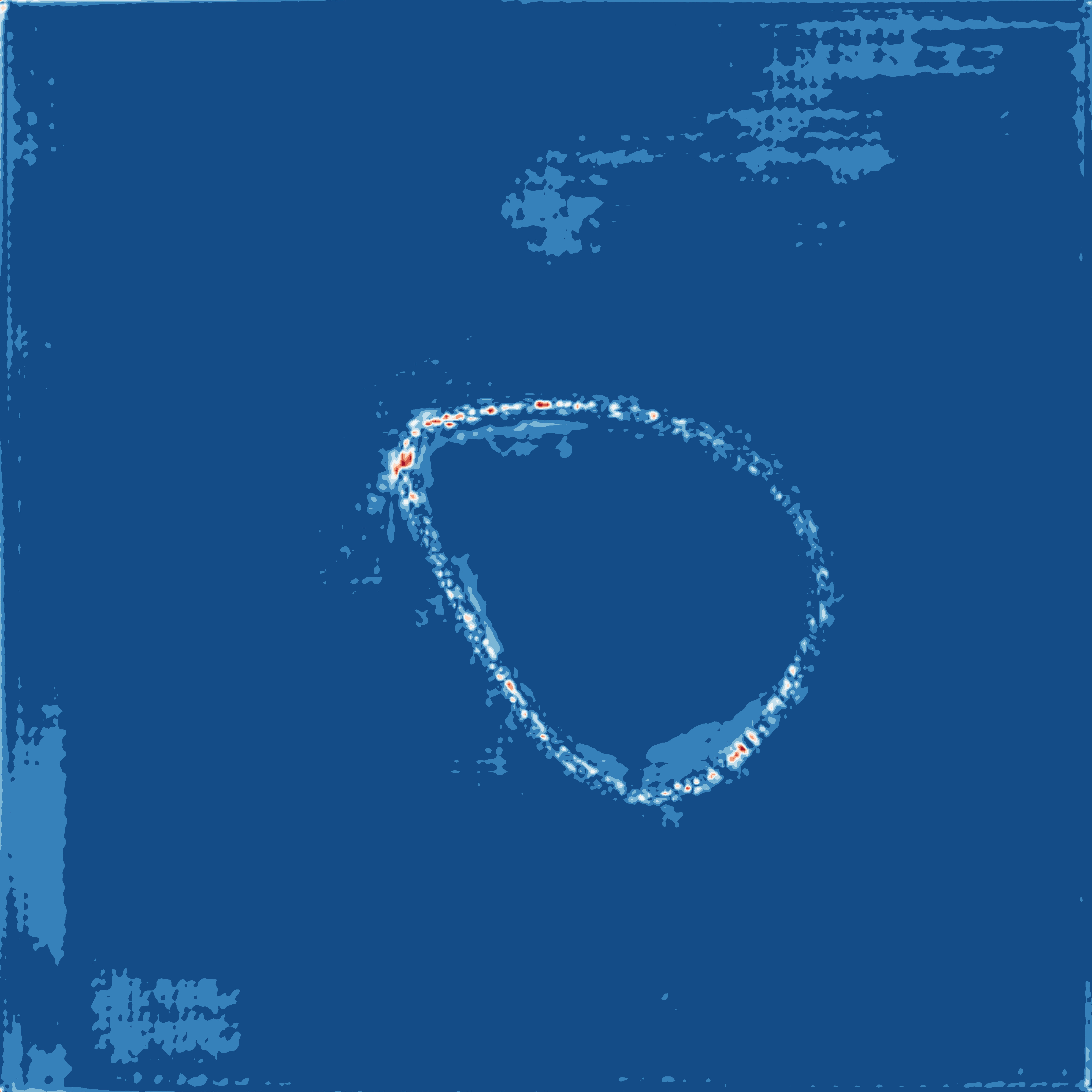}}
	\caption{$p$, absolute error}
	\label{fig:unet_p_err}
\end{subfigure}%
\begin{subfigure}[t]{.1\textwidth}
	\centering
	\raisebox{-2.7mm}{\includegraphics{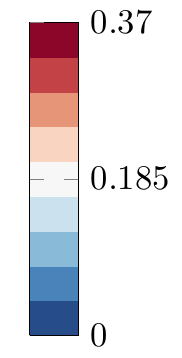}}
	%\raisebox{-2.7mm}{\colorbar{0}{0.369598}{3.18}}
\end{subfigure}%

\caption{\textbf{Flow predictions on a B\'ezier shape from the test set using a U-net model.} The horizontal velocity (\ref{fig:unet_u_ref}, \ref{fig:unet_u_pred}, \ref{fig:unet_u_err}), vertical velocity (\ref{fig:unet_v_ref}, \ref{fig:unet_v_pred}, \ref{fig:unet_v_err}) and pressure fields (\ref{fig:unet_p_ref}, \ref{fig:unet_p_pred}, \ref{fig:unet_p_err}) display similar error levels. The trained U-net has 8 filters in its first convolutional layer, leading to \num{1944763} trainable parameters in total. The error is mostly concentrated in the border of the obstacle.}
\label{fig:predUnet}
\end{figure}
%%%%%%%%%%%%

%%%%%%%%%%%%
\begin{figure}
\centering
\begin{subfigure}[t]{.25\linewidth}
	\centering
	\fbox{\includegraphics[width=.8\linewidth]{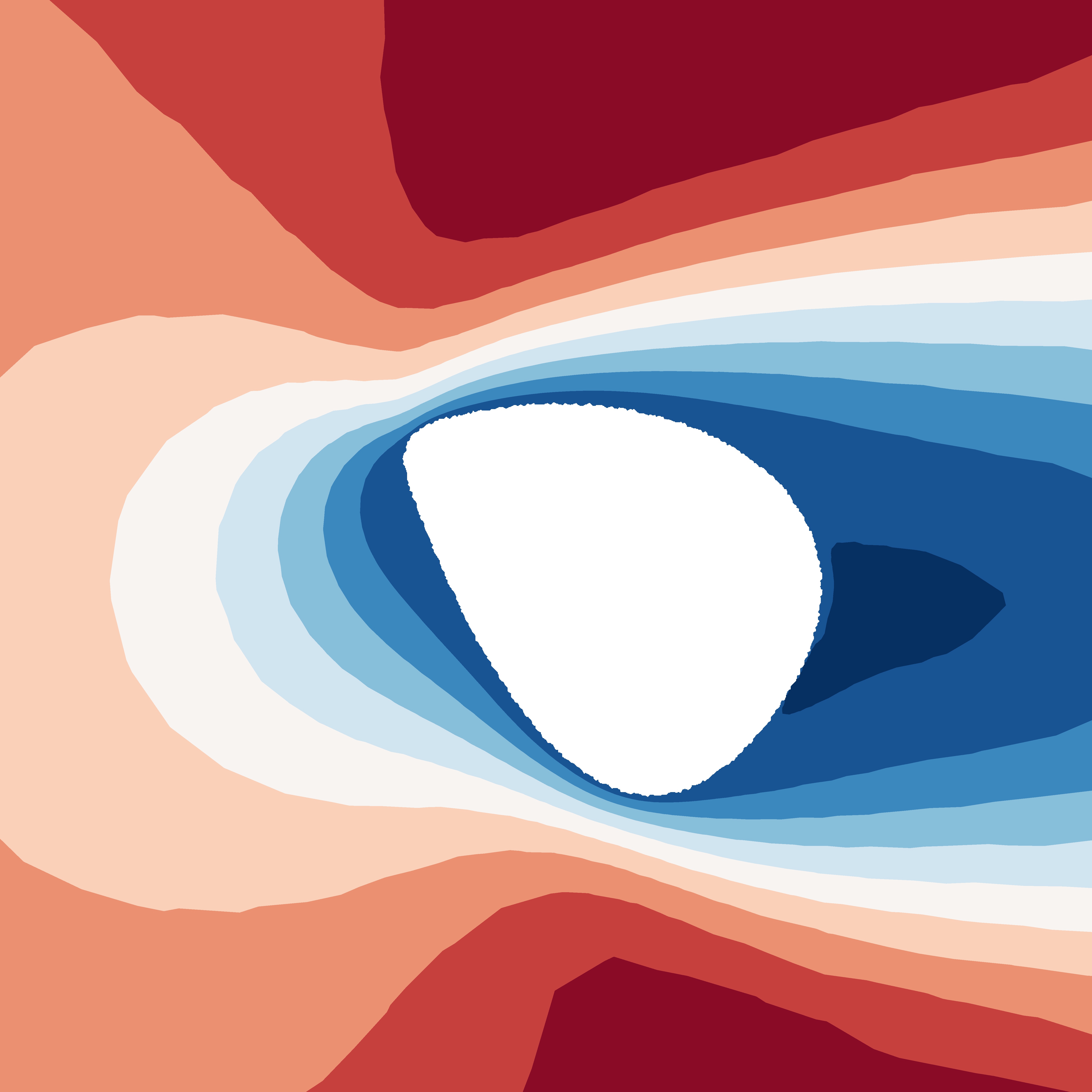}}
	\caption{$u$, reference}
	\label{fig:gcnn_u_ref}
\end{subfigure}%
\begin{subfigure}[t]{.25\linewidth}
	\centering
	\fbox{\includegraphics[width=.8\linewidth]{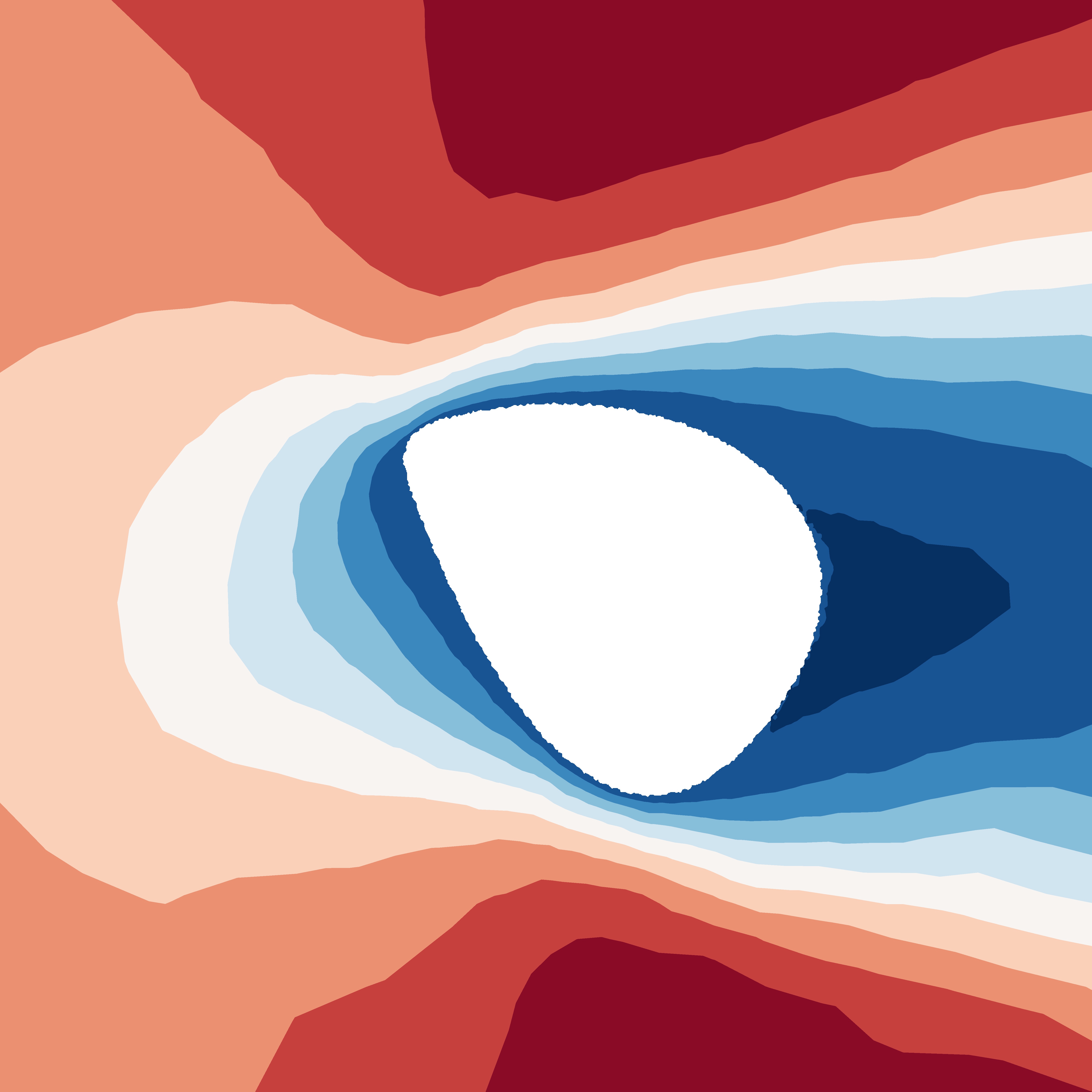}}
	\caption{$u$, predicted}
	\label{fig:gcnn_u_pred}
\end{subfigure}%
\begin{subfigure}[t]{.1\textwidth}
	\centering
	\raisebox{-2.7mm}{\includegraphics{main-figure37.pdf}}
	%\raisebox{-2.7mm}{\colorbar{-0.014665}{1.34147}{3.18}}
\end{subfigure}\qquad%
\begin{subfigure}[t]{.25\linewidth}
	\centering
	\fbox{\includegraphics[width=.8\linewidth]{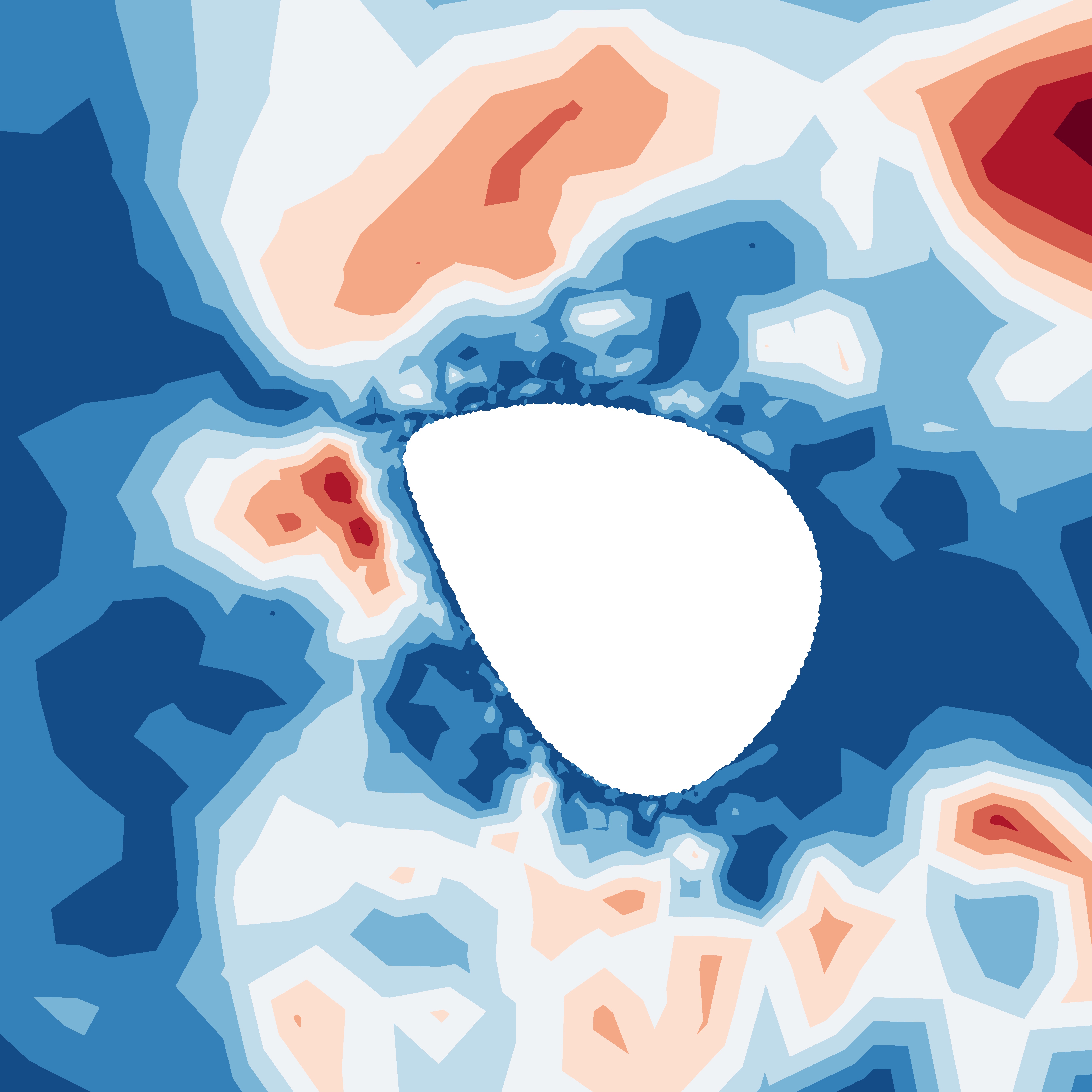}}
	\caption{$u$, absolute error}
	\label{fig:gcnn_u_err}
\end{subfigure}%
\begin{subfigure}[t]{.1\textwidth}
	\centering
	\raisebox{-2.7mm}{\includegraphics{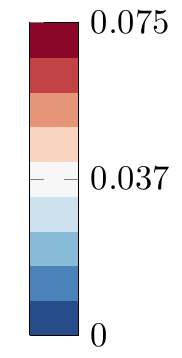}}
	%\raisebox{-2.7mm}{\colorbar{0}{0.075}{3.18}}
\end{subfigure}%

\medskip

\begin{subfigure}[t]{.25\linewidth}
	\centering
	\fbox{\includegraphics[width=.8\linewidth]{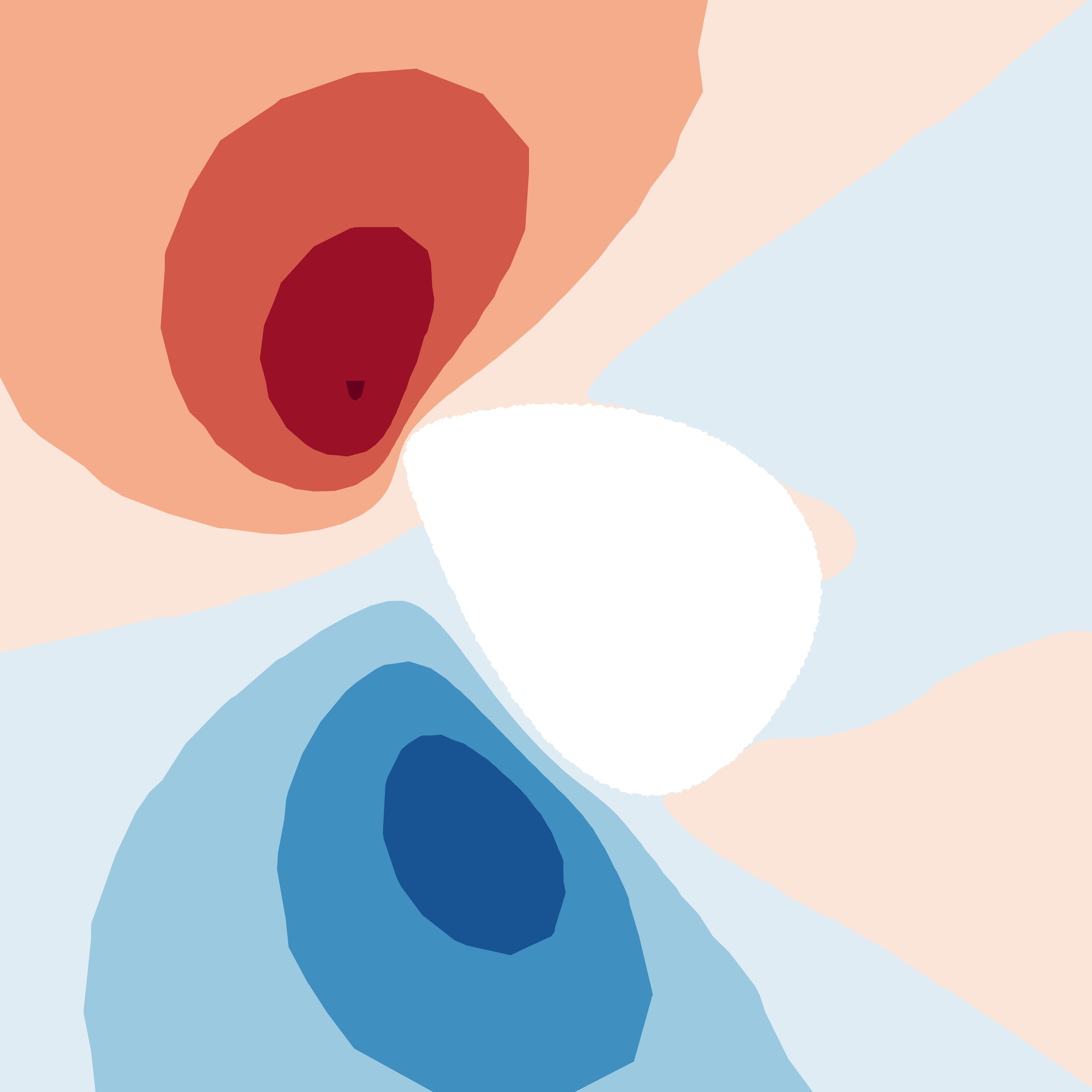}}
	\caption{$v$, reference}
	\label{fig:gcnn_v_ref}
\end{subfigure}%
\begin{subfigure}[t]{.25\linewidth}
	\centering
	\fbox{\includegraphics[width=.8\linewidth]{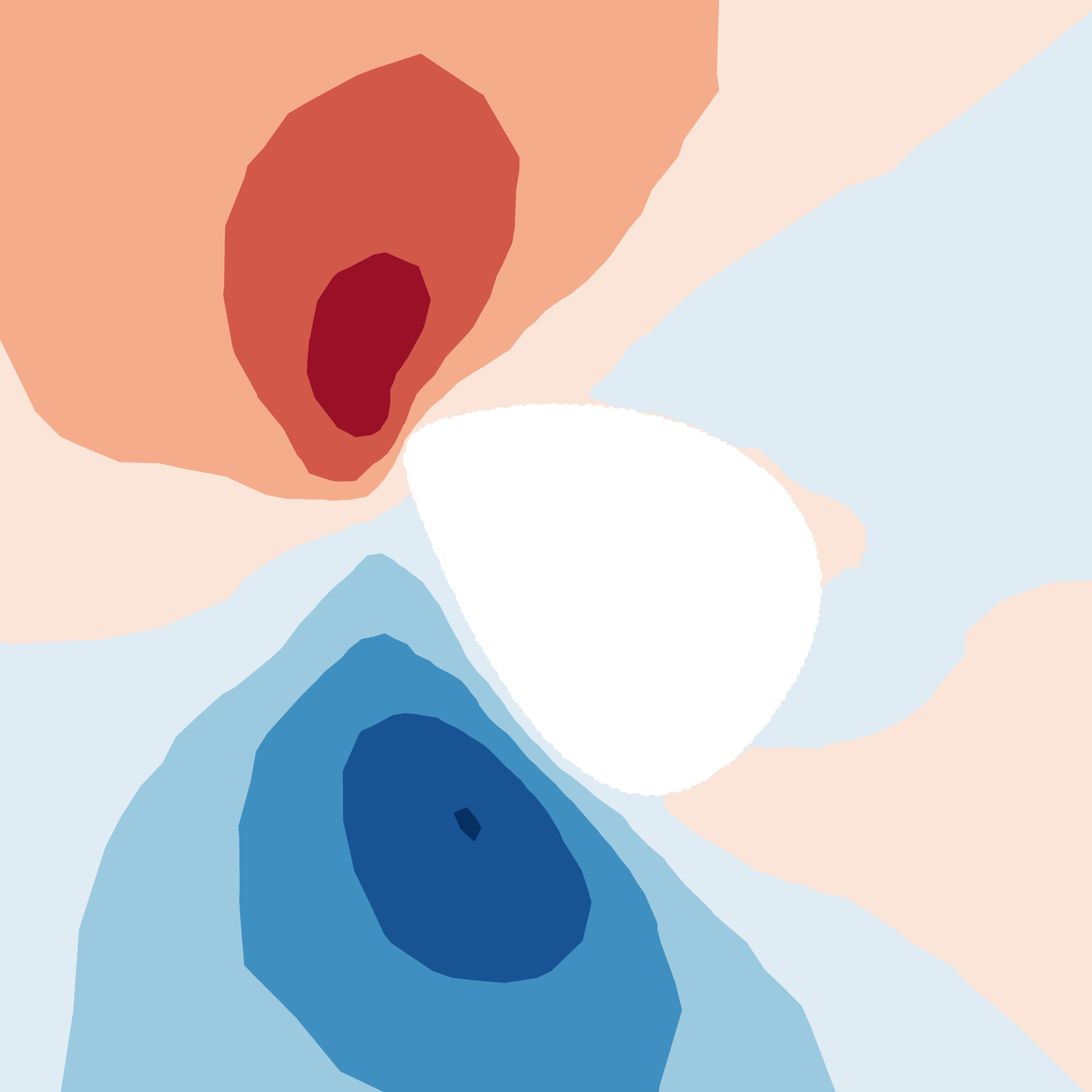}}
	\caption{$v$, predicted}
	\label{fig:gcnn_v_pred}
\end{subfigure}%
\begin{subfigure}[t]{.1\textwidth}
	\centering
	\raisebox{-2.7mm}{\includegraphics{main-figure39.pdf}}
	%\raisebox{-2.7mm}{\colorbar{-0.54483}{0.609577}{3.18}}
\end{subfigure}\qquad%
\begin{subfigure}[t]{.25\linewidth}
	\centering
	\fbox{\includegraphics[width=.8\linewidth]{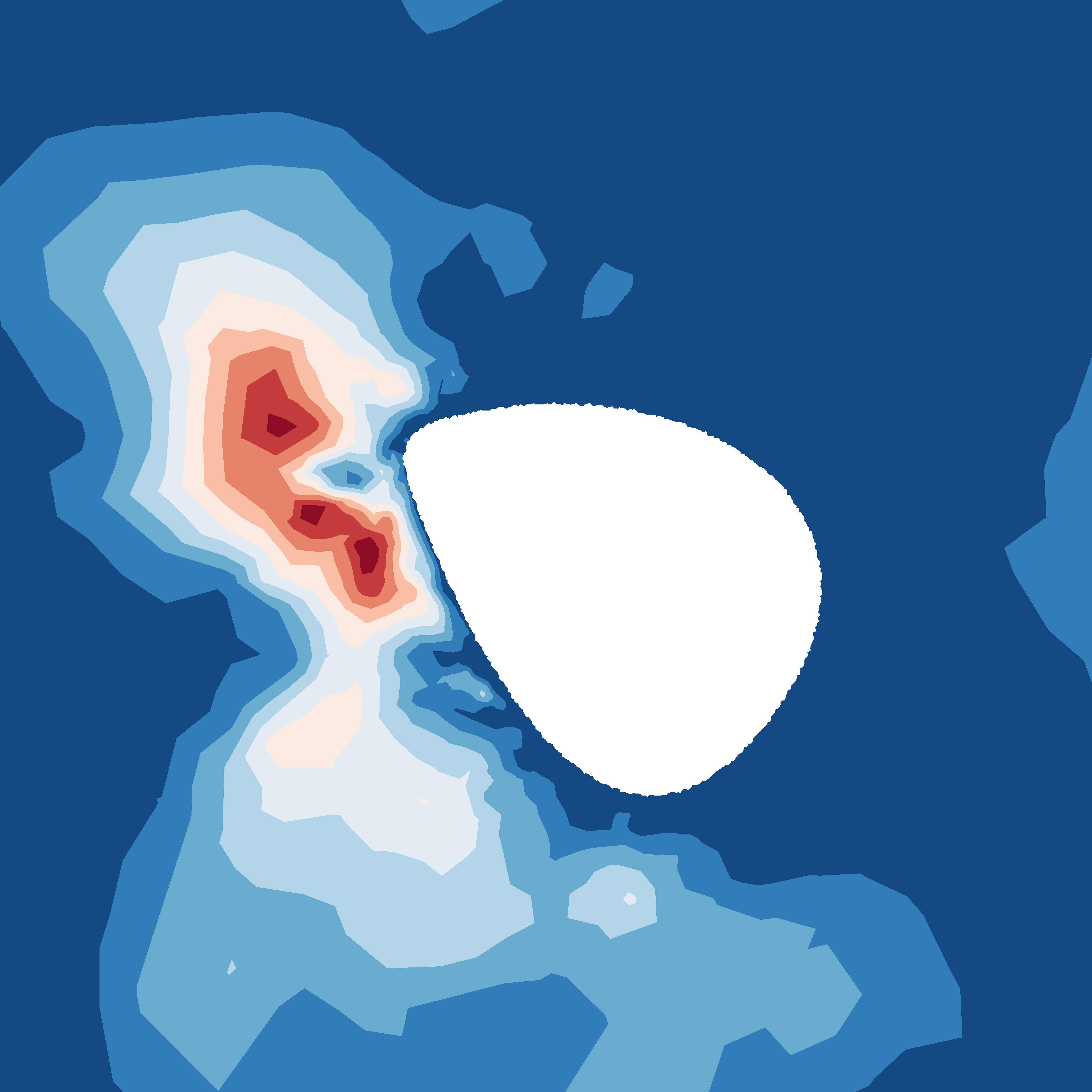}}
	\caption{$v$, absolute error}
	\label{fig:gcnn_v_err}
\end{subfigure}%
\begin{subfigure}[t]{.1\textwidth}
	\centering
	\raisebox{-2.7mm}{\includegraphics{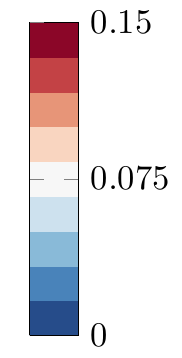}}
	%\raisebox{-2.7mm}{\colorbar{0}{0.15}{3.18}}
\end{subfigure}%

\medskip

\begin{subfigure}[t]{.25\linewidth}
	\centering
	\fbox{\includegraphics[width=.8\linewidth]{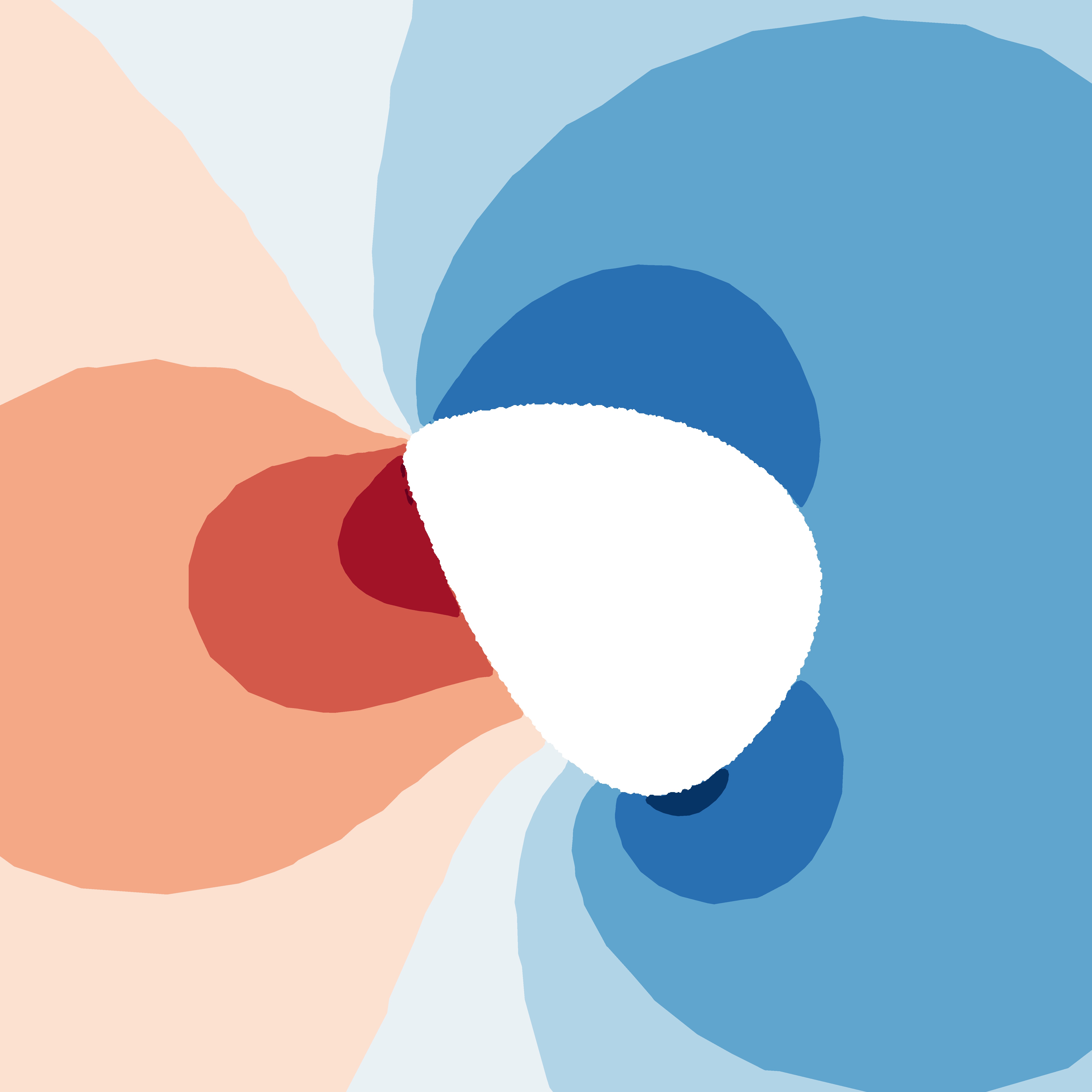}}
	\caption{$p$, reference}
	\label{fig:gcnn_p_ref}
\end{subfigure}%
\begin{subfigure}[t]{.25\linewidth}
	\centering
	\fbox{\includegraphics[width=.8\linewidth]{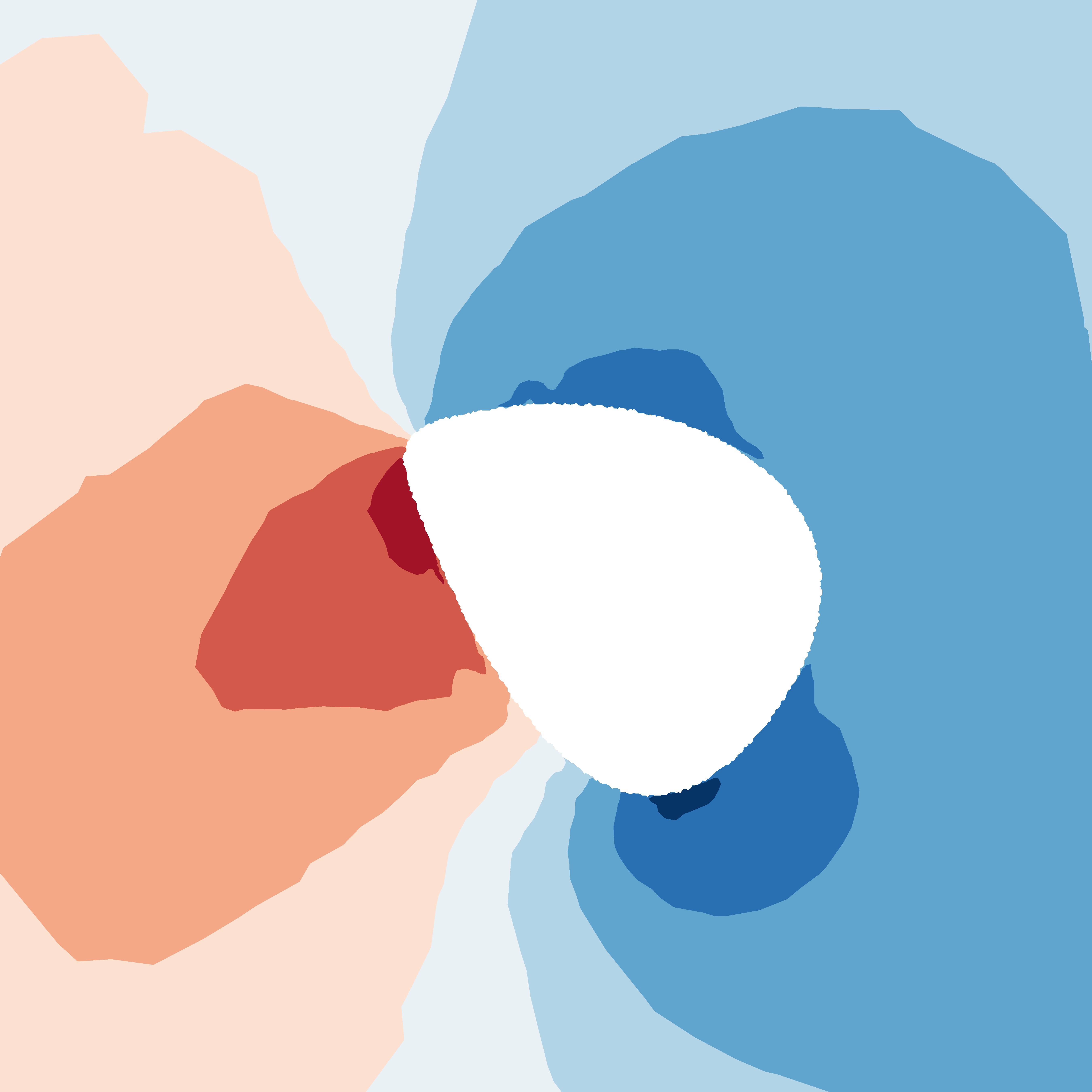}}
	\caption{$p$, predicted}
	\label{fig:gcnn_p_pred}
\end{subfigure}%
\begin{subfigure}[t]{.1\textwidth}
	\centering
	\raisebox{-2.7mm}{\includegraphics{main-figure41.pdf}}
	%\raisebox{-2.7mm}{\colorbar{-0.70241}{1.023732}{3.18}}
\end{subfigure}\qquad%
\begin{subfigure}[t]{.25\linewidth}
	\centering
	\fbox{\includegraphics[width=.8\linewidth]{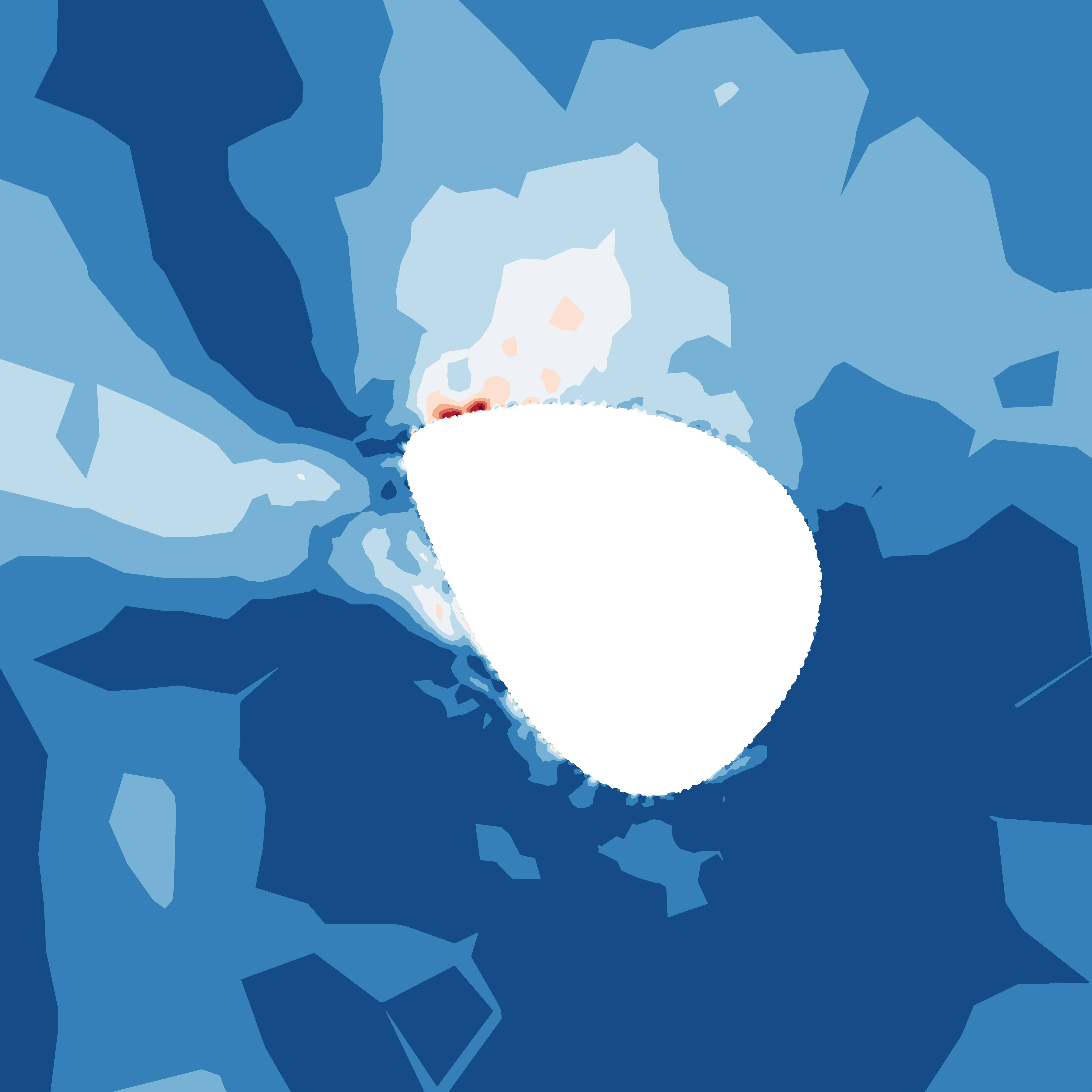}}
	\caption{$p$, absolute error}
	\label{fig:gcnn_p_err}
\end{subfigure}%
\begin{subfigure}[t]{.1\textwidth}
	\centering
	\raisebox{-2.7mm}{\includegraphics{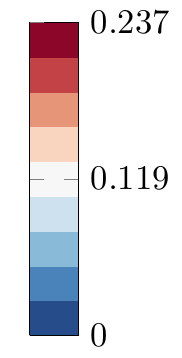}}
	%\raisebox{-2.7mm}{\colorbar{0}{0.237}{3.18}}
\end{subfigure}%

\caption{\rred{\textbf{Flow predictions on a B\'ezier shape from the test set using a GCNN model.} The horizontal velocity (\ref{fig:gcnn_u_ref}, \ref{fig:gcnn_u_pred}, \ref{fig:gcnn_u_err}), vertical velocity (\ref{fig:gcnn_v_ref}, \ref{fig:gcnn_v_pred}, \ref{fig:gcnn_v_err}) and pressure fields (\ref{fig:gcnn_p_ref}, \ref{fig:gcnn_p_pred}, \ref{fig:gcnn_p_err}) display similar error levels. The concerned GCNN has \num{217853} trainable parameters in total. The maximum absolute error is much smaller than the U-net method, with no specific pattern displayed.}}
\label{fig:predUnet}
\end{figure}
%%%%%%%%%%%%

%%%%%%%%%%%%
%%%%%%%%%%%%

%%%%%%%%%%%%%%%%%%%%%%%%%%%%%%%%%%%%%%%%%%%%%%%%%%%%%%%%%%
%%%%%%%%%%%%%%%%%%%%%%%%%%%%%%%%%%%%%%%%%%%%%%%%%%%%%%%%%%
\section{Conclusion}
\label{section:conclusion}

In the present contribution, a graph neural network architecture for 2D incompressible laminar flow prediction was presented. Compared to traditional convolution neural networks, the graph based method works directly on body-fitted triangular meshes, making it a method of choice for coupling with finite element numerical simulations. The proposed convolution method preserves permutation invariance, and is therefore insensitive to node and edge indexing. The GCNN network was trained on a dataset of 2000 laminar flows around random 2D shapes, generated using B\'ezier curves, resulting in an averaged MAE on the test set at \num{7.70e-3}, with no evidence of overfitting. The resulting model was proved to efficiently reconstruct velocity and pressure fields around a cylinder, respecting the symmetry induced by the input geometry. In particular, the predicted flow remained highly accurate even in the boundary layer of the obstacle. The predicted flow around a significantly different shape, namely a NACA0012 airfoil, presented higher MAE levels and ill-estimated amplitudes, while still respecting the physics of the reference flow. The predicted velocity and pressure fields were further used to compute the drag force by integration of the fields in the boundary layer, resulting in an average relative error equal to $3.43\%$ on the test set. Finally, the GCNN model performance was compared to that of a standard U-net model. The GCNN architecture proved its superiority in terms of accuracy and \red{model complexity}, although its required training time was significantly longer. Yet, the versatility of GCNN and its capabilities to work on body-fitted meshes with arbitrarily complex geometric characteristics make it a method of choice for the generation of accurate, fast prediction models coupled to computational fluid dynamics.

%%%%%%%%%%%%%%%%%%%%%%%%%%%
%%%%%%%%%%%%%%%%%%%%%%%%%%%
\section*{Acknowledgements} 
This work is supported by the Carnot M.I.N.E.S. Institute through the M.I.N.D.S. project.

%%%%%%%%%%%%%%%%%%%%%%%%%%%
%%%%%%%%%%%%%%%%%%%%%%%%%%%
\section*{Author declarations} 

%%%%%%%%%%%%%%%%%%%%%%%%%%%
\subsection*{Conflicts of interest} 

The authors have no conflicts to disclose.

%%%%%%%%%%%%%%%%%%%%%%%%%%%%%%%%%%%%%%%%%%%%%%%%%%%%%%%%%%
%%%%%%%%%%%%%%%%%%%%%%%%%%%%%%%%%%%%%%%%%%%%%%%%%%%%%%%%%%
%%%%%%%%%%%%%%%%%%%%%%%%%%%%%%%%%%%%%%%%%%%%%%%%%%%%%%%%%%
\appendix

%%%%%%%%%%%%%%%%%%%%%%%%%%%%%%%%%%%%%%%%%%%%%%%%%%%%%%%%%%
%%%%%%%%%%%%%%%%%%%%%%%%%%%%%%%%%%%%%%%%%%%%%%%%%%%%%%%%%%
%%%%%%%%%%%%%%%%%%%%%%%%%%%%%%%%%%%%%%%%%%%%%%%%%%%%%%%%%%
\section{Data availability statement}
\label{section:open_source}

The code of this project is available on the following github repository: \url{https://github.com/jviquerat/gnn_laminar_flow}\footnote{The code will be released upon publication of the present manuscript}.

%%%%%%%%%%%%%%%%%%%%%%%%%%%
%%%%%%%%%%%%%%%%%%%%%%%%%%%
%%%%%%%%%%%%%%%%%%%%%%%%%%%
\bibliographystyle{unsrt}
\bibliography{bib}

\begin{thebibliography}{10}

\bibitem{Guo2016}
X.~Guo, W.~Li, and F.~Iorio.
\newblock Convolutional neural networks for steady flow approximation.
\newblock In {\em 22nd ACM SIGKDD International Conference on Knowledge
  Discovery and Data Mining}, pages 481--490, 2016.

\bibitem{Jin2018}
X.~Jin, P.~Cheng, W.~L. Chen, and H.~Li.
\newblock Prediction model of velocity field around circular cylinder over
  various reynolds numbers by fusion convolutional neural networks based on
  pressure on the cylinder.
\newblock {\em Physics of Fluids}, 30(4), 2018.

\bibitem{Lee2019}
S.~Lee and D.~You.
\newblock Data-driven prediction of unsteady flow over a circular cylinder
  using deep learning.
\newblock {\em Journal of Fluid Mechanics}, 879:217–254, 2019.

\bibitem{chen2021twindecoder}
Junfeng Chen, Jonathan Viquerat, Frederic Heymes, and Elie Hachem.
\newblock A twin-decoder structure for incompressible laminar flow
  reconstruction with uncertainty estimation around 2d obstacles, 2021.

\bibitem{Thuerey2020}
N.~Thuerey, K.~Weißenow, L.~Prantl, and X.~Hu.
\newblock Deep learning methods for reynolds-averaged navier–stokes
  simulations of airfoil flows.
\newblock {\em AIAA Journal}, 58(1):25--36, 2020.

\bibitem{Fukami2018}
K.~Fukami, K.~Fukagata, and K.~Taira.
\newblock Super-resolution reconstruction of turbulent flows with machine
  learning.
\newblock {\em arXiv preprints}, 2018.

\bibitem{patil2021robust}
Aakash Patil, Jonathan Viquerat, George~El Haber, and Elie Hachem.
\newblock Robust deep learning for emulating turbulent viscosities, 2021.

\bibitem{Zhang2018}
Y.~Zhang, W.~J. Sung, and D.~N. Mavris.
\newblock Application of convolutional neural network to predict airfoil lift
  coefficient.
\newblock In {\em 2018 AIAA/ASCE/AHS/ASC Structures, Structural Dynamics, and
  Materials Conference}, 2018.

\bibitem{Viquerat2020}
J.~Viquerat and E.~Hachem.
\newblock A supervised neural network for drag prediction of arbitrary 2d
  shapes in laminar flows at low reynolds number.
\newblock {\em Computers \& Fluids}, 210:104645, 2020.

\bibitem{Lee2020}
K.~Lee and K.~T. Carlberg.
\newblock Model reduction of dynamical systems on nonlinear manifolds using
  deep convolutional autoencoders.
\newblock {\em Journal of Computational Physics}, 404:108973, 2020.

\bibitem{Bukka2020}
S.~R. Bukka, A.~R. Magee, and R.~K. Jaiman.
\newblock Deep convolutional recurrent autoencoders for flow field prediction,
  2020.

\bibitem{Gonzalez2018}
F.~J. Gonzalez and M.~Balajewicz.
\newblock Deep convolutional recurrent autoencoders for learning
  low-dimensional feature dynamics of fluid systems, 2018.

\bibitem{pathak2020using}
Jaideep Pathak, Mustafa Mustafa, Karthik Kashinath, Emmanuel Motheau, Thorsten
  Kurth, and Marcus Day.
\newblock Using machine learning to augment coarse-grid computational fluid
  dynamics simulations.
\newblock {\em arXiv preprint arXiv:2010.00072}, 2020.

\bibitem{kochkov2021machine}
Dmitrii Kochkov, Jamie~A Smith, Ayya Alieva, Qing Wang, Michael~P Brenner, and
  Stephan Hoyer.
\newblock Machine learning--accelerated computational fluid dynamics.
\newblock {\em Proceedings of the National Academy of Sciences}, 118(21), 2021.

\bibitem{messner2020convolutional}
Mark~C Messner.
\newblock Convolutional neural network surrogate models for the mechanical
  properties of periodic structures.
\newblock {\em Journal of Mechanical Design}, 142(2):024503, 2020.

\bibitem{LeCun1995-LECCNF}
Yann LeCun and Yoshua Bengio.
\newblock Convolutional networks for images, speech, and time series.
\newblock In Michael~A. Arbib, editor, {\em Handbook of Brain Theory and Neural
  Networks}, page 3361. MIT Press, 1995.

\bibitem{GAO2021110079}
Han Gao, Luning Sun, and Jian-Xun Wang.
\newblock Phygeonet: Physics-informed geometry-adaptive convolutional neural
  networks for solving parameterized steady-state pdes on irregular domain.
\newblock {\em Journal of Computational Physics}, 428:110079, 2021.

\bibitem{kipf2017semisupervised}
Thomas~N. Kipf and Max Welling.
\newblock Semi-supervised classification with graph convolutional networks,
  2017.

\bibitem{pmlr-v119-de-avila-belbute-peres20a}
Filipe De~Avila Belbute-Peres, Thomas Economon, and Zico Kolter.
\newblock Combining differentiable {PDE} solvers and graph neural networks for
  fluid flow prediction.
\newblock In Hal~Daumé III and Aarti Singh, editors, {\em Proceedings of the
  37th International Conference on Machine Learning}, volume 119 of {\em
  Proceedings of Machine Learning Research}, pages 2402--2411. PMLR, 13--18 Jul
  2020.

\bibitem{pmlr-v70-gilmer17a}
Justin Gilmer, Samuel~S. Schoenholz, Patrick~F. Riley, Oriol Vinyals, and
  George~E. Dahl.
\newblock Neural message passing for quantum chemistry.
\newblock In Doina Precup and Yee~Whye Teh, editors, {\em Proceedings of the
  34th International Conference on Machine Learning}, volume~70 of {\em
  Proceedings of Machine Learning Research}, pages 1263--1272. PMLR, 06--11 Aug
  2017.

\bibitem{ogoke2020graph}
Francis Ogoke, Kazem Meidani, Amirreza Hashemi, and Amir~Barati Farimani.
\newblock Graph convolutional neural networks for body force prediction, 2020.

\bibitem{NIPS2017_5dd9db5e}
Will Hamilton, Zhitao Ying, and Jure Leskovec.
\newblock Inductive representation learning on large graphs.
\newblock In I.~Guyon, U.~V. Luxburg, S.~Bengio, H.~Wallach, R.~Fergus,
  S.~Vishwanathan, and R.~Garnett, editors, {\em Advances in Neural Information
  Processing Systems}, volume~30. Curran Associates, Inc., 2017.

\bibitem{battaglia2018relational}
Peter~W. Battaglia, Jessica~B. Hamrick, Victor Bapst, Alvaro Sanchez-Gonzalez,
  Vinicius Zambaldi, Mateusz Malinowski, Andrea Tacchetti, David Raposo, Adam
  Santoro, Ryan Faulkner, Caglar Gulcehre, Francis Song, Andrew Ballard, Justin
  Gilmer, George Dahl, Ashish Vaswani, Kelsey Allen, Charles Nash, Victoria
  Langston, Chris Dyer, Nicolas Heess, Daan Wierstra, Pushmeet Kohli, Matt
  Botvinick, Oriol Vinyals, Yujia Li, and Razvan Pascanu.
\newblock Relational inductive biases, deep learning, and graph networks, 2018.

\bibitem{sanchezgonzalez2020learning}
Alvaro Sanchez-Gonzalez, Jonathan Godwin, Tobias Pfaff, Rex Ying, Jure
  Leskovec, and Peter~W. Battaglia.
\newblock Learning to simulate complex physics with graph networks, 2020.

\bibitem{pfaff2021learning}
Tobias Pfaff, Meire Fortunato, Alvaro Sanchez-Gonzalez, and Peter Battaglia.
\newblock Learning mesh-based simulation with graph networks.
\newblock In {\em International Conference on Learning Representations}, 2021.

\bibitem{8100059}
F.~{Monti}, D.~{Boscaini}, J.~{Masci}, E.~{Rodolà}, J.~{Svoboda}, and M.~M.
  {Bronstein}.
\newblock Geometric deep learning on graphs and manifolds using mixture model
  cnns.
\newblock In {\em 2017 IEEE Conference on Computer Vision and Pattern
  Recognition (CVPR)}, pages 5425--5434, 2017.

\bibitem{doi:10.1063/5.0044093}
Mengfei Xu, Shufang Song, Xuxiang Sun, and Weiwei Zhang.
\newblock A convolutional strategy on unstructured mesh for the adjoint vector
  modeling.
\newblock {\em Physics of Fluids}, 33(3):036115, 2021.

\bibitem{Bruchon2009}
J.~Bruchon, H.~Digonnet, and T.~Coupez.
\newblock Using a signed distance function for the simulation of metal forming
  processes: Formulation of the contact condition and mesh adaptation.
\newblock {\em International Journal for Numerical Methods in Engineering},
  78(8):980--1008, 2009.

\bibitem{Hachem2013}
E.~Hachem, S.~Feghali, R.~Codina, and T.~Coupez.
\newblock Immersed stress method for fluid structure interaction using
  anisotropic mesh adaptation.
\newblock {\em International Journal for Numerical Methods in Engineering},
  94:805--825, 2013.

\bibitem{Tezduyar1992}
T.E. Tezduyar, S.~Mittal, S.E. Ray, and R.~Shih.
\newblock Incompressible flow computations with stabilized bilinear and linear
  equal-order-interpolation velocity-pressure elements.
\newblock {\em Computer Methods in Applied Mechanics and Engineering},
  95(2):221--242, 1992.

\bibitem{Bazilevs2007}
Y.~Bazilevs, V.M. Calo, J.A. Cottrell, T.J.R. Hughes, A.~Reali, and
  G.~Scovazzi.
\newblock Variational multiscale residual-based turbulence modeling for large
  eddy simulation of incompressible flows.
\newblock {\em Computer Methods in Applied Mechanics and Engineering},
  197(1):173--201, 2007.

\bibitem{Takizawa2018}
K.~Takizawa, T.~E. Tezduyar, and Y.~Otoguro.
\newblock Stabilization and discontinuity-capturing parameters for space–time
  flow computations with finite element and isogeometric discretizations.
\newblock {\em Computational Mechanics}, 62(5):1169--1186, 2018.

\bibitem{Otoguro2019}
Y.~Otoguro, K.~Takizawa, T.~E. Tezduyar, K.~Nagaoka, R.~Avsar, and Y.~Zhang.
\newblock Space–time vms flow analysis of a turbocharger turbine with
  isogeometric discretization: computations with time-dependent and
  steady-inflow representations of the intake/exhaust cycle.
\newblock {\em Computational Mechanics}, 64(5):1403--1419, 2019.

\bibitem{Otoguro2020}
Y.~Otoguro, K.~Takizawa, and T.~E. Tezduyar.
\newblock Element length calculation in b-spline meshes for complex geometries.
\newblock {\em Computational Mechanics}, 65:1085--1103, 2020.

\bibitem{ramachandran2018}
Prajit Ramachandran, Barret Zoph, and Quoc~V. Le.
\newblock Searching for activation functions, 2018.

\bibitem{pmlr-v9-glorot10a}
Xavier Glorot and Yoshua Bengio.
\newblock Understanding the difficulty of training deep feedforward neural
  networks.
\newblock In Yee~Whye Teh and Mike Titterington, editors, {\em Proceedings of
  the Thirteenth International Conference on Artificial Intelligence and
  Statistics}, volume~9 of {\em Proceedings of Machine Learning Research},
  pages 249--256, Chia Laguna Resort, Sardinia, Italy, 13--15 May 2010. PMLR.

\bibitem{10.1007/978-3-030-61616-8_20}
Takato Otsuzuki, Hideaki Hayashi, Yuchen Zheng, and Seiichi Uchida.
\newblock Regularized pooling.
\newblock In {\em Artificial Neural Networks and Machine Learning -- ICANN
  2020}, pages 241--254, Cham, 2020. Springer International Publishing.

\bibitem{kingma2017adam}
Diederik~P. Kingma and Jimmy Ba.
\newblock Adam: A method for stochastic optimization, 2017.

\bibitem{UnetRonneberger}
Olaf Ronneberger, Philipp Fischer, and Thomas Brox.
\newblock U-net: Convolutional networks for biomedical image segmentation.
\newblock In {\em Medical Image Computing and Computer-Assisted Intervention --
  MICCAI 2015}, pages 234--241, 2015.

\end{thebibliography}

\end{document}